\documentclass[reprint, showpacs, preprintnumbers, amsmath, amssymb, aps, nofootinbib]{revtex4}

\pdfoutput=1

\usepackage{color}
\usepackage{graphicx}% Include figure files
\usepackage{dcolumn}% Align table columns on decimal point
\usepackage{overpic}
\usepackage{psfrag}
\usepackage{bm}% bold math

\def\ddd{{\partial}}

\def\00{\mathbf{0}}
\def\AAA{\mathbf{A}}
\def\BB{\mathbf{B}}
\def\CC{\mathbf{C}}

\def\s2{G_{opt}^2}
\def\DUs2{\bnabla_{\UU_c} \s2}
\def\EE{\mathbf{E}}

\def\KK{\mathbf{K}}
\def\LL{\mathbf{L}}
\def\MM{\mathbf{M}}
\def\NN{\mathbf{N}}

\def\Pa{P^\dag} 
\def\PP{\mathbf{P}}
\def\QQa{\mathbf{Q}^\dag}
\def\QQ{\mathbf{Q}}
\def\SSS{\mathbf{S}}
\def\Ua{U^\dag} 	
\def\UUa{\mathbf{U}^\dag}		
\def\UU{\mathbf{U}}
\def\Va{V^\dag}

\def\aa{\mathbf{a}}
\def\bb{\mathbf{b}}

\def\ev{\lambda}
\def\ff{\mathbf{f}}
\def\qq{\mathbf{q}}
\def\qqa{\mathbf{q}^\dag}

\def\uu{\mathbf{u}}
\def\uua{\mathbf{u}^\dag}

\def\xx{\mathbf{x}}
\def\yy{\mathbf{y}}
\def\zz{\mathbf{z}}
 
\def\nnu{\Rey^{-1}}

\def\cbz{\cos(\beta z)}
\def\sbz{\sin(\beta z)}
\def\c2bz{\cos(2 \beta z)}
\def\s2bz{\sin(2 \beta z)}

\def\be{\begin{equation}}
\def\ee{\end{equation}}

\providecommand\bnabla{\boldsymbol{\nabla}}
\providecommand\bcdot{\boldsymbol{\cdot}}

\newcommand{\ps}   [2]{\ensuremath{\left(\left. {#1} \, \right| \, {#2} \right) }}
\newcommand{\pps}   [2]{\ensuremath{\left(\left(\left. {#1} \, \right| \, {#2} \right)\right) }}
\newcommand{\psw}  [2]{\ensuremath{\left\langle\left. {#1} \, \right| \, {#2} \right\rangle }}
\newcommand{\ppsw}  [2]{\ensuremath{\left\langle\left\langle\left. {#1} \, \right| \, {#2} \right\rangle\right\rangle }}

\newcommand\Rey{\mbox{\textit{Re}}}

\begin{document}

%\preprint{APS/123-QED}

\title{Second-order sensitivity in the cylinder wake: \\
optimal spanwise-periodic wall actuation and wall deformation}

\author{E. Boujo}
\affiliation{LadHyX, UMR CNRS 7646, \'Ecole Polytechnique, 91128 Palaiseau, France}

\author{A. Fani}%
%\affiliation{%

\author{F. Gallaire}
\affiliation{
LFMI, \'Ecole Polytechnique F\'ed\'erale de Lausanne, CH-1015 Lausanne, Switzerland
}%

\date{June 2019}

%---------------------------------------
%---------------------------------------
%---------------------------------------
\begin{abstract}
Two-dimensional (2D) flows can be controlled efficiently using spanwise ``waviness'', i.e. a control  (e.g. wall blowing/suction or wall deformation) that is periodic in the spanwise direction. This study tackles the global linear stability of 2D flows subject to small-amplitude 3D spanwise-periodic control. Building on previous work for parallel flows, an adjoint method is proposed for computing the second-order sensitivity of eigenvalues. Since such control has indeed a zero net first-order (linear) effect, the second-order (quadratic) effect prevails. The sensitivity operator allows one (i)~to predict the effect of any control without actually computing the controlled flow, and (ii)~to compute the optimal control (and an orthogonal set of sub-optimal controls) for stabilization/destabilization or frequency modification. The proposed method  takes advantage of the very spanwise-periodic nature of the control to reduce computational complexity (from a fully 3D problem to a 2D problem). The approach is applied to the leading eigenvalue of the laminar flow around a circular cylinder, and two kinds of spanwise-harmonic control are explored: wall actuation via blowing/suction, and wall deformation. Decomposing the eigenvalue variation, it is found that the 3D contribution (from the spanwise-periodic first-order flow modification) is generally larger than the 2D contribution from the mean flow correction (spanwise-invariant second-order flow modification). Over a wide range of control spanwise wavenumber, the optimal control for flow stabilization is  
symmetric about the wake centerline, 
leading to varicose streaks in the cylinder wake. Analyzing the competition between amplification and stabilization shows that optimal varicose streaks are not significantly more amplified than sinuous streaks but have a stronger stabilizing effect. The optimal wall deformation induces a flow modification very similar to that induced by the optimal wall actuation. In general, spanwise and tangential actuation have a  small contribution to the optimal control, so normal-only actuation is a good trade-off between simplicity and effectiveness. Our method opens the way to the systematic design of optimal spanwise-periodic control for a variety of control objectives other than linear stability properties.
\end{abstract}

\pacs{Valid PACS appear here}% PACS, the Physics and Astronomy

%---------------------------------------
%---------------------------------------
%---------------------------------------
\maketitle

%
%----------------------------------------
%----------------------------------------
%----------------------------------------
\section{Introduction}

A large body of studies is devoted to the control the flow over bluff bodies, well known to produce a significant amount of mean aerodynamics drag as well as unsteady vortex shedding. 
Several approaches have been proposed, from open-loop control featuring either passive appendices (e.g. end plate, splitter plate, small cylinder or flexible tail) or actuating devices (e.g. plasma actuation or steady/unsteady base bleeding), to closed-loop control (e.g. via transverse motion or wall blowing/suction, all relying on an appropriate sensing of flow variables). 

We focus here on open-loop bluff-body flow control, and more specifically on spanwise waviness, i.e. the steady spanwise-periodic control of a nominally 2D flow. This type of control has proven efficient to suppress or attenuate vortex shedding in the wake of bluff bodies thanks to 3D perturbations created with devices ranging from wrapped helical cables \cite{Zdravkovich1981} to indentations of the trailing or leading edges (\cite{Tanner1972, TombazisBearman1997, BearmanOwen1998} and many others), even at high Reynolds numbers, as discussed by \cite{Choi08}. 
Similarly, azimuthally-periodic chevrons were shown to reduce low-frequency noise emitted by axisymmetric jets (e.g. \cite{Bridges2004,Zaman2011}).
This noise reduction was interpreted by \cite{Gudmundsson2007, Sinha2016} as a reduction in the growth rate of the Kelvin--Helmholtz instability.

Many other studies have investigated the effect of spanwise waviness on stability and/or aerodynamic performance in both laminar and turbulent regimes, for instance wavy circular cylinders \cite{Ahmed92, Ahmed93, Lee07, Lam08, Zhang2016}, twisted circular cylinders \cite{Kim16}, circular cylinders with wavy wall actuation \cite{Kim05, Rocco2015}, wavy rectangular cylinders \cite{Lam2012}, and wavy airfoils \cite{Lin2013,Serson2017}.

In these studies, the shape of the disturbance and its spanwise (or azimuthal) wavelength were chosen arbitrarily, or improved by iterative trial and error. 
More recently, inspired by the success of spanwise-periodic control in the context of streaky boundary layers, 
Del~Guercio, Cossu and Pujals \cite{DelGuercio2014globalcyl}  proposed to optimize for the spatial amplification of 3D disturbances in a nominally 2D cylinder wake, thereby targeting the largest possible flow modification. 
They observed, without however ensuring it, that this flow modification had a stabilizing effect on the leading eigenmode.

The optimal growth rate reduction in 2D flows by a 2D steady distributed force or boundary flow can be easily determined through the calculation of a linear sensitivity map, obtained with an adjoint formulation  \cite{Hill92AIAA, Marquet08cyl}. 
The scalar product of this linear sensitivity map with any arbitrary control yields the variation of the growth rate. 
It is well known that for the spanwise-periodic control of nominally 2D flows, this scalar product vanishes, and that, at leading order the variation of the eigenvalue with respect to the uncontrolled (2D) case depends quadratically upon the 3D modulation amplitude \cite{Hinch1991, Cossu14secondorder, Tammisola2014, Boujo15b}. 
This quadratic dependence on the control amplitude was first shown to apply for absolute growth rates in parallel wakes \cite{Hwang2013}, and later for temporal growth rates in parallel wakes \cite{DelGuercio2014parallel} and for the sensitivity of global modes in non-parallel wakes \cite{DelGuercio2014globalwake, DelGuercio2014globalcyl}. 

The second-order sensitivity tensor has been explicitly computed in \cite{Cossu14secondorder} for the Ginzburg-Landau equation, and in \cite{Tammisola2014} and \cite{Boujo15b} for parallel flows. 
More specifically, in our previous study \cite{Boujo15b} we computed optimal spanwise-periodic base flow modifications for parallel flows by rewriting and manipulating the second-order perturbation system into a Hessian matrix form. 
This enabled us to extract, from the Hessian's extremal eigenvalues, the most stabilizing and destabilizing flow modifications. 
These manipulations involved forming explicitly the inverse of a matrix, which was possible because the flow was parallel with 1D eigenmodes. 
In the case of 2D flows, like the flow around a circular cylinder, Tammisola \cite{Tammisola2017} proposed an algorithm to compute the second-order eigenvalue variation caused by a given spanwise-periodic perturbation by solving two consecutive linear problems.
This approach enabled her to optimize the most stabilizing wall-blowing by repeatedly applying the previous algorithm to a finite basis of functions spanning the cylinder wall. 

The objective of this paper is to generalize the framework of \cite{Boujo15b} to 2D flows and thereby obtain a continuous formulation of the second-order sensitivity tensor. 
When applied to wall actuation, this formulation reveals the superimposition of two contributions: 
(i)~from the interaction between the spanwise-periodic base flow modification and the spanwise-periodic eigenmode modification, 
and (ii)~from the spanwise-invariant base flow modification or ``mean flow correction'' \cite{Maurel95, Mantic2014, Marant2018} (see also sections \ref{sec:sensit_general} and \ref{sec:compet_3D_2D}).

The paper is organized as follows. 
Section \ref{sec:pb} describes the problem of spanwise-periodic control in non-parallel 2D flows, the formulation of the second-order eigenvalue sensitivity tensor, and the optimization procedure used to compute the optimal control. 
Numerical details are given in section \ref{sec:num}.
The method is illustrated with the spanwise-periodic control of the flow around a circular cylinder:
optimal wall blowing/suction and optimal wall deformation for stabilization are presented in sections \ref{sec:growth-wallactu}-\ref{sec:growth-walldef}, and optimal control for frequency modification in section \ref{sec:freq}.
Finally, section~\ref{sec:conclusion} summarizes the main conclusions.
Appendices~A-C provide technical details about second-order sensitivity, the specific case of spanwise-periodic control, and the optimization procedure.
Appendix~D briefly comments on results for spanwise-periodic volume control, drawing links with 2D volume control and with spanwise-periodic wall control.
Appendix~E discusses the validity of the results in the limit of small spanwise wavenumbers, based on a 3D linear stability analysis of the uncontrolled 2D flow.

%----------------------------------------
%----------------------------------------
%----------------------------------------
\section{Problem formulation}
\label{sec:pb}

%-----------------------------------------------
%-----------------------------------------------
\subsection{Uncontrolled flow}

We consider an incompressible flow, e.g.
the flow past a spanwise-infinite circular cylinder.
We use the notation $\QQ(x,y,z,t)=(\UU,P)^T$ for the flow state,
with $\UU=(U,V,W)^T$ the  velocity field of components $U$, $V$  and $W$ in the streamwise $x$, cross-stream $y$ and spanwise $z$ directions, and  $P$ the pressure field.
The flow is governed by the nonlinear Navier--Stokes (NS) equations in the domain $\Omega$ and no-slip boundary condition on the cylinder wall $\Gamma$,
\begin{align} 
& \partial_t \UU + \UU \bcdot \bnabla\UU + \bnabla  P - \nnu \bnabla^2  \UU = \00,
\quad
\bnabla \bcdot \UU = 0
\quad \mbox{ in } \Omega,
\\
& \UU=\00\quad \mbox{ on } \Gamma,
\end{align}
which we note in compact form 
\be  
\EE \, \partial_t  \QQ + \NN(\QQ) = \00
\ee
after introducing the  operator $\EE$ such that $\EE\,\QQ=(\UU,0)^T$, and the NS operator $\NN(\QQ)$.
In our example flow, the Reynolds number $\Rey=U_\infty D/\nu$ is defined with the cylinder diameter $D$, free-stream velocity $U_\infty$ and fluid kinematic viscosity $\nu$.

We focus in particular on the 
two-dimensional (2D) steady flow  
$\QQ(x,y)=(U,V,0,P)^T$  solution of the  steady NS equations
\begin{align} 
& \UU \bcdot \bnabla\UU + \bnabla  P - \nnu \bnabla^2  \UU = \00,
 \quad \bnabla \bcdot \UU = 0
\quad \mbox{ in } \Omega,
\\
& \UU=\00\quad \mbox{ on } \Gamma,
\label{eq:baseflow}
\end{align}
and hereafter denoted ``base flow''.
The dynamics of small-amplitude perturbations
$\qq'(x,y,z,t) = \qq(x,y,z) e^{\ev t} = (\uu,p)^T  e^{\ev t}$ 
are  governed by the linearized NS equations
\begin{align} 
& \ev \uu +  \UU \bcdot \bnabla\uu  + \uu \bcdot \bnabla\UU    +   \bnabla  p - \nnu \bnabla^2  \uu = \00,
\quad \bnabla \bcdot \uu = 0 
\quad \mbox{ in } \Omega,
\\
& \uu=\00\quad \mbox{ on } \Gamma,
\end{align}
which we  note in compact form as an eigenvalue problem,
\begin{align} 
(\ev\EE+\AAA) \qq&=\00,
\end{align}
after introducing  $\AAA$ the  NS operator linearized around the base flow.
The set of growth rates $\ev_r$ and frequencies $\ev_i$ determine the linear stability properties of the flow.
Specifically, the cylinder flow becomes unstable at $\Rey=47$ via a supercritical Hopf bifurcation: a pair of complex conjugate eigenvalues crosses the $\ev_r=0$ axis, and time invariance is broken by the shedding of 2D vortices at a frequency close to $\ev_i/(2\pi)=0.12$.
The flow remains 2D up to $\Rey=189$~\cite{barkley96}.

%
%-----------------------------------------------
%-----------------------------------------------
\subsection{Eigenvalue variation induced by spanwise-periodic control}
\label{sec:22}

%--- codes/sketch_wallctrl_walldef.m
\begin{figure}
\centerline{  
   \begin{overpic}[trim=25mm 15mm 28mm 25mm, clip, width=8.2cm,tics=10]{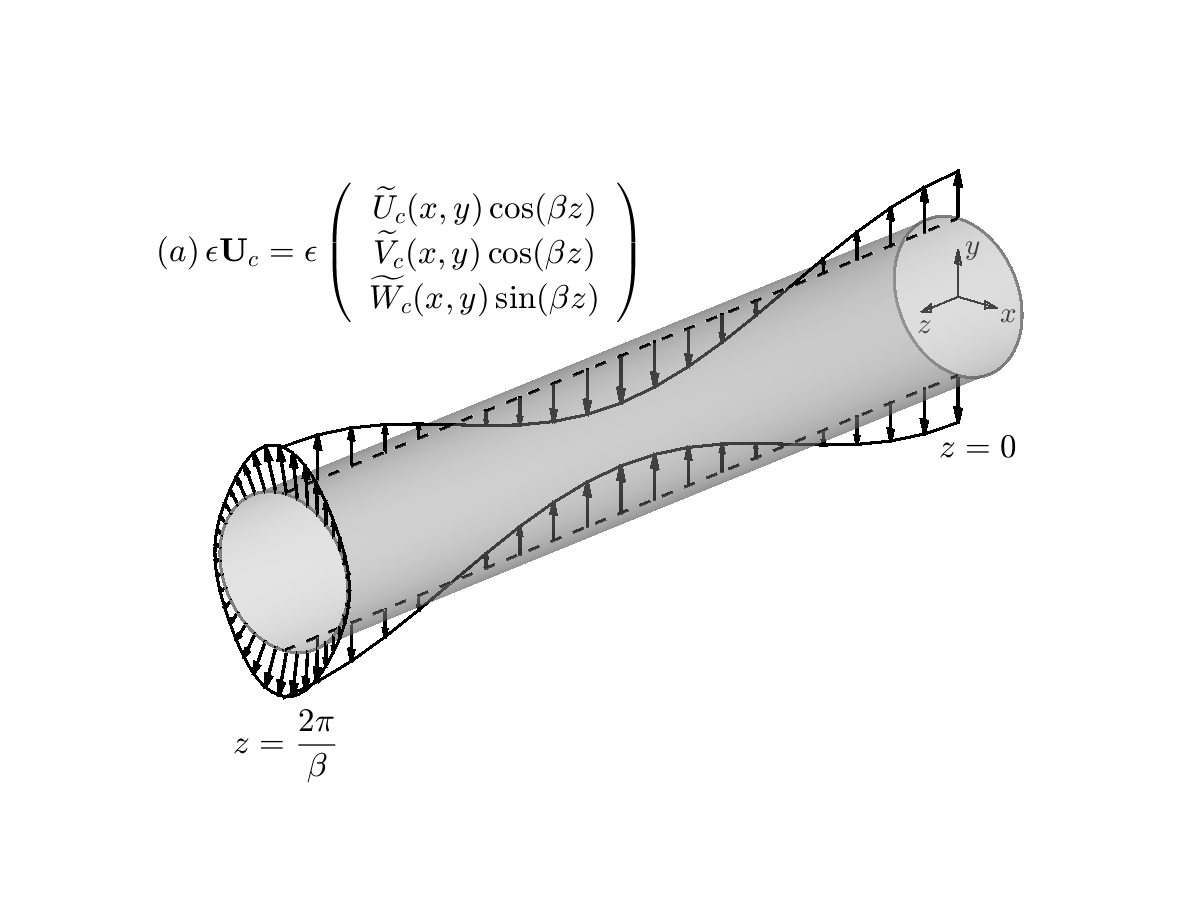}        
   \end{overpic}  
   \hspace{0.5cm}
   \begin{overpic}[trim=25mm 15mm 28mm 25mm, clip=true, width=8.2cm,tics=10]{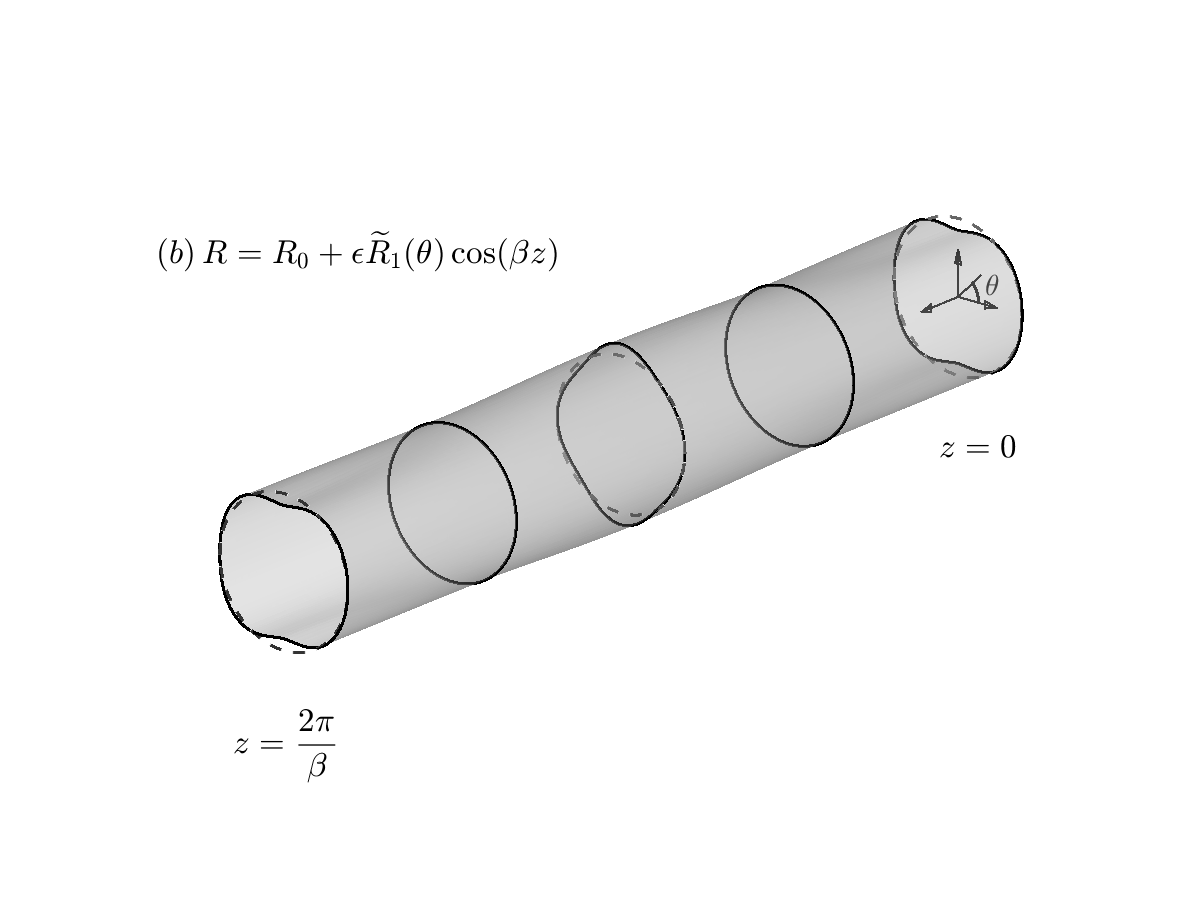}       
   \end{overpic}  
}
\caption{
Sketch of the steady spanwise-periodic control considered in this study:
$(a)$~wall blowing/suction,
$(b)$~wall deformation.
}  
\label{fig:sketch_control}
\end{figure}

We now assume that a small-amplitude steady control is applied to the flow, by means of a volume force 
$\CC(x,y,z)=(C_x,C_y,C_z)^T$ in the domain, and 
blowing/suction $\UU_c(x,y,z)=(U_{c},V_{c},W_{c})^T$ at the cylinder wall (see sketch in Fig.~\ref{fig:sketch_control}$(a)$):
\begin{align}
&	   \UU \bcdot  \bnabla\UU + \bnabla  P - \nnu \bnabla^2  \UU = \epsilon\CC,
\quad \bnabla \bcdot \UU = 0 \quad \mbox{ in } \Omega
\label{eq:total_controlled_BF_mom}
\\
& \UU = \epsilon \UU_c \quad \mbox{ on } \Gamma.
\label{eq:total_controlled_BF_div}
\end{align}
The effect of the control is a modification of the  base flow,  eigenvectors and eigenvalues, which can be expressed with a power series expansion in the small amplitude $\epsilon$:
\begin{align} 
& \QQ = \QQ_0+\epsilon\QQ_1+\epsilon^2\QQ_2+\ldots,
\label{eq:exp_Q}
\\
& \qq = \qq_0+\epsilon\qq_1+\epsilon^2\qq_2+\ldots,
\label{eq:exp_q}
\\
& \ev = \ev_0+\epsilon\ev_1+\epsilon^2\ev_2+\ldots.
%\label{eq:exp_ev}
\label{eq:exp}
\end{align} 
We are interested in particular in the eigenvalue variation induced by the control.

We first observe that the base flow at order $\epsilon^0$, $\epsilon^1$ and $\epsilon^2$ is solution of
\begin{align} 
& \NN(\QQ_0) =\00 \qquad \qquad \qquad \quad\,\,\,  \mbox{ in } \Omega,
\qquad
 \UU_0=\00 \quad \,\,\,\,  \mbox{ on } \Gamma,
   \label{eq:Q0}
\\
& \AAA_0 \QQ_1 = (\CC,0)^T \qquad \qquad  \quad\,\, \mbox{ in } \Omega,
\qquad
 \UU_1=\UU_c \quad \mbox{ on } \Gamma, 
   \label{eq:Q1}
\\
& \AAA_0 \QQ_2 = (- \UU_1 \bcdot \bnabla \UU_1,0)^T  \quad \mbox{ in } \Omega,
\qquad
 \UU_2=\00 \quad \,\,\,\, \mbox{ on } \Gamma,
\label{eq:Q2}
\end{align}
where $\AAA_0$ is the NS operator linearized around the uncontrolled base flow $\QQ_0$.
Note that the control modifies the base flow  at all orders $\epsilon^n$, $n\geq1$, due to forcing terms similar to 
$-\UU_1 \bcdot \bnabla \UU_1$ in (\ref{eq:Q2}) for $n=2$. 
This contrasts with the case of a prescribed base flow modification (without control), where $\QQ$ is exactly equal to $\QQ_0+\epsilon\QQ_1$.

%--- plot_3D_isosurface.m
\begin{figure}
\centerline{  
   \begin{overpic}[trim=55mm 49mm 42mm 33mm, clip, width=9cm,tics=10]{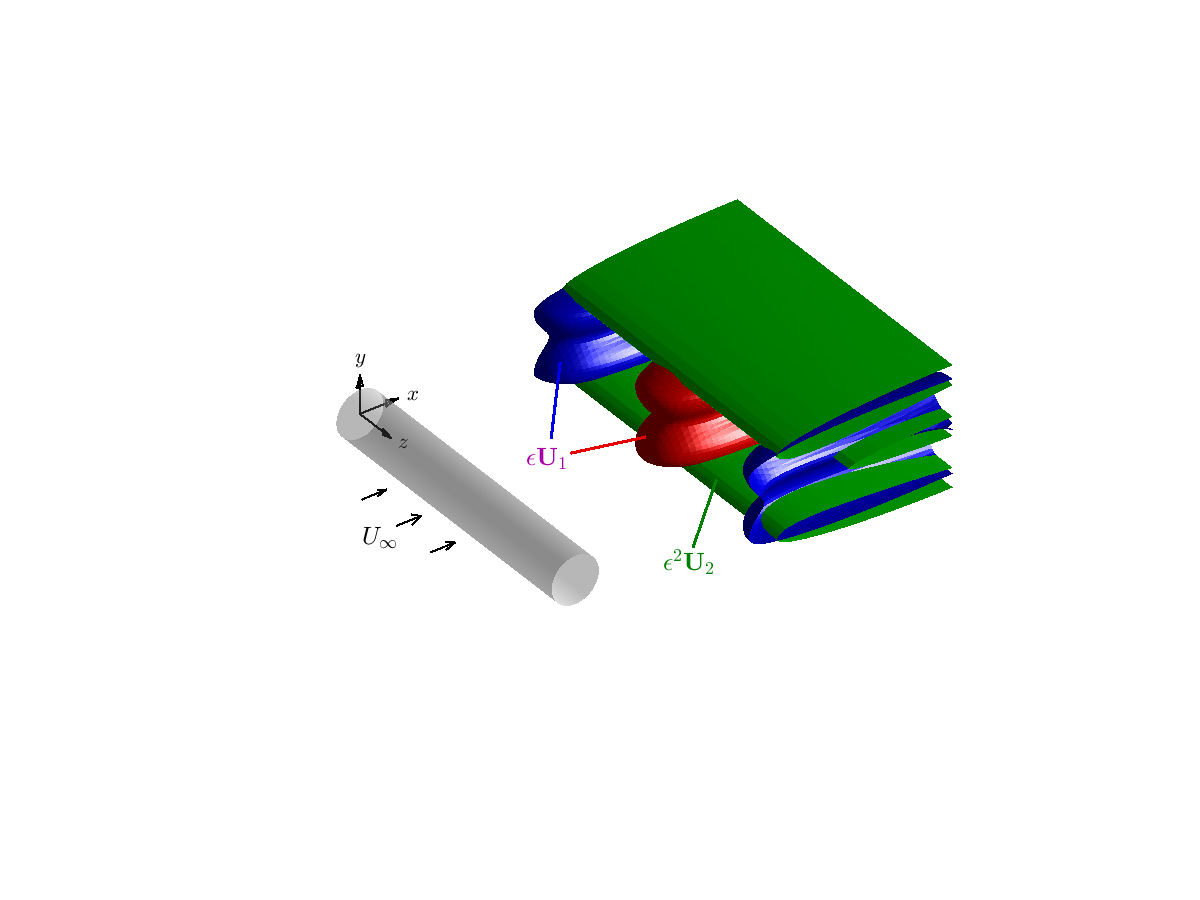}   
   \end{overpic}  
}
\caption{
Sketch of the flow modification (velocity isosurfaces) induced by a steady spanwise-periodic control (Fig.~\ref{fig:sketch_control}):
first-order spanwise-periodic modification (red and blue; here corresponding to low- and high-speed streamwise streaks),
and spanwise-invariant second-order modification (mean flow correction, green).
(The spanwise-periodic component of the second-order modification is not shown.)
The net linear effect of $\UU_1$ on the  eigenvalue is zero, $\epsilon\lambda_1=0$.
By contrast, $\UU_1$ and $\UU_2$ have a non-zero quadratic effect $\epsilon^2\lambda_2$.
}  
\label{fig:sketch_U1_U2}
\end{figure}

Next, we turn our attention to the eigenvalue problem. At leading order $\epsilon^0$, we obtain
\begin{align} 
(\ev_0\EE+\AAA_0) \qq_0&=\00.
  \label{eq:evp0}
\end{align}
We focus on the leading (most unstable) eigenmode 
$\qq_0(x,y)=(u_0,v_0,0,p_0)^T$, which is two dimensional, like $\QQ_0$.
At first and second orders $\epsilon^1$, $\epsilon^2$, we obtain 
\begin{align} 
& (\ev_0\EE+\AAA_0) \qq_1 = -(\ev_1\EE+\AAA_1) \qq_0,
  \label{eq:evp1}
\\
& (\ev_0\EE+\AAA_0) \qq_2 =
-(\ev_1\EE+\AAA_1) \qq_1 
-(\ev_2\EE+\AAA_2) \qq_0,
 \label{eq:evp2}
\end{align}
where the linear operators $\AAA_1$ and $\AAA_2$ depend only on $\UU_1$ and $\UU_2$, respectively (see appendix~A).
We introduce 2D and 3D Hermitian inner products in the domain and on the wall
\begin{align}
&\displaystyle \ps{\aa}{\bb} = 
\iint
\overline\aa\bcdot\bb \,\mathrm{d}x \mathrm{d}y,
&\displaystyle \pps{\aa}{\bb} = \lim_{L_z \rightarrow \infty} 
\frac{1}{L_z}
\int_{-L_z/2}^{L_z/2} \ps{\aa}{\bb} \,\mathrm{d}z,
\\ %-------------------------------------
&\displaystyle \psw{\aa}{\bb} = 
\int_\Gamma
\overline\aa\bcdot\bb \,\mathrm{d}\Gamma,
&\displaystyle \ppsw{\aa}{\bb} = \lim_{L_z \rightarrow \infty} 
\frac{1}{L_z}
\int_{-L_z/2}^{L_z/2} \psw{\aa}{\bb} \,\mathrm{d}z,
\end{align}
where the overbar stands for the conjugate of a complex quantity.
In all cases, the induced norm will be denoted $||\cdot||$.
We denote $\BB^\dag$ the adjoint operator of an operator $\BB$  such that
\begin{align}
\pps{\aa}{\BB\bb}=\pps{\BB^\dag\aa}{\bb} \quad \forall\,\aa,\bb.
\label{eq:adjoint}
\end{align}
For instance, the adjoint NS operator is defined by 
$\AAA_0^\dag \qq_0^\dag = \uua_0\bcdot\bnabla\UU_0^T - \UU_0\bcdot\bnabla\uua_0 -\bnabla p_0^\dag - \nnu \bnabla^2 \uua_0$.
Projecting (\ref{eq:evp1}) on the leading adjoint eigenmode $\qqa_0$ 
solution of
\be 
(\overline\ev_0 \EE + \AAA_0^\dag) \qq_0^\dag = \00
 \label{eq:evpadj}
\ee
and normalized according to $\ps{ {\bf q}_0^\dag } {\EE {\bf q}_0}=1$,
allows us to obtain the first-order eigenvalue variation \cite{Hinch1991, Trefethen93, Chomaz05, Giannetti07}:
\be 
 \lambda_1  =  \pps{ {\bf q}_0^\dag } {- \AAA_1 {\bf q}_0}.
 \label{eq:ev1}
\ee
Although we will never need to compute it explicitly,  we note that the first-order eigenmode modification can be expressed from (\ref{eq:evp1}) as 
\be 
\qq_1 = -(\ev_0\EE+\AAA_0)^{-1} (\ev_1\EE+\AAA_1) \qq_0.
 \label{eq:q1}
\ee
In general the operator $(\ev_0\EE+\AAA_0)$  is not invertible  since (\ref{eq:evp0}) has a non-trivial solution, but the inverse is taken in the subspace orthogonal to $\qq_0$, and $\qq_1$ is defined up to any constant component in the direction of $\qq_0$ \cite{Hinch1991}.
This is made possible by the solvability condition (Fredholm theorem) to be satisfied by (\ref{eq:evp1}):
the forcing term $(\lambda_1 {\bf E} +{\bf A}_1){\bf q}_0$ is orthogonal to the solution $\qqa_0$ of  the  adjoint equation (\ref{eq:evpadj})
associated with (\ref{eq:evp0}), as  expressed precisely by (\ref{eq:ev1}).
The second-order eigenvalue variation is obtained in a similar way: 
\be 
\ev_2 
= \pps{ \qq_0^\dag }{-\AAA_2\qq_0} 
+ \pps{ \qq_0^\dag }{(\ev_1\EE+\AAA_1)(\ev_0\EE+\AAA_0)^{-1}(\ev_1\EE+\AAA_1) \qq_0}.
  \label{eq:ev2}
\ee

In this study we focus on spanwise-periodic control. In this  case, the first-order flow modification $\QQ_1$ and  the operator ${\bf A}_1$ are periodic in $z$ (see Fig.~\ref{fig:sketch_U1_U2}), therefore the inner product (\ref{eq:ev1}) vanishes 
(an effect of averaging in the $z$ direction)
and
the first-order eigenvalue variation is zero,
 $\ev_1=0$. 
In other words, similar to spanwise-periodic flow modification \cite{Hwang2013, DelGuercio2014parallel, Cossu14secondorder}, spanwise-periodic control has no first-order effect on stability properties.
Accordingly, the second-order variation simplifies to 
\be 
\ev_2 
= \pps{ \qq_0^\dag }{-\AAA_2\qq_0} 
+ \pps{ \qq_0^\dag }{\AAA_1(\ev_0\EE+\AAA_0)^{-1} \AAA_1 \qq_0}.
  \label{eq:ev2}
\ee 

The second term is similar to the expression obtained for parallel flows \cite{Boujo15b} and more generally for prescribed base flow modification (without control).
The first term results from the second-order flow modification $\QQ_2$ induced by the control, as recently revealed in the different context of time-dependent parallel flows by Marant and Cossu \cite{Marant2018}. 
It turns out that the spanwise-periodic component of $\QQ_2$ has no effect on $\ev_2$, due to the above-mentioned averaging effect. 
However, $\QQ_2$ also contains a spanwise-invariant component that does have an effect on $\ev_2$ (see Fig.~\ref{fig:sketch_U1_U2} and appendix~B), and therefore the first term in (\ref{eq:ev2}) must be retained.
This term, which corresponds to a mean flow correction \cite{Maurel95, Mantic2014, Marant2018}, does not seem to have been considered in \cite{Tammisola2017}. 
Its influence is systematically evaluated  in the following sections.

%-----------------------------------------------
%-----------------------------------------------
\subsection{Second-order sensitivity}
\label{sec:sensit}

%-----------------------------------------------
\subsubsection{General expression}
\label{sec:sensit_general}

The expression of the second-order eigenvalue variation (\ref{eq:ev2}) is useful to highlight the contributions of the first-order and second-order flow modifications.
It is also of practical use for evaluating the effect of a given control since it does not require solving the linear stability of the controlled flow
or even computing the eigenmode modification $\qq_1$. However, it requires solving for the flow modifications $\QQ_1$ and $\QQ_2$. 
A more useful alternative lies in sensitivity operators $\SSS_{2,*}$ that allow one to evaluate $\ev_2$ directly for any volume control or wall control with simple inner products:
\begin{align}
\ev_2 = \pps{\CC}{\SSS_{2,\CC} \CC}
+ \ppsw{\UU_c}{\SSS_{2,\UU_c} \UU_c}.
\end{align}

A series of manipulations (detailed in appendix~A) leads to the explicit expressions of such second-order sensitivity operators:
\begin{align}
\SSS_{2,\CC} &=  \PP^T {\AAA_{0,\CC}^\dag}^{-1} \SSS_{2,\QQ_1} {\AAA_{0,\CC}}^{-1} \PP,
\label{eq:S2C_0}
\\
\SSS_{2,\UU_c} &=  \PP^T {\AAA_{0,\UU_c}^\dag}^{-1} \SSS_{2,\QQ_1} {\AAA_{0,\UU_c}}^{-1} \PP.
\label{eq:S2Uc_0}
\end{align}
In the expressions, $\AAA_{0,\CC}$ and $\AAA_{0,\UU_c}$ are defined by the wall-actuation-only and volume-control-only versions of (\ref{eq:Q1}), respectively:
\begin{align}
& \AAA_{0,\CC} \QQ_1 = (\CC,0)^T 
 \, \mbox{ in } \Omega,
\qquad
 \UU_1=\00 \,\,\,\, \mbox{ on } \Gamma,
\\
& \AAA_{0,\UU_c} \QQ_1 = \00 \,\,  \qquad \mbox{ in } \Omega,
\qquad
\UU_1=\UU_c \, \mbox{ on } \Gamma;
\end{align}
$\PP$ is the prolongation operator to velocity-pressure space from velocity-only space (all three velocity components in the most general case, or one or two components when considering specific wall controls; see e.g. section~\ref{sec:simplified_wall});
and
$\SSS_{2,\QQ_1}$ is the second-order sensitivity to flow modification such that $\ev_2 = \pps{\QQ_1}{\SSS_{2,\QQ_1} \QQ_1}$:
\begin{align}
\SSS_{2,\QQ_1} &= 
\KK
+
\MM^\dag (\ev_0\EE+\AAA_0)^{-1} \LL,
\end{align}
where $\KK$, $\LL$ and $\MM$ only depend on the direct and global eigenmodes $\qq_0$, $\qqa_0$ of the uncontrolled flow.
The second term is similar to the second-order sensitivity operator for parallel flows \cite{Boujo15b},
whereas the first term is new and results from the control-induced second-order flow modification $\QQ_2$ or, more specifically, from the spanwise-invariant component of $\QQ_2$ (mean flow correction: see Fig.~\ref{fig:sketch_U1_U2}).

%-----------------------------------------------
\subsubsection{Spanwise-periodic expression}
\label{sec:sensit_spanwise_periodic}

At this stage, the operators (\ref{eq:S2C_0})-(\ref{eq:S2Uc_0}) do not depend on control-specific flow modifications $\QQ_1$, $\QQ_2$; in principle, they can therefore be computed once for all.
However, they still depend on the spanwise coordinate $z$.
We now derive \textit{reduced} $z$-independent, yet exact, expressions of the sensitivity operators. 
As shown below, this  makes it possible to  evaluate the eigenvalue variation $\ev_2$ and determine the optimal spanwise-periodic controls $\CC$ and $\UU_c$ using only 2D fields, making these operations significantly more computationally affordable than with 3D fields.

As detailed in appendix~B, we consider the following harmonic wall forcing on $\Gamma$ and harmonic volume forcing in $\Omega$:
\be 
\UU_c(x,y,z) =  \left( \begin{array}{c}
\widetilde U_{c}(x,y) \cbz \\ \widetilde V_{c}(x,y) \cbz \\ \widetilde W_{c}(x,y) \sbz
\end{array} \right),
\quad
\CC(x,y,z) =  \left( \begin{array}{c}
\widetilde C_x(x,y) \cbz \\ \widetilde C_y(x,y) \cbz \\ \widetilde C_z(x,y) \sbz 
\end{array} \right).
\label{eq:harm_U_C}
\ee
With this control, the first-order flow modification is also spanwise-harmonic, of same wavenumber:
\begin{align}
\QQ_1=
\left(\begin{array}{c}
\widetilde U_1(x,y) \cbz \\
\widetilde V_1(x,y) \cbz \\
\widetilde W_1(x,y) \sbz \\
\widetilde P_1(x,y) \cbz
\end{array}\right).
\end{align}
Therefore, the forcing term  
$- \UU_1 \bcdot \bnabla\UU_1^T$ in (\ref{eq:Q2}) is the sum of a 2D term (wavenumber 0) and of a 3D term (wavenumber $2\beta$).
In turn, the second-order flow modification is the sum 
\begin{align}
\QQ_2 &= \QQ_2^{2D}(x,y) + \QQ_2^{3D}(x,y,z)
\end{align}
of a 3D spanwise-periodic component $\QQ_2^{3D}(x,y,z)$ of wavenumber $2\beta$ that does not contribute to $\ev_2$, and of a 2D spanwise-invariant component $\QQ_2^{2D}(x,y)=(U^{2D},V^{2D},0,P^{2D})^T$, the mean flow correction.
Taking advantage of the specific form of $\QQ_1$ and $\QQ_2$ allows us to simplify the sensitivity operators:
\begin{align}
\widetilde \SSS_{2,\widetilde\QQ_1} &= 
\widetilde \KK
+
\widetilde{\MM}^\dag (\ev_0\EE+\widetilde\AAA_0)^{-1} \widetilde\LL,
\\
\widetilde \SSS_{2,\widetilde\CC} &= 
\PP^T 
\left. \widetilde{\AAA}^\dag_{0,\CC} \right.^{-1} 
\widetilde \SSS_{2,\widetilde\QQ_1} 
\left. \widetilde{\AAA}_{0,\CC} \right.^{-1} 
\PP,
\label{eq:S2C_red_0}
\\
\widetilde \SSS_{2,\widetilde\UU_c} 
&=
\PP^T 
\left. \widetilde{\AAA}_{0,\UU_c}^\dag \right.^{-1} \widetilde \SSS_{2,\widetilde\QQ_1} 
\left. \widetilde{\AAA}_{0,\UU_c} \right.^{-1} 
\PP,
\label{eq:S2Uc_red_0}
\end{align}
where all the operators involved are purely $z$-independent versions of their 3D counterparts.
The eigenvalue modification induced by any spanwise-periodic control can now be evaluated with simple 2D inner products:
\begin{align}
\ev_2 = \ps{\widetilde \CC}{\widetilde\SSS_{2,\widetilde\CC} \widetilde\CC} +
\psw{\widetilde \UU_c}{\widetilde \SSS_{2,\widetilde\UU_c} \widetilde \UU_c}.
\end{align}

%-----------------------------------------------
%-----------------------------------------------
\subsection{Optimal spanwise-periodic control}

Second-order sensitivity operators are useful not only for predicting the eigenvalue variation induced by a \textit{specific} control, but also for computing the \textit{optimal} control, i.e. the control that induces the largest eigenvalue variation (in a broad sense, i.e. the largest increase or decrease of the eigenmode's growth rate $\ev_{2r}$ or frequency $\ev_{2i}$).
Section~\ref{sec:opt_wall_act} presents the method for computing optimal volume control and wall control (blowing/suction).
Section~\ref{sec:opt_wall_def} explains how the method can be slightly modified for  optimal wall deformation.

Note that in this study we focus on optimal wall blowing/suction and optimal wall deformation, motivated by their ease of implementation and by the vast existing body of literature.
For the sake of completeness, some results for optimal volume control are briefly mentioned in appendix~D, drawing interesting links with 2D volume control and with spanwise-periodic wall control.

%-----------------------------------------------
\subsubsection{Optimal spanwise-periodic volume  and wall control}
\label{sec:opt_wall_act}

Given the expressions of the second-order sensitivity operators (\ref{eq:S2C_red_0})-(\ref{eq:S2Uc_red_0}),
the optimal unit-norm control is defined by the maximization problem
\begin{align}
\max_{||\widetilde \CC||=1} (\ev_{2r}) 
&=  
\max_{||\widetilde\CC||=1}  
\ps{\widetilde\CC}{ \frac{1}{2} 
\left( \widetilde\SSS_{2,\widetilde\CC,r}
 +\widetilde\SSS_{2,\widetilde\CC,r}^T \right) \widetilde\CC}
=
\max_{\widetilde\CC}  
\dfrac{ 
\ps{\widetilde\CC}{ \frac{1}{2} 
\left( \widetilde\SSS_{2,\widetilde\CC,r}
+\widetilde\SSS_{2,\widetilde\CC,r}^T \right) \widetilde\CC} 
}
{ \ps{\widetilde\CC}{\widetilde\CC} }
\nonumber
\\
&=  
\ev_{max} \left\{ 
\frac{1}{2}
\left( \widetilde\SSS_{2,\widetilde\CC,r}
 +\widetilde\SSS_{2,\widetilde\CC,r}^T \right)
\right\},
\label{eq:maxlam2rC}
\\
\max_{||\widetilde\UU_c||=1} (\ev_{2r}) 
&= 
\max_{||\widetilde\UU_c||=1}  \psw{\widetilde\UU_c}
{ \frac{1}{2}  
\left( \widetilde\SSS_{2,\widetilde\UU_c,r}
+\widetilde\SSS_{2,\widetilde\UU_c,r}^T \right) \widetilde\UU_c}
= 
\max_{\widetilde\UU_c}  
\dfrac{
\psw{\widetilde\UU_c}
{ \frac{1}{2}  
\left( \widetilde\SSS_{2,\widetilde\UU_c,r}
+\widetilde\SSS_{2,\widetilde\UU_c,r}^T \right) \widetilde\UU_c}
}
{ \psw{\widetilde\UU_c}{\widetilde\UU_c} }
\nonumber
\\
&= \ev_{max} \left\{  \frac{1}{2} 
\left( \widetilde\SSS_{2,\widetilde\UU_c,r}
 +\widetilde\SSS_{2,\widetilde\UU_c,r}^T \right)
 \right\}, 
\label{eq:maxlam2rUc}
\end{align}
where $\widetilde \SSS_{2,*,r}$ 
and   $\widetilde \SSS_{2,*,i}$ stand for the real and imaginary part of $\widetilde \SSS_{2,*}$, respectively.
Similar expressions hold for 
$\min(\ev_{2r})$, 
$\max(\ev_{2i})$, and
$\min(\ev_{2i})$.
In each case,  the last equality comes from the operators
$\widetilde\SSS_{2,*,r}+\widetilde\SSS_{2,*,r}^T$ and 
$\widetilde\SSS_{2,*,i}+\widetilde\SSS_{2,*,i}^T$ being real symmetric, so that the Rayleigh quotient is maximal for the largest eigenvalue (resp. minimal for the smallest eigenvalue). Thus, the maximization can be solved as an eigenvalue problem (see also appendix~C).
The optimal control $\widetilde\CC$ or $\widetilde\UU_c$ corresponding to the largest (resp. smallest) eigenvalue variation $\ev_{2r}$ or $\ev_{2i}$
is the eigenvector associated with $\ev_{max}$ (resp. $\ev_{min}$).
In addition to the largest (resp. smallest) eigenvalue, solving for the 
2$^{nd}$, 3$^{rd}$, $\ldots$ $k^{th}$ 
largest (resp. smallest) eigenvalues 
yields an orthogonal set of optimal controls.

At this point, two important differences with  optimal 2D control should be stressed (see also \cite{Boujo15b}).
Recall that for spanwise-invariant control, sensitivity operators are fields defined by 
$\ev_1 = \ps{  \SSS_{1, \CC}  }{ \CC}
       + \psw{ \SSS_{1, \UU_c}}{ \UU_c}$ 
\cite{Hill92AIAA, Marquet08cyl}. 
One can show that the optimal control for flow destabilization, for instance, is the real part of the sensitivity field itself;
conversely, the optimal control for flow stabilization is \textit{minus} the real part of the sensitivity field.
The situation is different for spanwise-periodic control. 
First, sensitivity operators are not fields but tensors. 
Second, changing the sign of a spanwise-periodic control does not change the eigenvalue variation (because the latter is quadratic in $\widetilde \CC$ and $\widetilde \UU_c$; or alternatively because changing the sign of the control is equivalent to shifting  the $z$ origin by $\pi/\beta$). 
This shows that the optimal controls associated with $\max(\ev_{2r})$ and $\min(\ev_{2r})$ (or with $\max(\ev_{2i})$ and $\min(\ev_{2i})$) are \textit{not} simply of opposite signs but must necessarily have different spatial structures.

%-----------------------------------------------
\subsubsection{Optimal spanwise-periodic wall deformation}

\label{sec:opt_wall_def}

The formalism introduced so far allows one to compute the optimal wall control for modifying the growth rate or the frequency of a global eigenmode.
A minor modification allows one to compute the optimal wall deformation as well, as outlined in this section.
This is highly relevant to open-loop control applications, where passive strategies (such as shape modification) may be preferred to active strategies (such as wall blowing/suction).

Consider a small-amplitude deformation of the cylinder [see sketch in Fig.~\ref{fig:sketch_control}$(b)$].
If at any spanwise location $z$ we denote the  undeformed radius $R_0$ and the deformed radius $R(\theta)=R_0+\epsilon R_1(\theta)$, the no-slip condition at the deformed wall $\UU(R)=\00$ can be Taylor expanded in $\UU$ and $R$: 
\begin{align}
\UU(R_0+\epsilon R_1) &= \UU_0(R_0+\epsilon R_1) + \epsilon \UU_1(R_0+\epsilon R_1) + \ldots 
\nonumber
\\
&= \UU_0(R_0) 
+ \epsilon \left[
 R_1 \left.\partial_R \UU_0\right|_{R_0} +  \UU_1(R_0) 
 \right] + \ldots = \00,
 \label{eq:taylor}
\end{align}
where $\left.\partial_R \UU_0\right|_{R_0} = \left.\partial_n \UU_0\right|_{R_0}$ is the normal derivative of the original flow at the undeformed wall (``flattened'' boundary condition).
Since the original flow satisfies the no-slip condition $\UU_0(R_0)=\00$, it follows at first order:
\begin{align}
 R_1 \left.\partial_n \UU_0\right|_{R_0} +  \UU_1(R_0) = \00.
\label{eq:flattenedBC}
\end{align}

From this relation, it is straightforward to obtain the flow modification $\UU_1(R_0)$ induced at the wall by a given deformation $R_1(\theta)$.
Here we are interested in the inverse problem: is there a wall deformation $R_1(\theta)$ equivalent to a given wall actuation $\UU_1(R_0)$? 
The answer is not obvious at first glance since, at each azimuthal location $\theta$, (\ref{eq:flattenedBC}) is  an over-determined system of three equations for one single unknown $R_1$.
However, we note that 
(i)~$U_{0z}=0$ everywhere since the unperturbed flow is 2D, and thus 
$\left.\partial_n U_{0z}\right|_{R_0}=0$;
(ii)~continuity ensures that 
$\left.\partial_n U_{0n}\right|_{R_0} = 0$\footnote{For instance, the continuity equation in cylindrical coordinates,
$\partial_r U_{0r} + U_{0r}/r + \partial_\theta U_{0\theta} / r + \partial_z U_{0z} = 0$, reduces to $\partial_r U_{0r} = 0$ at the wall owing to the no-slip condition, to the wall radius $R_0$ being constant, and to the unperturbed flow being 2D.}; 
so finally:
\begin{align}
U_{1n}(R_0)=0, 
\quad
R_1 \left.\partial_n U_{0t}\right|_{R_0} + U_{1t}(R_0)=0, \label{eq:U1ntz}
\quad
U_{1z}(R_0)=0.
\end{align}
We  therefore observe that, at first order, a wall deformation $R_1(\theta)$ induces at the original wall a purely tangential velocity $U_{1t}$ and no normal velocity:
\begin{align}
U_{1t}(R_0) = -R_1 \left.\partial_n U_{0t}\right|_{R_0}.
\label{eq:U1t}
\end{align}
Conversely, a given tangential wall actuation $U_{1t}$ is equivalent, at first order, to a well-defined wall deformation $R_1$.

Given the equivalence (\ref{eq:U1t}) between wall deformation and tangential wall blowing/suction,
for a spanwise-periodic wall deformation
$R=R_0+\epsilon \widetilde R_1(\theta) \cbz$
the sensitivity to wall deformation can be defined as 
\begin{align}
\widetilde \SSS_{2,\widetilde R_1} 
&=
\PP_R^T 
\left. \widetilde{\AAA}_{0,\UU_c}^\dag \right.^{-1} \widetilde \SSS_{2,\widetilde\QQ_1} 
\left. \widetilde{\AAA}_{0,\UU_c} \right.^{-1} 
\PP_R,
\quad \mbox{ such that }
\ev_2 = 
\psw{\widetilde R_1}{\widetilde \SSS_{2,\widetilde R_1} \widetilde R_1},
\label{eq:S2R_red_0}
\end{align}
where $\PP_R$ is the prolongation operator to velocity-pressure space from tangential velocity-only space
\textit{weighted} by $\left.-\partial_n U_{0t}\right|_{R_0}$.
The optimal wall deformation is defined by 
\begin{align}
\max_{||\widetilde R_1||=1} (\ev_{2r}) 
&= 
\max_{||\widetilde R_1||=1}  \psw{\widetilde R_1}
{ \frac{1}{2}  
\left( \widetilde\SSS_{2,\widetilde R_1,r}
+\widetilde\SSS_{2,\widetilde R_1,r}^T \right) \widetilde R_1}
= 
\max_{\widetilde R_1}  
\dfrac{
\psw{\widetilde R_1}
{ \frac{1}{2}  
\left( \widetilde\SSS_{2,\widetilde R_1,r}
+\widetilde\SSS_{2,\widetilde R_1,r}^T \right) \widetilde R_1}
}
{ \psw{\widetilde R_1}{\widetilde R_1} }
\nonumber
\\
&= \ev_{max} \left\{  \frac{1}{2} 
\left( \widetilde\SSS_{2,R_1,r}
 +\widetilde\SSS_{2,R_1,r}^T \right)
 \right\}, 
\label{eq:maxlam2rR}
\end{align}
with similar expressions for 
$\min(\ev_{2r})$, 
$\max(\ev_{2i})$, and
$\min(\ev_{2i})$.

We note that 
wall deformation has a twofold effect on the eigenmode:
(i)~it modifies the base flow $\UU$ on which the eigenmode develops,
(ii)~it modifies the no-slip boundary for the eigenmode.
While only the first effect is present for wall blowing/suction, 
rigorously speaking the second effect too must be taken into account for wall deformation.
Just as the no-slip boundary condition was Taylor-expanded and flattened for $\UU$, it can be Taylor-expanded  and flattened for $\uu$: 
\begin{align}
 R_1 \left.\partial_n \uu_0\right|_{R_0} +  \uu_1(R_0) = \00,
\label{eq:flattenedBC_eigmode}
\end{align}
which specifies the boundary value of $\uu_1$ as a function of the known $\uu_0$ and $R_1$.
In section~\ref{sec:growth-walldef}, we will show that this effect from the boundary condition is actually much smaller than the effect from the base flow.

%----------------------------------------
%----------------------------------------
%----------------------------------------
\section{Numerical method}
\label{sec:num}

%-----------------------------------------------
%-----------------------------------------------
\subsection{Sensitivity analysis and optimization}
\label{sec:num2D}

All calculations are performed using the methods described in \cite{Bou14,Boujo15a}. 
A two-dimensional triangulation
of the domain 
\be 
\Omega = \{ (x,y) \,|\, 
-10\leq x \leq 50,\, 
|y|\leq 10,\, 
\sqrt{x^2+y^2}\geq 0.5 \} 
\ee 
with mesh points strongly clustered close to the cylinder wall is generated using the finite-element software \textit{FreeFem++} \cite{freefem}, resulting in approximately 9100  elements.
Velocity and pressure fields are discretized with
P2 and P1 Taylor--Hood elements, respectively, yielding
a total of approximately 41000 degrees of freedom.
All discrete operators involved in the calculation of
base flow, leading eigenmode and optimal control are built from their continuous counterparts expressed
in variational form. 
Steady base flows are obtained by solving (\ref{eq:baseflow}) with an iterative Newton method.
The linear stability eigenvalue problem (for the unperturbed leading eigenmode $\qq_0$) and all  optimization eigenvalue problems (for the optimal controls $\CC$ and $\UU_c$ and optimal wall deformation $R_1$) are solved with an implicitly restarted Arnoldi method. 
Second-order sensitivity operators contain inverse operators and are therefore not formed explicitly; rather, the optimization is performed iteratively  and only requires evaluating matrix-vector products and solving linear systems of equations (see also appendix~C).
We generally computed the largest/smallest 3 eigenvalues in the optimization eigenvalue problems;
we checked that they were well converged by computing up to 20 leading eigenvalues for some conditions.

Uncontrolled base flow calculations were validated  in the range $50 \leq \Rey \leq 100$ by comparing the length of the recirculation region to results reported in \cite{Giannetti07};
eigenvalue calculations were validated in the same range of Reynolds numbers by comparing the leading growth rate to results reported in \cite{Barkley06} and \cite{Giannetti07}. The agreement was very good in all cases.
Unless otherwise stated, we focus in the following on the flow at $\Rey=50$.

%-----------------------------------------------
%-----------------------------------------------
\subsection{Three-dimensional stability analysis}
\label{sec:num3D}

Fully three-dimensional calculations are performed for validation purposes with the open-source, massively parallel spectral-element code \textit{NEK5000}~\cite{nek5000}.
These calculations serve a fourfold purpose: 
\begin{enumerate}
\item
verify that our two-dimensional optimization is properly implemented;
\item
verify that the total controlled base flow $\QQ$ is  well captured by the expansion (\ref{eq:exp_Q}) truncated at order $\epsilon^2$;
\item
verify that higher-order eigenvalue variations $\epsilon^n \ev_n$, $n>2$, can be safely discarded for small amplitudes $\epsilon$;
\item
assess whether any other eigenmode is destabilized when then control stabilizes the leading eigenmode (an effect observed in the plane Poiseuille flow \cite{Boujo15b}).
\end{enumerate}

The numerical method is similar to that described in \cite{Fani2013-T}.
Lagrange polynomial interpolants of order $N=5 $ and $N-2 =3$ are used for velocity and pressure, respectively, based on Gauss-Lobatto-Legendre %(GLL)
quadrature points in each hexahedral element. 
The computational domain has the same extension as the 2D domain in the $x$ and $y$ directions, and $0\leq z \leq 4\pi$ in the spanwise direction.
The total 3D base flow $\QQ$ induced by the control is obtained
by solving the nonlinear NS equations (\ref{eq:total_controlled_BF_mom})-(\ref{eq:total_controlled_BF_div}) with a first order scheme. 
The unstable steady solution is obtained by using the \textit{BoostConv} algorithm  proposed by \cite{Citro2017}. 
The 3D eigenvalue problem associated with the linear stability of the total base flow is solved with an Arnoldi method, using the linearized version of \textit{NEK5000} as a time-stepper, and with a second-order scheme (BDF2).
For both nonlinear and linearized NS equations, convective terms are discretized in time with explicit backward-differentiation, and viscous terms with an implicit scheme.

%----------------------------------------
%----------------------------------------
%----------------------------------------
\section{Optimal wall actuation for stabilization}
\label{sec:growth-wallactu}

In this section we present and discuss the effect of optimal spanwise-periodic wall actuation (blowing/suction) for flow stabilization.
Optimal wall deformation for stabilization will be considered in section \ref{sec:growth-walldef}, and optimal control for frequency modification in section \ref{sec:freq}.
For the sake of brevity, optimal volume control is reported in appendix~D.

%----------------------------------------
%----------------------------------------
%----------------------------------------
\subsection{Optimal wall actuation}

\begin{figure}
\centerline{   
   \begin{overpic}[width=7cm, trim=5mm 65mm 20mm 60mm, clip=true, tics=10]{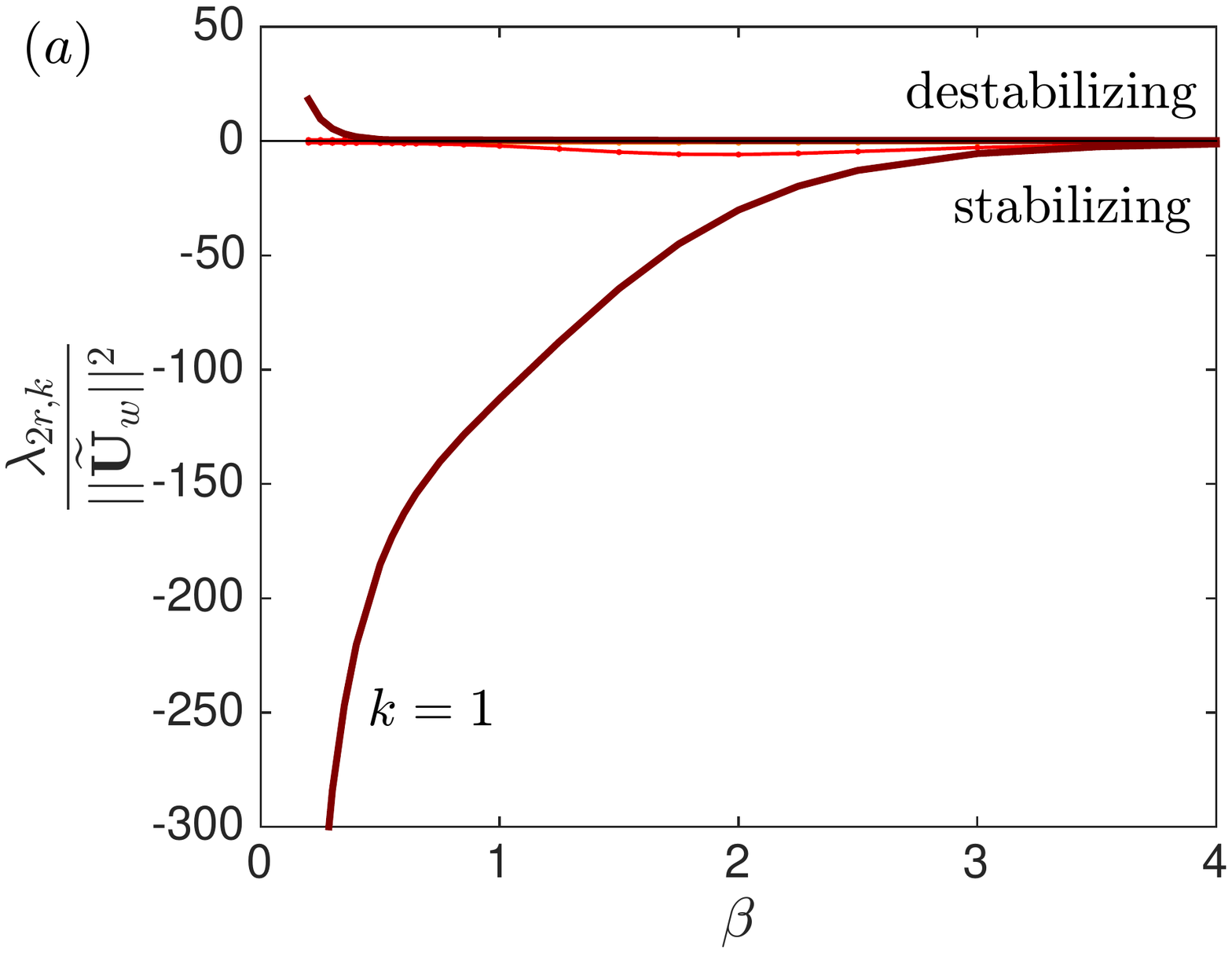}      
   \end{overpic}  
   \hspace{0.3cm}
   \begin{overpic}[width=7cm, trim=5mm 65mm 20mm 60mm, clip=true, tics=10]{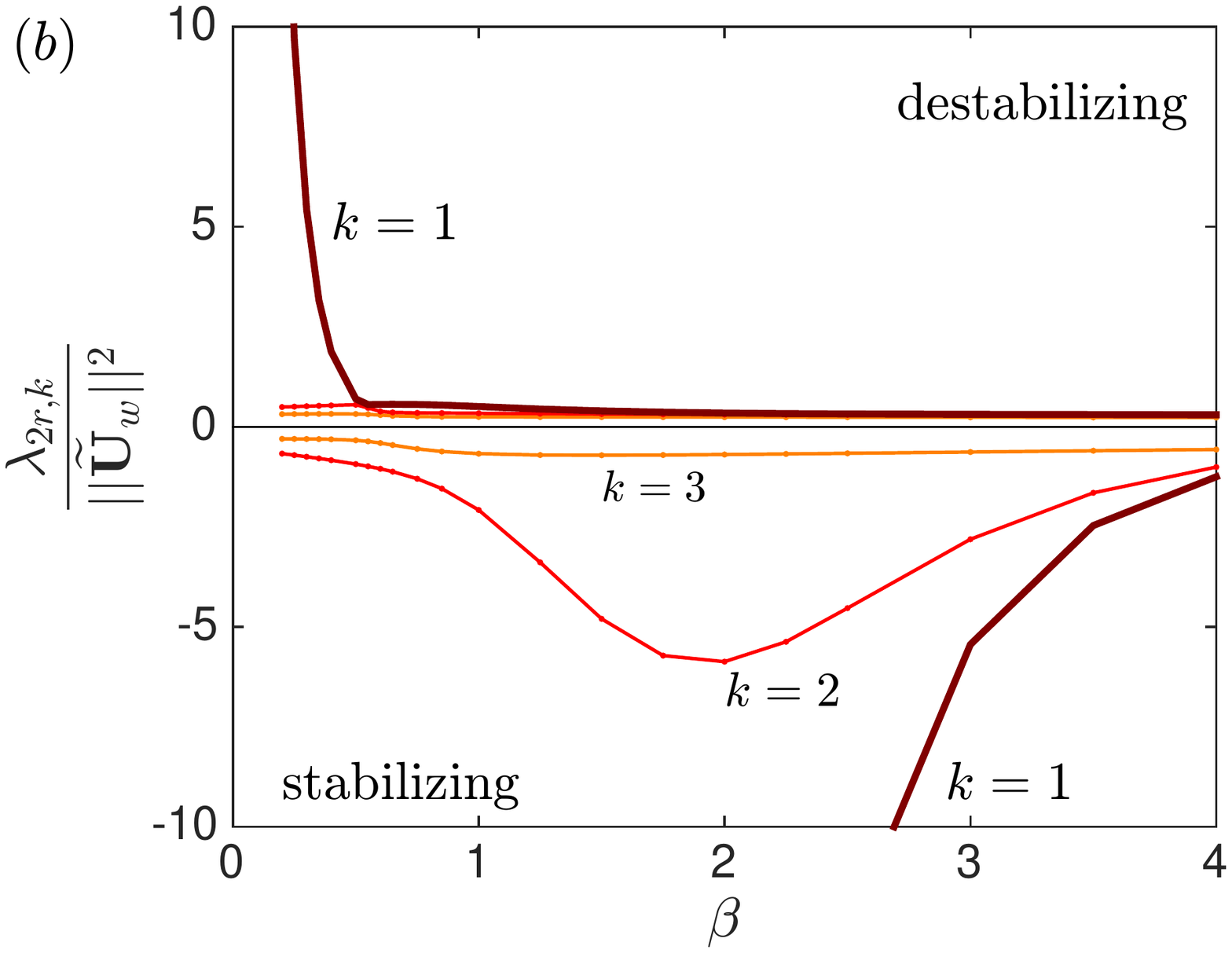}      
   \end{overpic} 
}
\caption{
$(a)$~Normalized growth rate variation
induced by the optimal wall control $\UU_w$ for  stabilization ($\lambda_{2r}<0$) and destabilization ($\lambda_{2r}>0$).
Optimal ($k=1$) and first sub-optimals ($k=2$, $3$).
$\Rey=50.$
$(b)$~Close-up view of sub-optimals.
In this and subsequent figures, increasing values of $k$  are shown with thinner lines and lighter hues.
} 
\label{fig:lam2r_opt_Uw_r} 
\end{figure}

We first investigate optimal flow stabilization with spanwise-periodic wall blowing/suction, using all velocity components $(U_n,U_t,W)^T$.
Figure~\ref{fig:lam2r_opt_Uw_r} shows the optimal stabilizing second-order growth rate variation ($\min(\ev_{2r})<0$) obtained with a unit-norm control at $\Rey=50$. 
The variation is decreasing with spanwise wavenumber $\beta$, which indicates that larger  wavelengths $2\pi/\beta$ are more efficient, as already observed in other shear flows \cite{DelGuercio2014parallel, Cossu14secondorder, Tammisola2014}. 
The stabilizing effect obtained with sub-optimal controls is much smaller in the range of $\beta$ of interest.
The optimal destabilizing effect ($\max(\ev_{2r})>0$) is also shown for reference and appears considerably smaller than its optimal stabilizing counterpart
(recall that, unlike the 2D case, 3D spanwise-harmonic optimal stabilizing and destabilizing controls are not simply related by opposite signs).

%--- plot_fig3.m

As analyzed in \cite{Tammisola2014} and \cite{Boujo15b},the strong divergence observed when $\beta \rightarrow 0$ results from a modal resonance between the 2D leading eigenmode and 3D eigenmodes of spanwise wavenumber $\beta_0=\pm \beta$.
As $\beta$ decreases, the minimal distance $d$ between those 3D eigenvalues and the 2D leading eigenvalue decreases too,
the modal resonance becomes stronger and results in a non-small eigenvalue variation $\ev_2$, and the expansion (\ref{eq:exp}) breaks down.
For this expansion to remain valid, the control amplitude $\epsilon$ must stay smaller than
the previously defined distance $d$ between the 2D eigenvalue and the closest $\beta$-periodic eigenvalue
\cite{Tammisola2014}.
A 3D linear stability analysis of the uncontrolled flow (see details in Appendix~E) shows that 
$d \simeq 0.035$, 0.075 and  0.185 for $\beta=0.4$, 0.6 and 1, respectively.
Therefore, our results are valid at those spanwise wavenumbers $\beta$ for control amplitudes $\epsilon$ smaller than those $d$ values
(equivalently, they are valid at those values of $\epsilon=d$ for wavenumbers larger than those $\beta$ values).
We note that, in an experimental setting,
the finite spanwise extension of the system sets a minimal value for $\beta$.

%--- plot_variation_vs_epsilon_withtext.m
\begin{figure}
\centerline{   
   \begin{overpic}[width=7cm, trim=5mm 65mm 20mm 60mm, clip=true, tics=10]{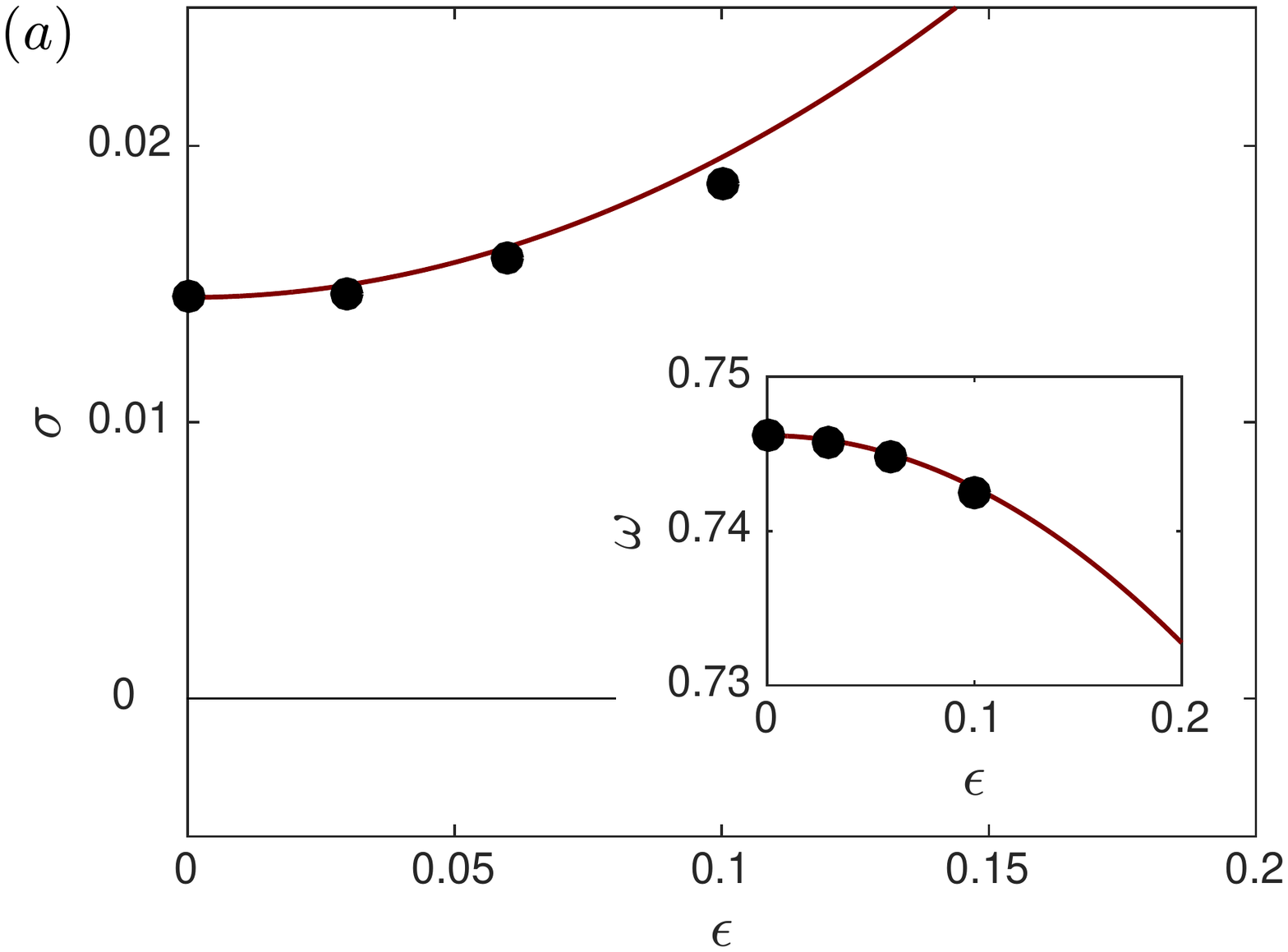} 
   \end{overpic}
   \hspace{0.8cm}
   \begin{overpic}[width=7cm, trim=5mm 65mm 20mm 60mm, clip=true, tics=10]{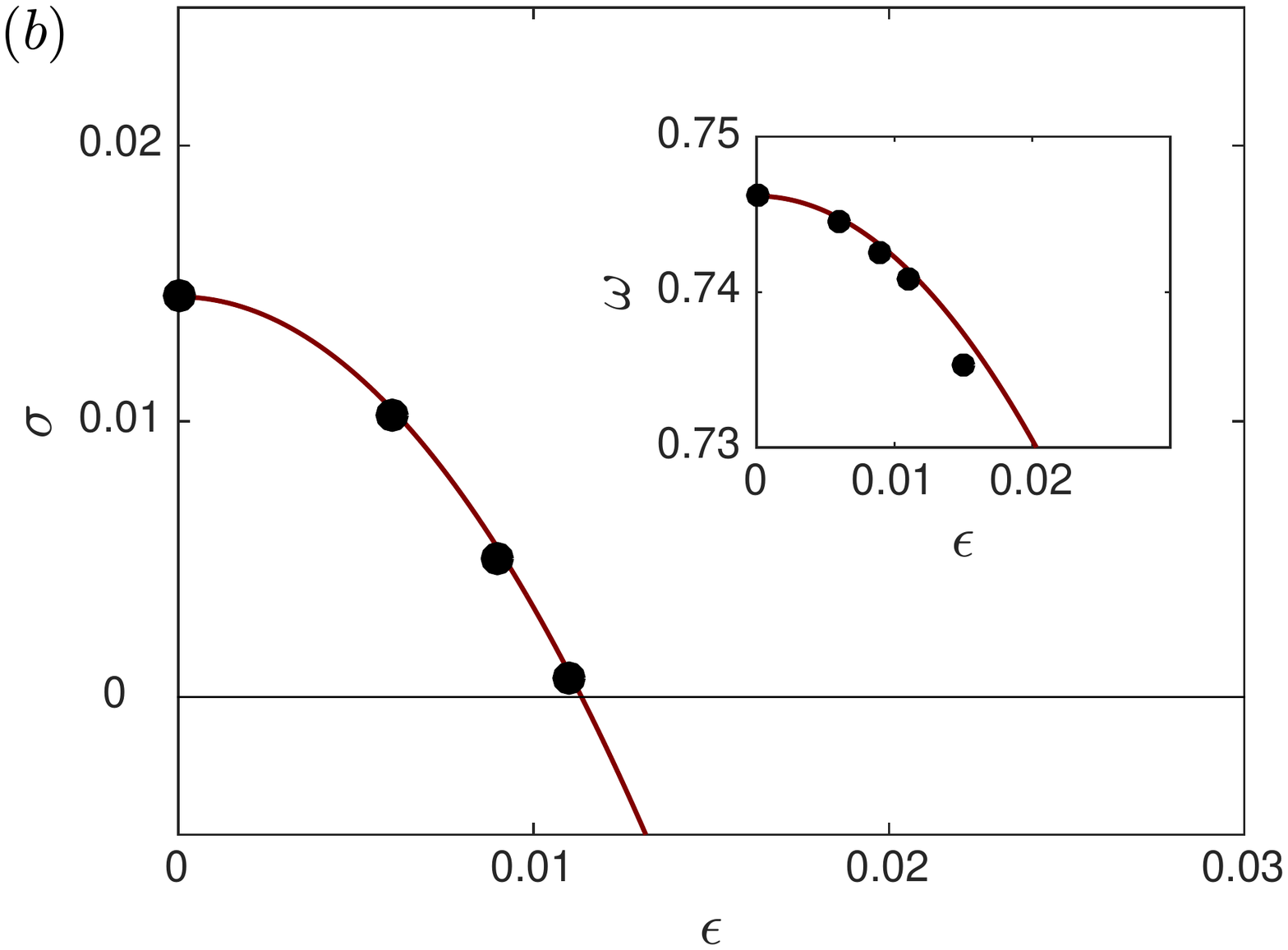} 
   \end{overpic}
}
\caption{
Effect of the optimal $(a)$~destabilizing and $(b)$~stabilizing wall blowing/suction on growth rate $\sigma$ and frequency $\omega$.
Lines: sensitivity prediction, symbols: full stability analysis.
$\Rey=50$, $\beta=1$.
}  
\label{fig:valid2-1}
\end{figure}

The eigenvalue variation induced by the optimal wall control for destabilization and stabilization for $\beta=1$ is shown in Fig.~\ref{fig:valid2-1}. The quadratic variation $\ev=\ev_0+\epsilon^2\ev_2$ from the 2D sensitivity prediction compares well with the total variation obtained with 3D stability analysis.

%--- plot_fig5.m
\begin{figure}
\centerline{   
   \begin{overpic}[width=5.7cm, trim=0mm 75mm 40mm 65mm, clip=true, tics=10]{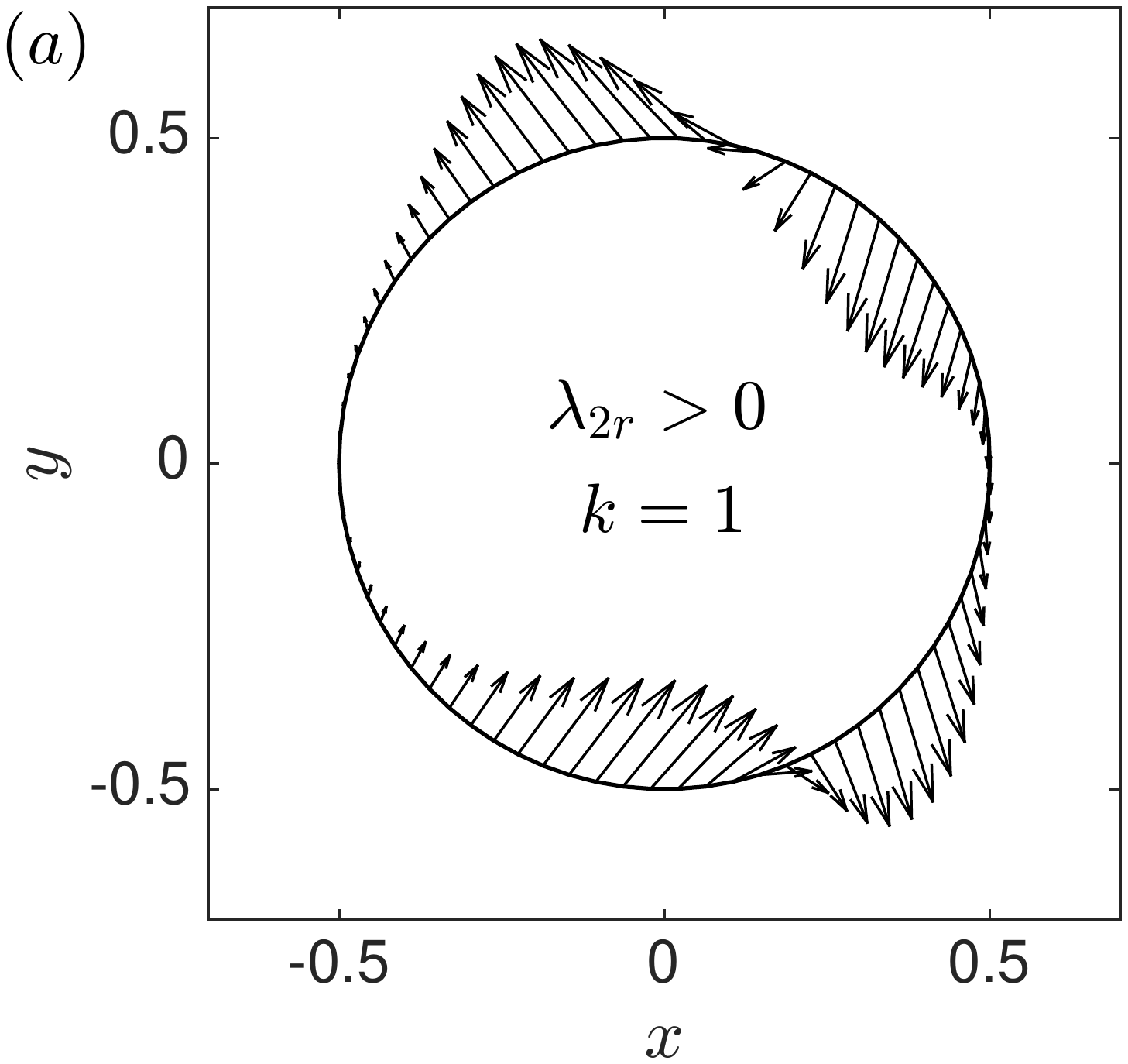} 
   \end{overpic} 
   \begin{overpic}[width=6.cm, trim=15mm 75mm 15mm 65mm, clip=true, tics=10]{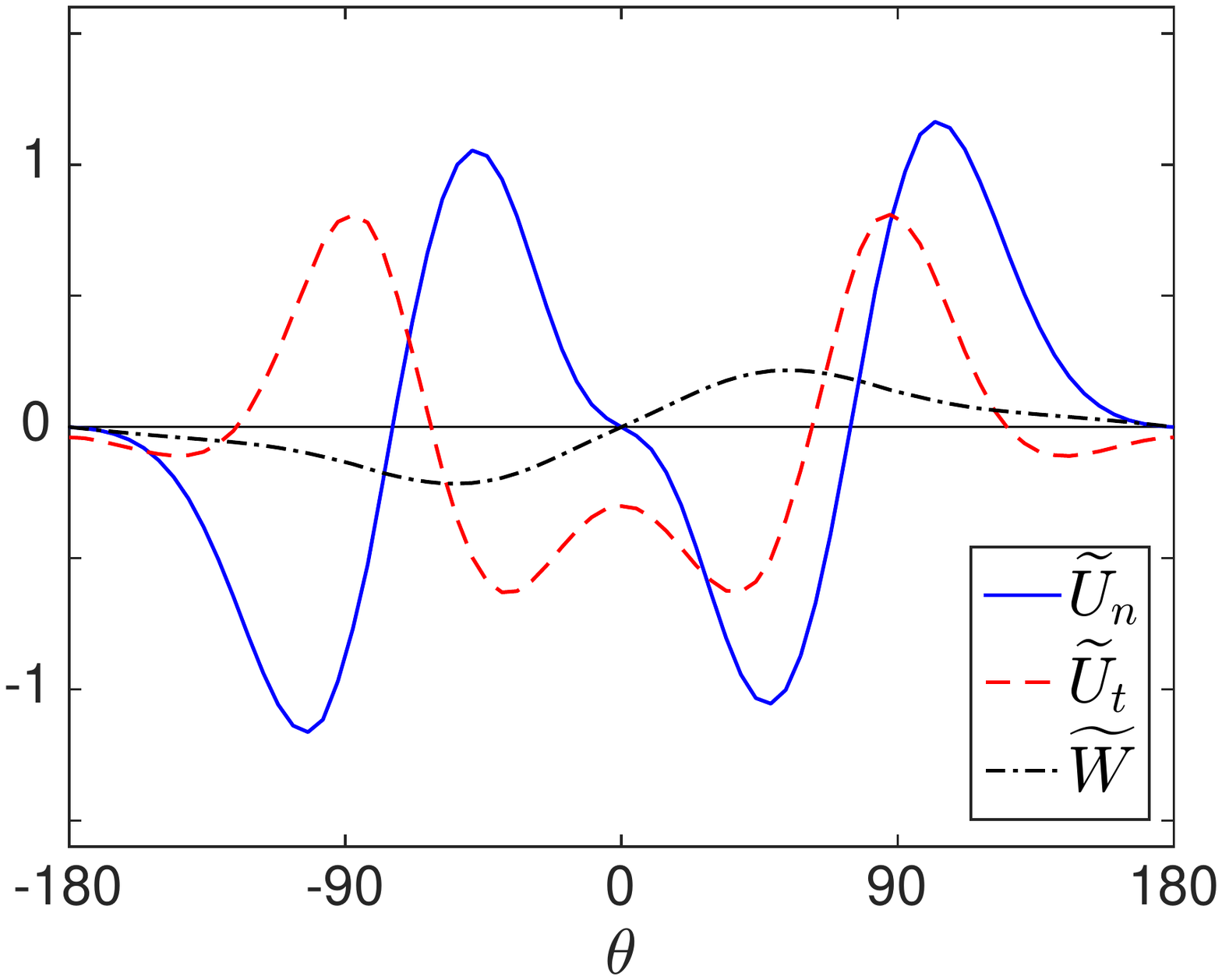}   
   \end{overpic}  
}
\centerline{   
   \begin{overpic}[width=5.7cm, trim=0mm 75mm 40mm 65mm, clip=true, tics=10]{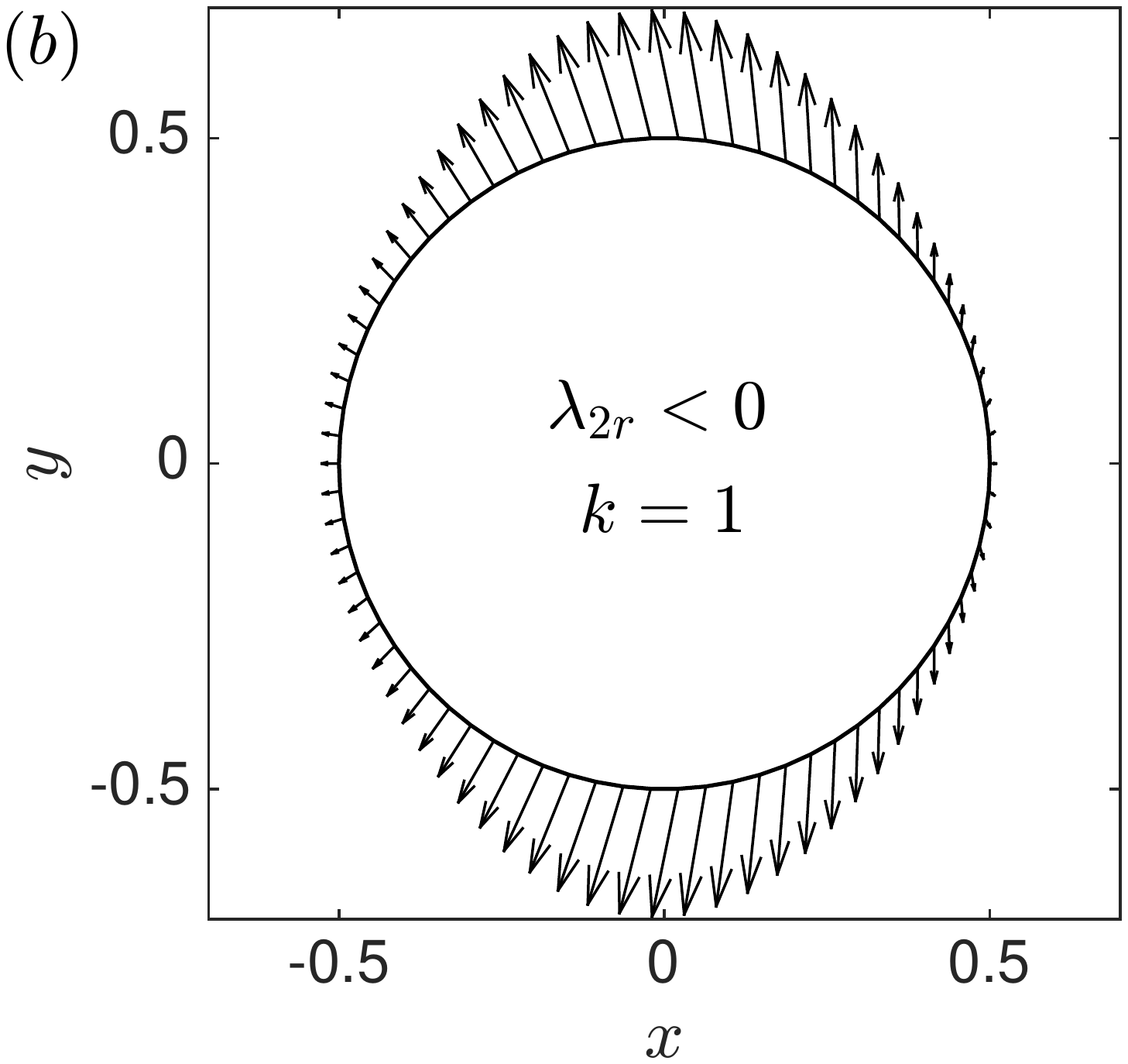} 
   \end{overpic} 
   \begin{overpic}[width=6.cm, trim=15mm 75mm 15mm 65mm, clip=true, tics=10]{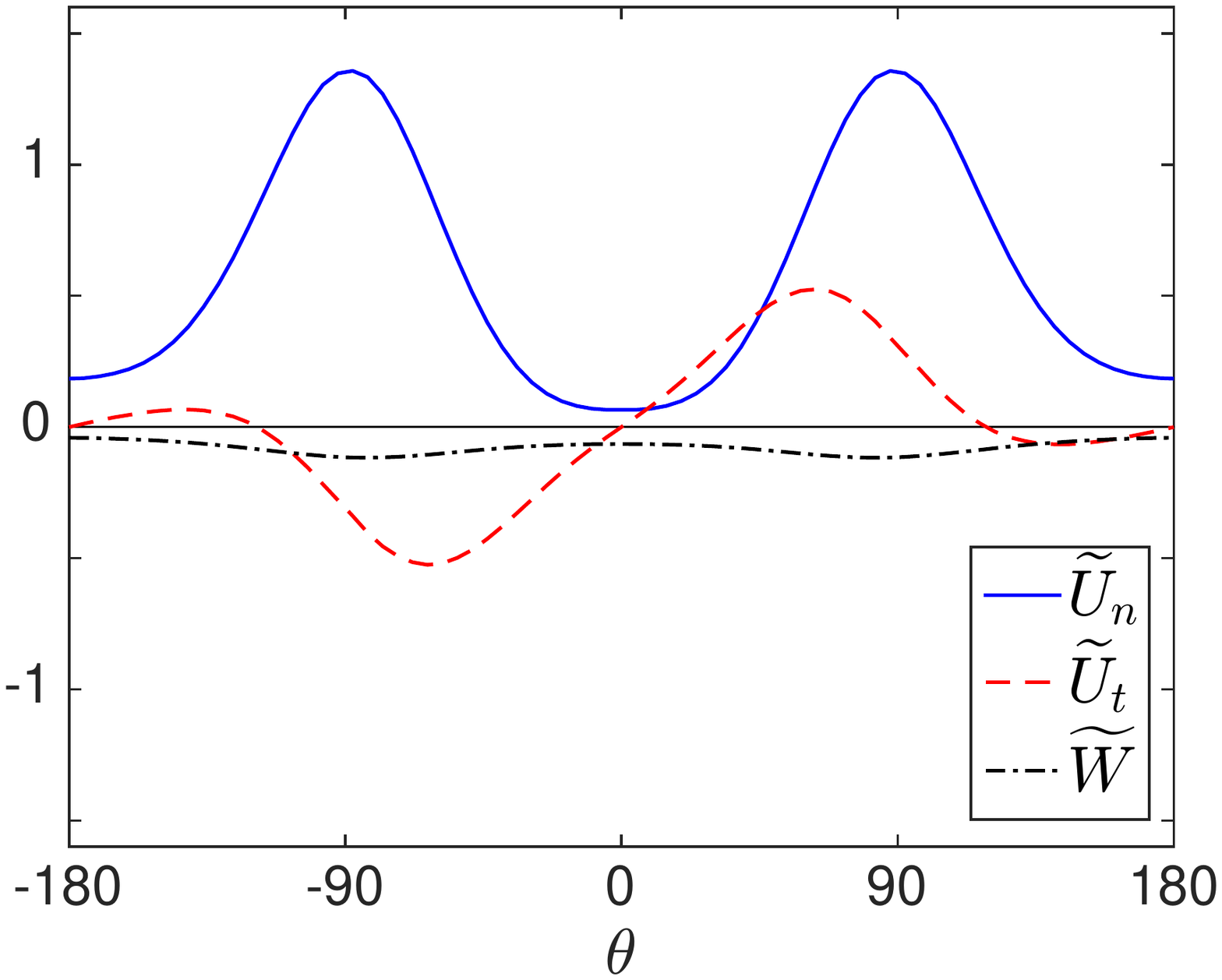}   
   \end{overpic}  
}
\centerline{   
   \begin{overpic}[width=5.7cm, trim=0mm 65mm 40mm 65mm, clip=true, tics=10]{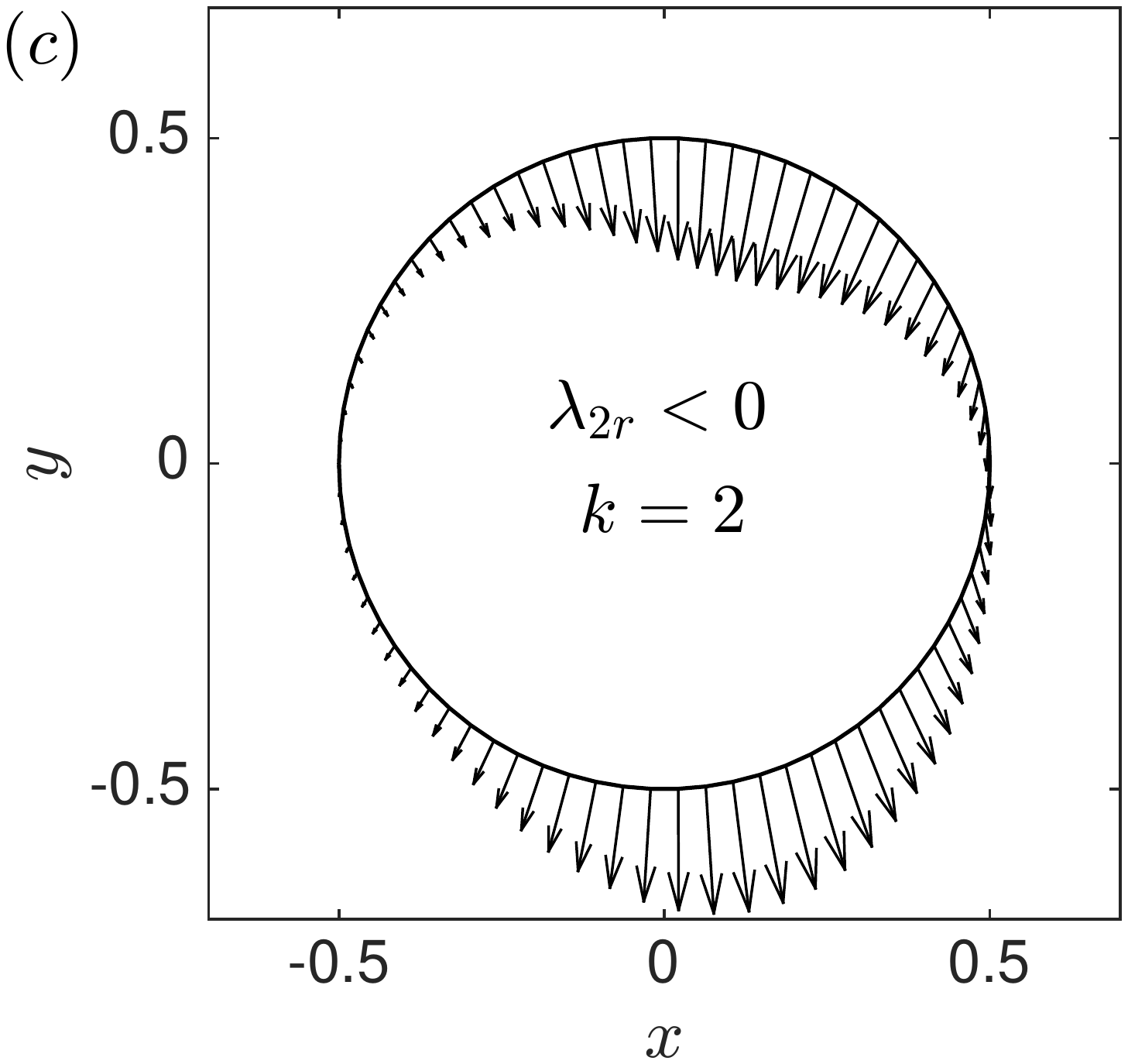} 
   \end{overpic} 
   \begin{overpic}[width=6.cm, trim=15mm 65mm 15mm 65mm, clip=true, tics=10]{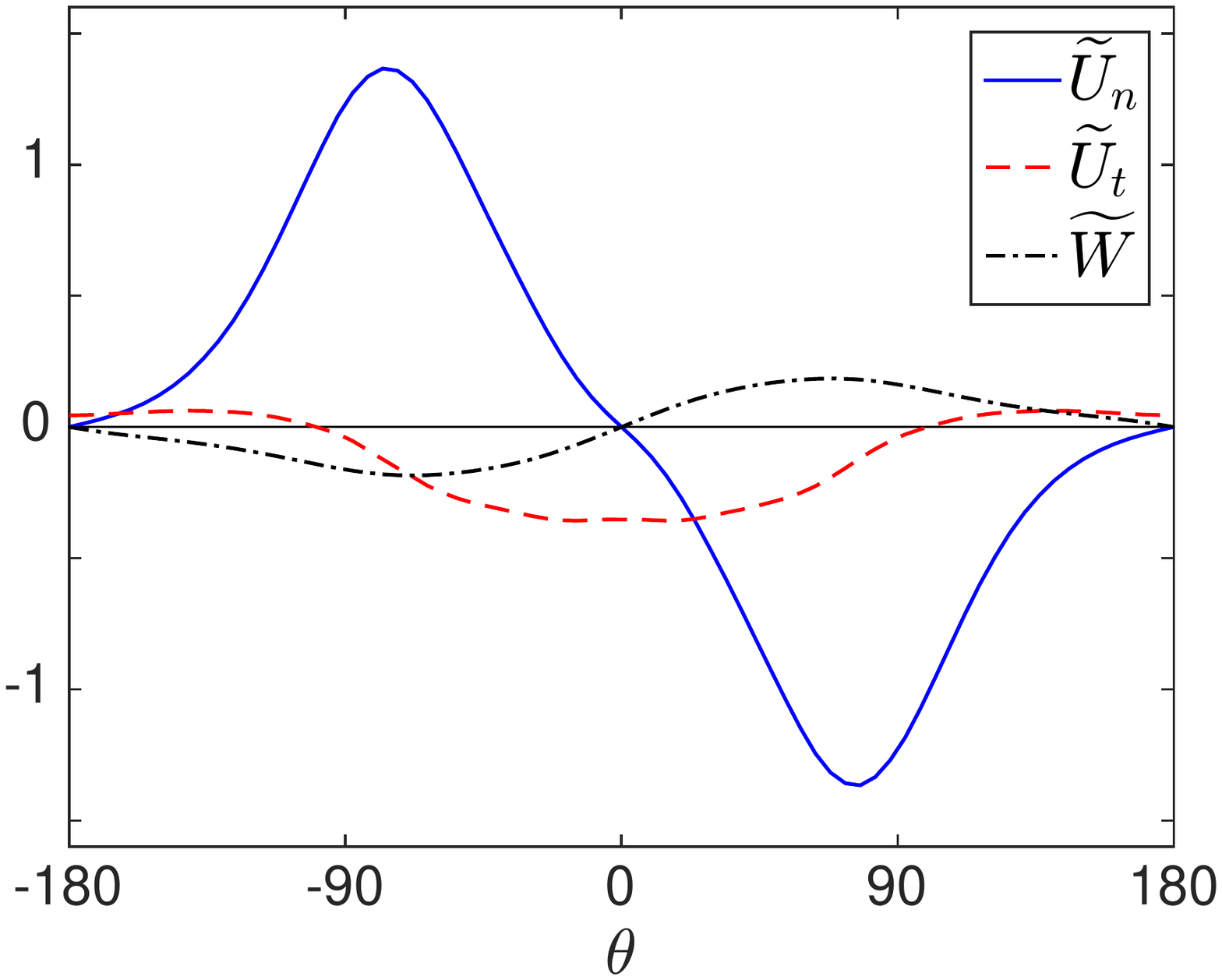}   
   \end{overpic}  
}
\caption{
Optimal wall control $\widetilde\UU_w$ for destabilization/stabilization: 
left, arrows of in-plane velocity components;
right, full velocity field variation with $\theta$.
$\Rey=50$, $\beta=1$.
$(a)$~Optimal destabilizing wall control;
$(b)$~Optimal stabilizing wall control;
$(c)$~First sub-optimal stabilizing wall control.
}  
\label{fig:opt_Uw_dest_stab}
\end{figure}

Figure~{\ref{fig:opt_Uw_dest_stab} shows optimal wall control for $\beta=1$.
The optimal destabilizing wall control in Fig.~{\ref{fig:opt_Uw_dest_stab}$(a)$ has a $y$-antisymmetric normal component that is maximal  at $\theta=\pm 45^\circ$ and $\pm 100^\circ$, and a $y$-symmetric tangential component that changes sign close to the separation points $\theta=\pm 60^\circ$. The spanwise component is rather small.
The optimal stabilizing wall control has a symmetric normal component that is maximal on the top and bottom sides of the cylinder ($\theta=\pm 90^\circ$) and minimal on the upstream and downstream faces. 
The orientation is mainly normal, although the antisymmetric  tangential component is substantial around the separation points ($\theta=\pm 60^\circ$). Again, the spanwise component is small.
Note that properties of top-down symmetry (symmetry about the wake centerline $y=0$)  alone do not explain the destabilizing or stabilizing character of the control, as illustrated by the first sub-optimal control.
Overall, little qualitative variation is observed for wavenumbers in the range $0.5 \leq \beta \leq 2$.

%-----------------------------------------------
%-----------------------------------------------
\subsection{Control-induced flow modification}

%--- plot_mono_C1_Q1_Q2_real.m
%--- plot_mono_C1_Q1_Q2_real_arrows_xy.m
%--- plot_mono_C1_Q1_Q2_real_arrows_yz.m
\begin{figure}
\centerline{   
   \begin{overpic}[trim=5mm 14mm 0mm 17mm, clip, height=4.1cm,tics=10]{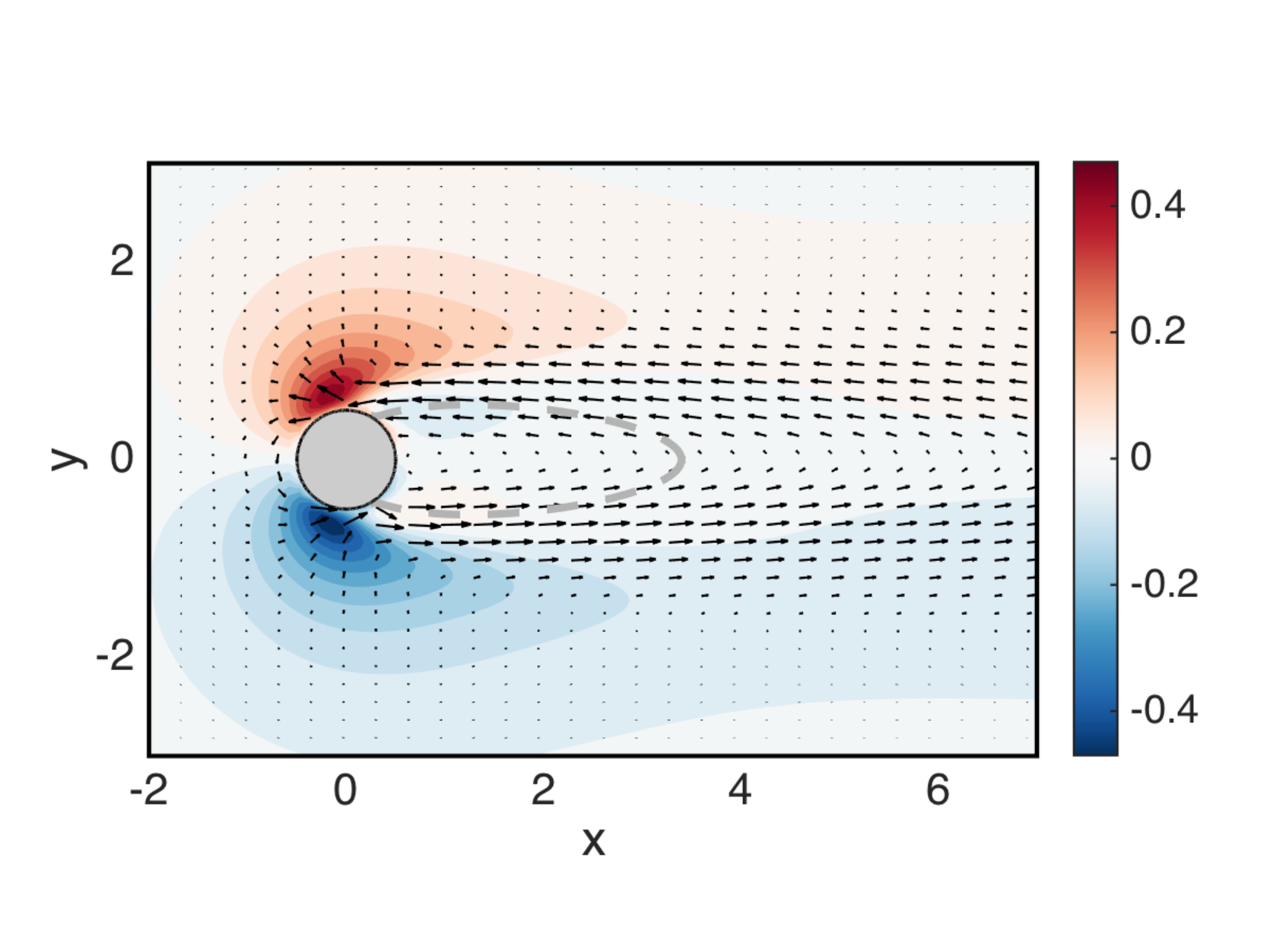}
      \put(-2,54){$(a)$} 
   \end{overpic}        
   \begin{overpic}[trim=23mm 0 5mm 0, clip, height=4.05cm,tics=10]{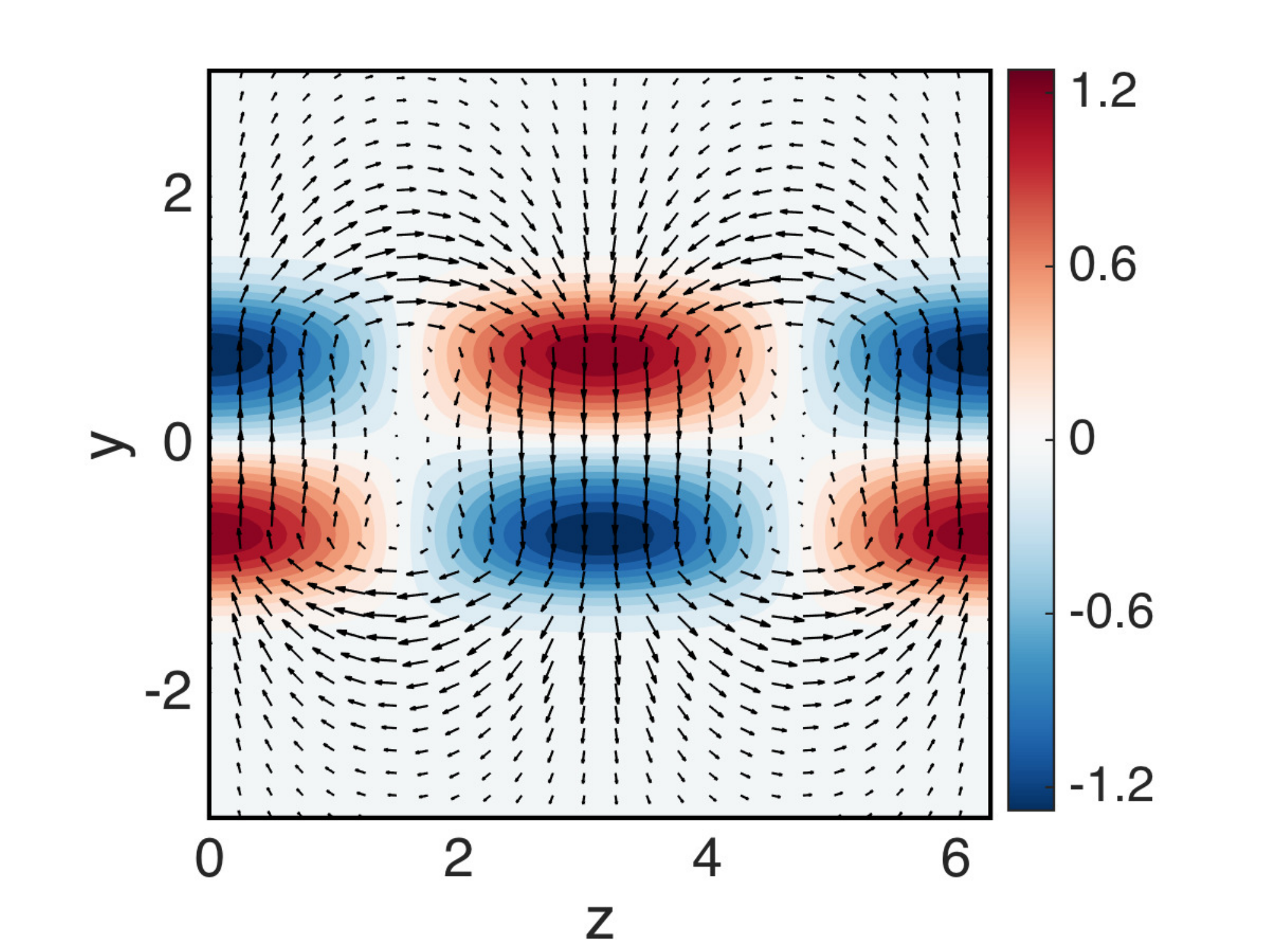}
   \end{overpic}   
   \begin{overpic}[trim=15mm 14mm 5mm 17mm, clip, height=4.1cm,tics=10]{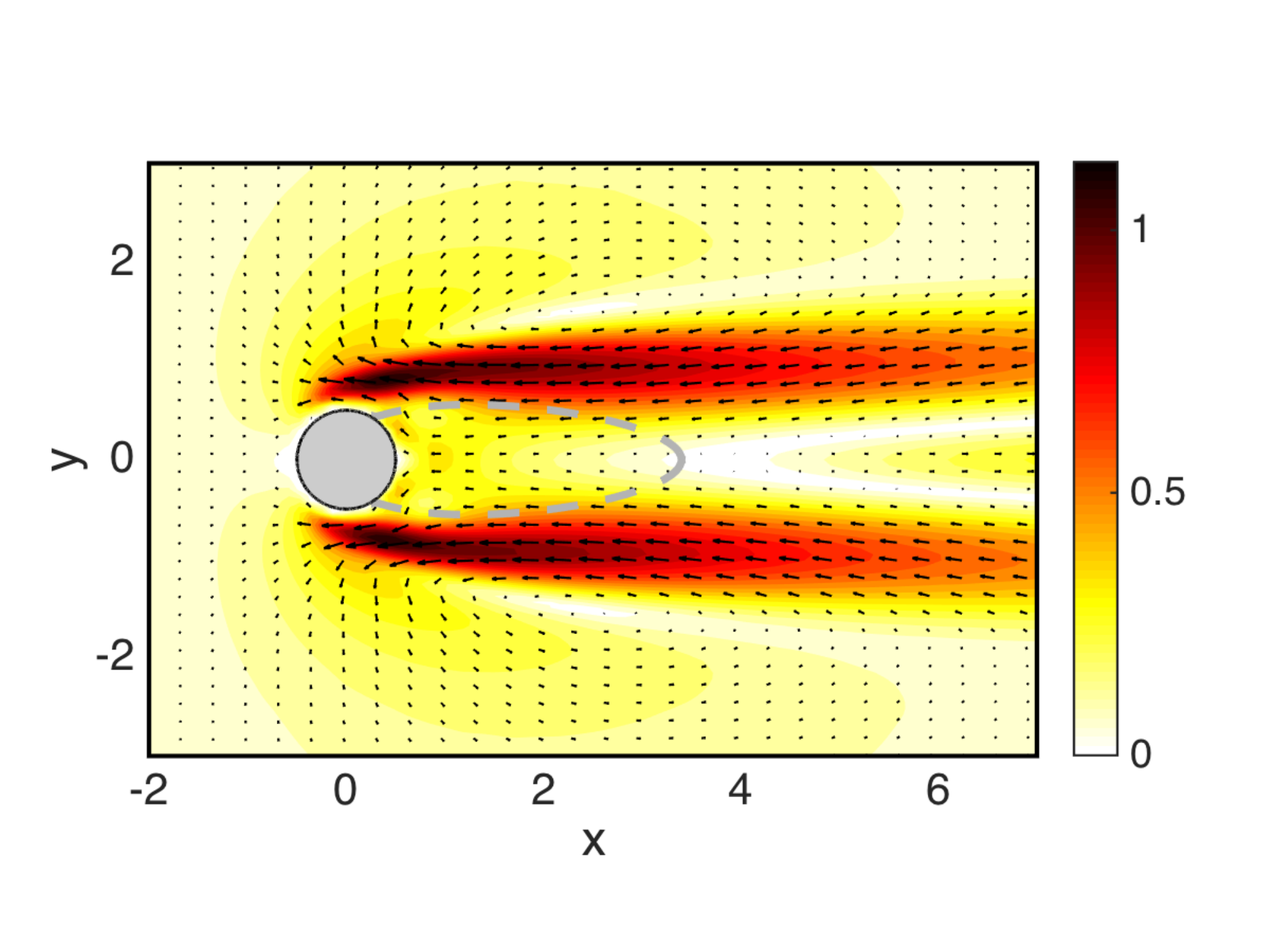}     
   \end{overpic}  
}
\centerline{   
   \begin{overpic}[trim=5mm 14mm 0mm 17mm, clip, height=4.1cm,tics=10]{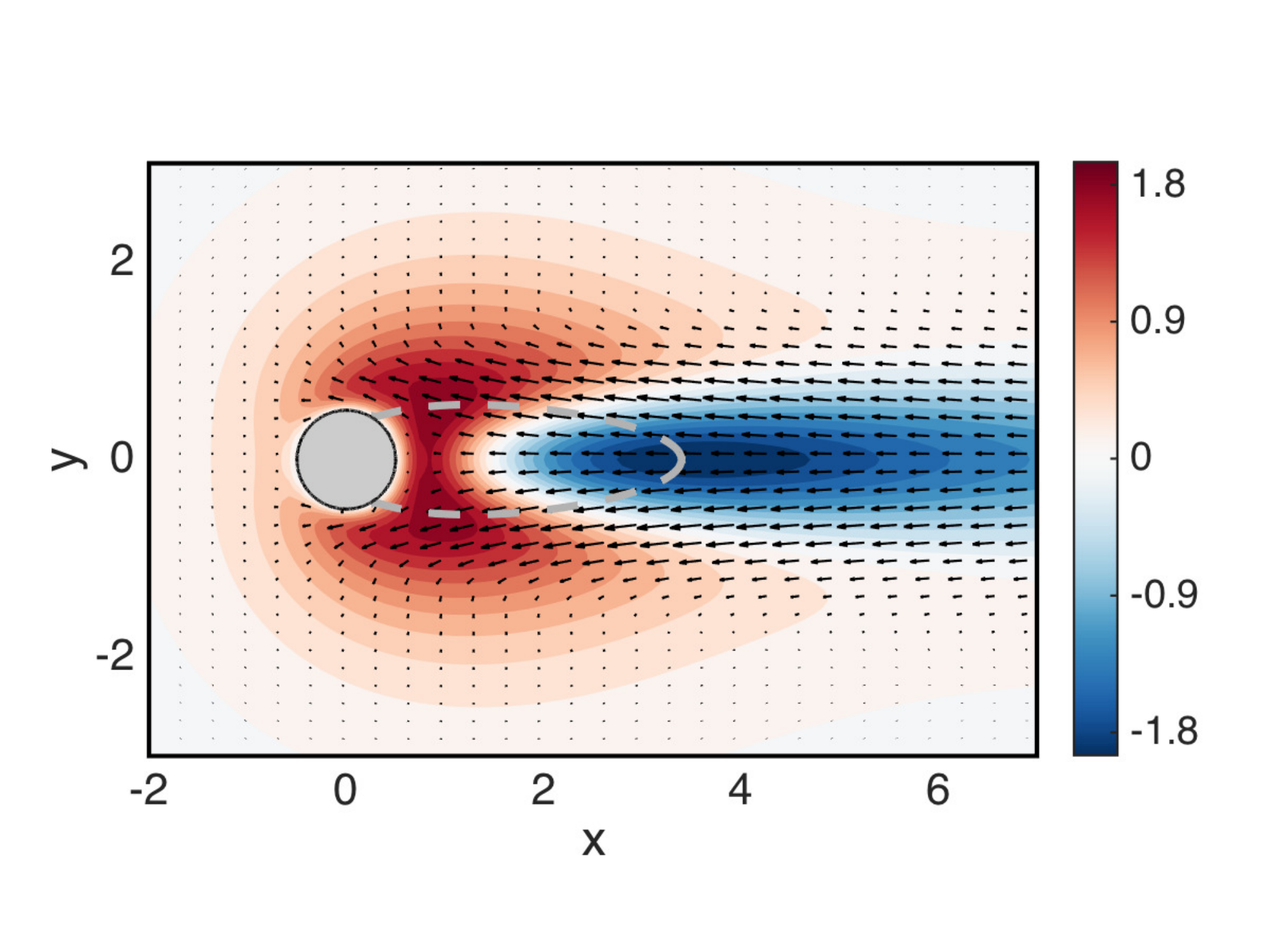}
      \put(-2,54){$(b)$}
   \end{overpic} 
   \begin{overpic}[trim=23mm 0 5mm 0, clip, height=4.05cm,tics=10]{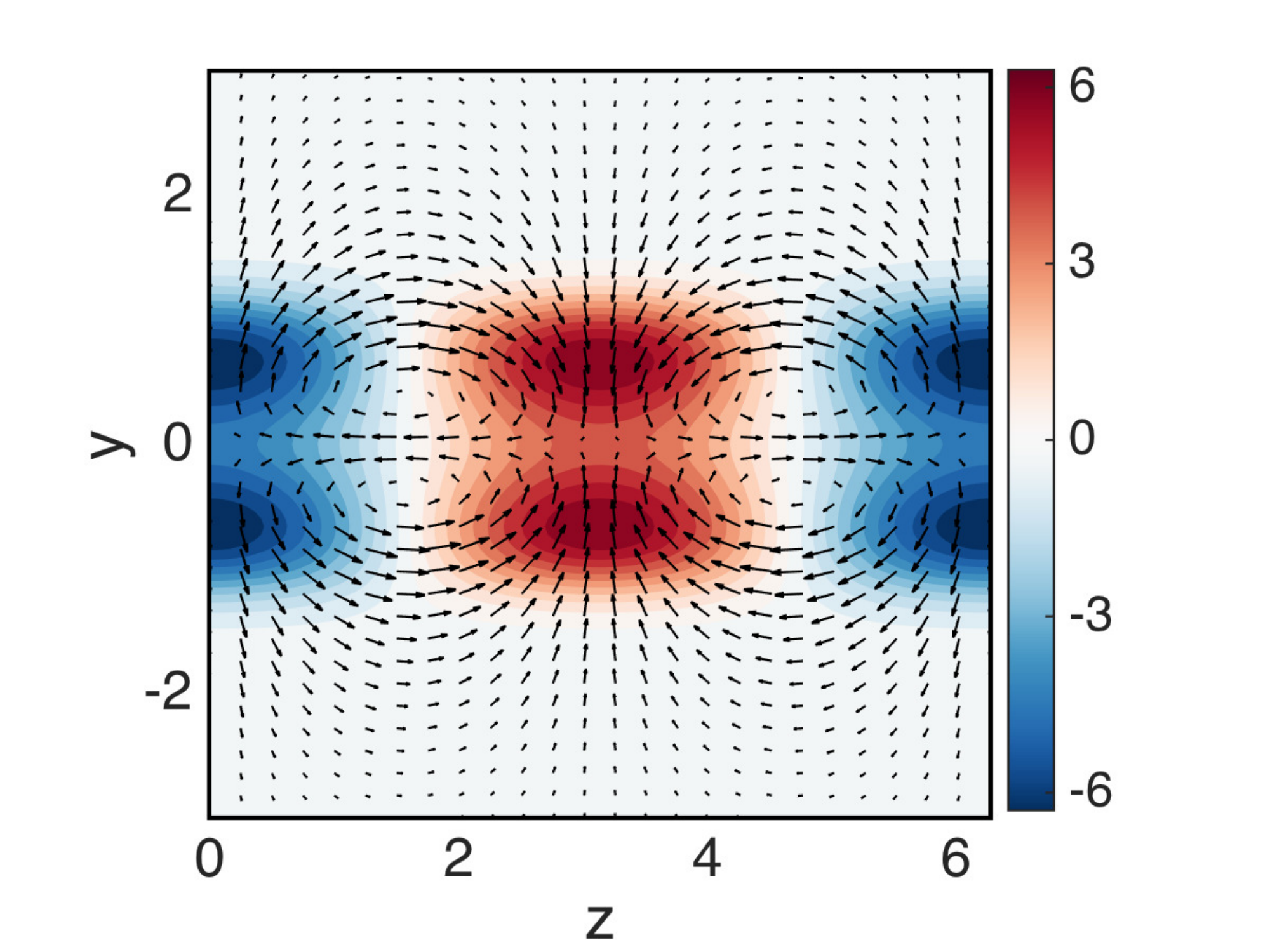}
   \end{overpic} 
   \begin{overpic}[trim=15mm 14mm 5mm 17mm, clip, height=4.1cm,tics=10]{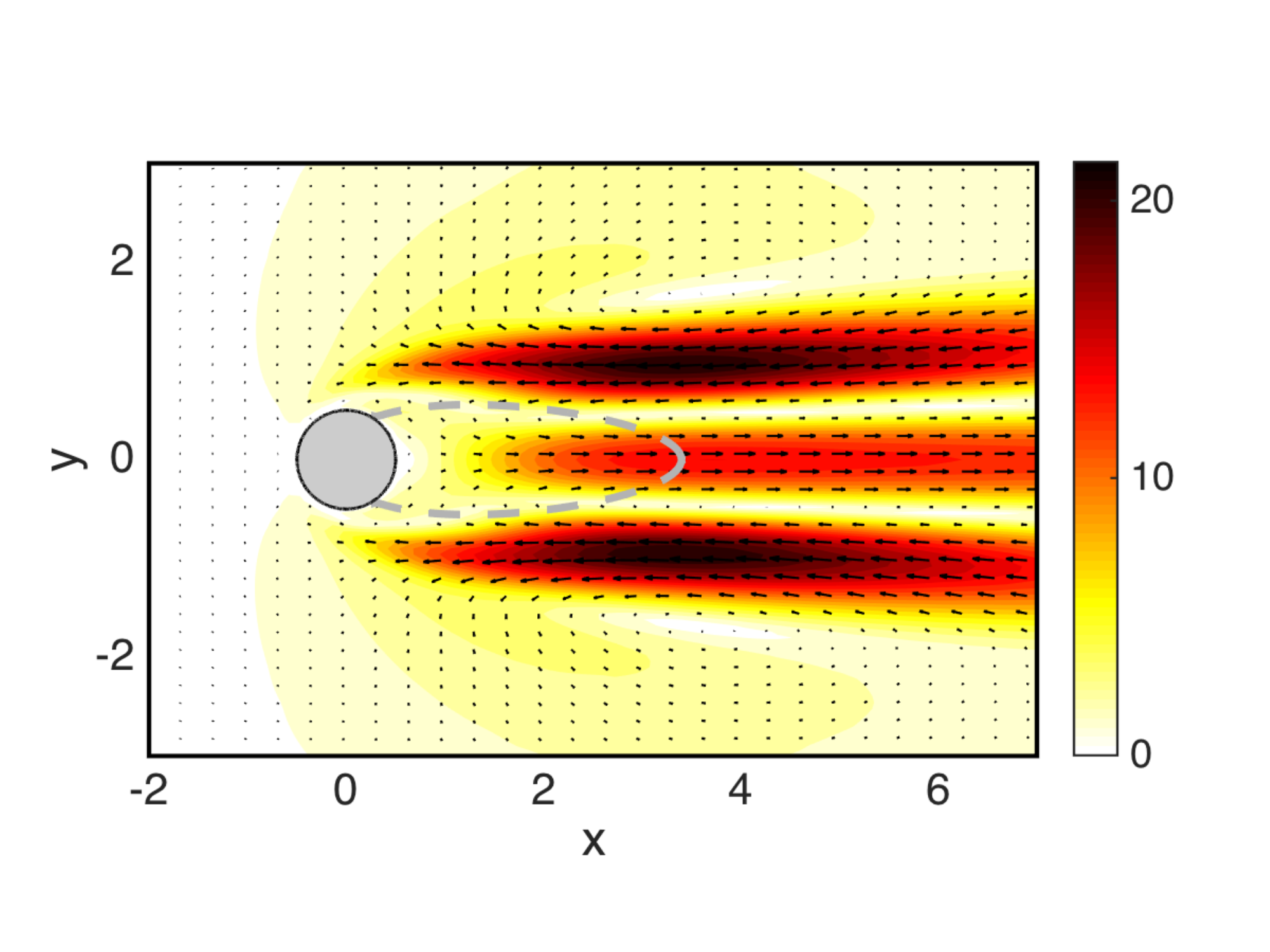}     
   \end{overpic}      
}
\centerline{   
   \begin{overpic}[trim=5mm 14mm 0mm 17mm, clip, height=4.1cm,tics=10]{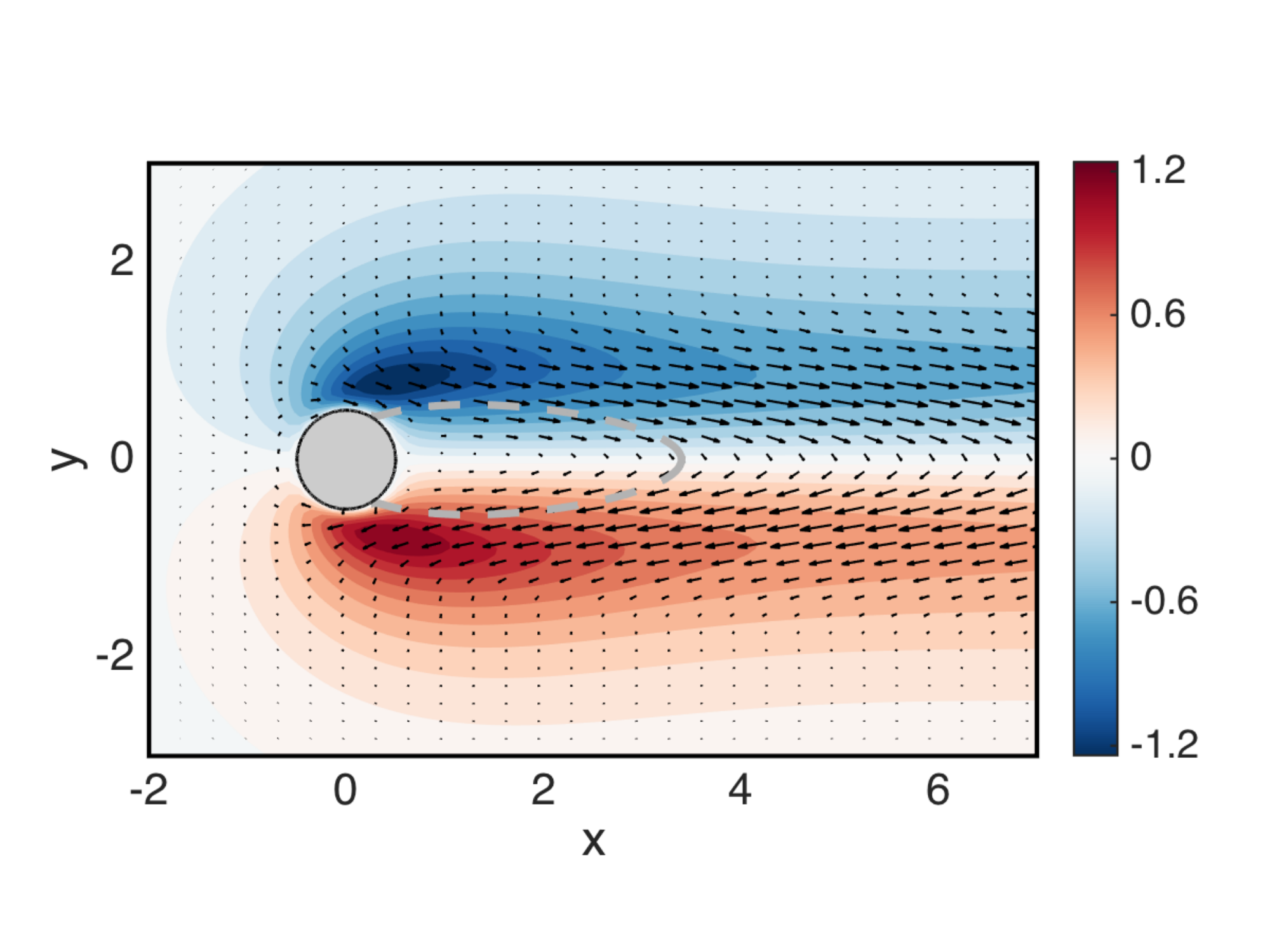}
      \put(-2,54){$(c)$} 
   \end{overpic}      
   \begin{overpic}[trim=23mm 0 5mm 0, clip, height=4.05cm,tics=10]{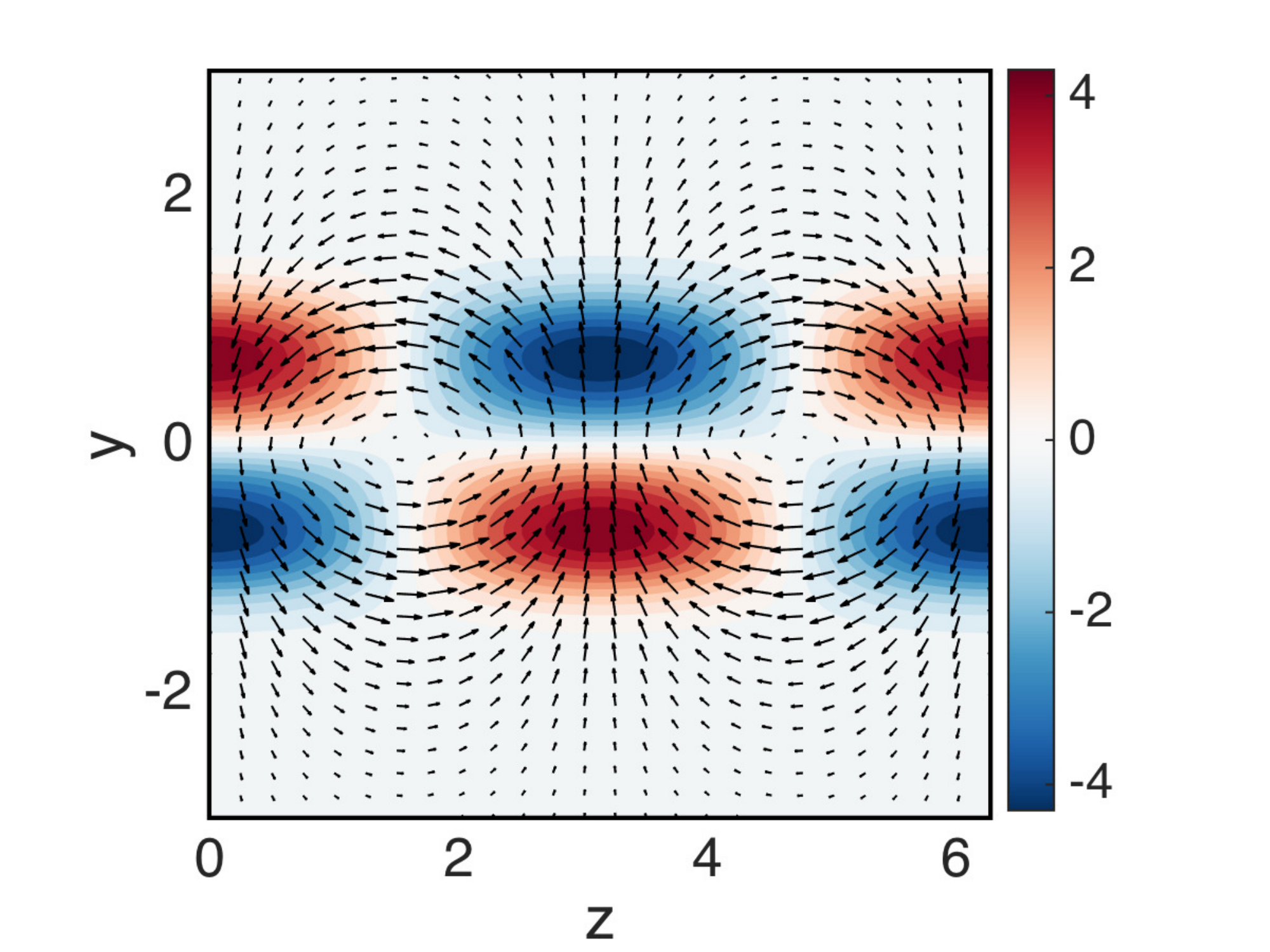}
   \end{overpic} 
   \begin{overpic}[trim=15mm 14mm 5mm 17mm, clip, height=4.1cm,tics=10]{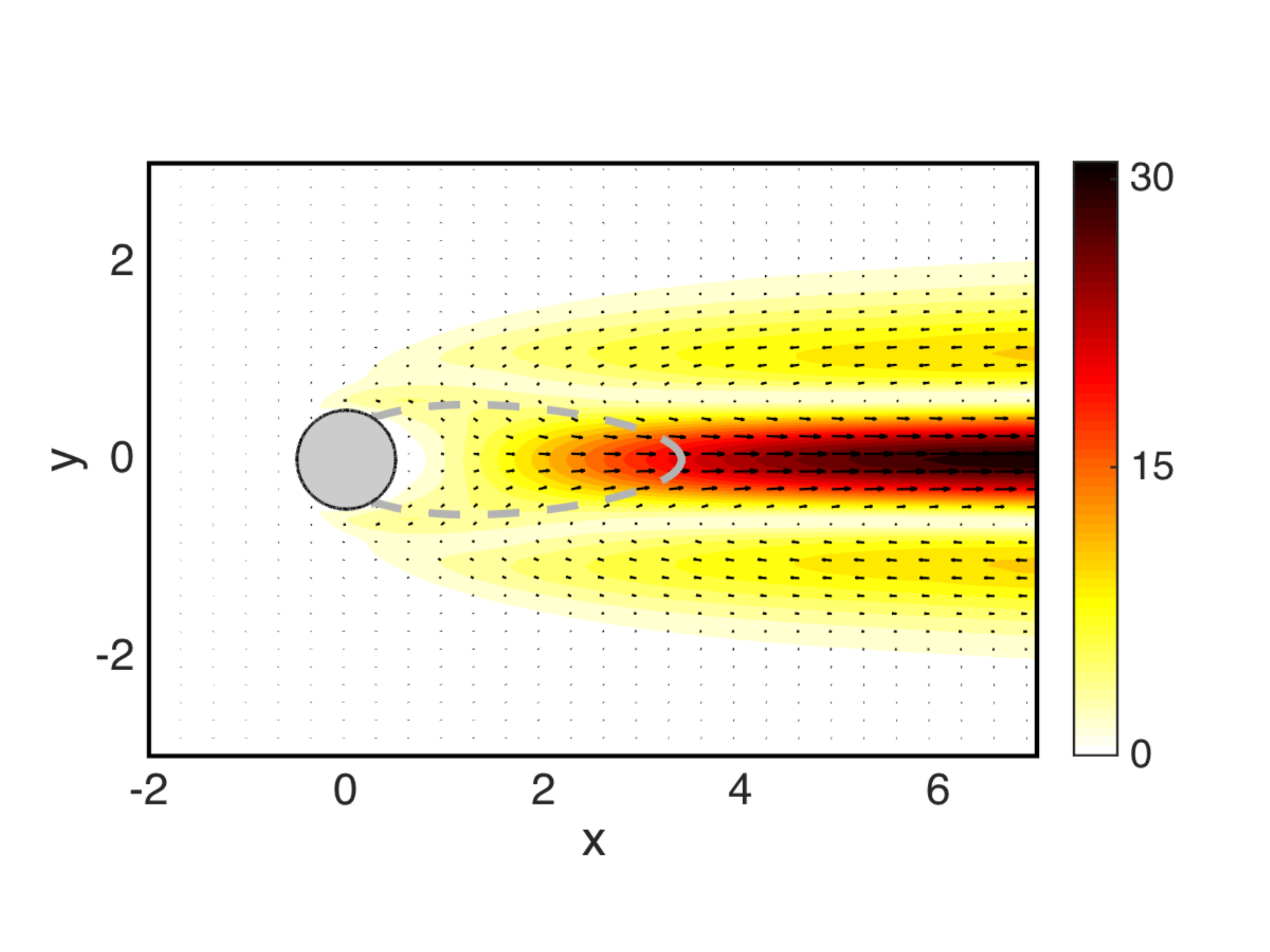}     
   \end{overpic}   
}
\caption{
First-order flow modification $\UU_1 = ( \widetilde U_1(x,y)\cbz, \widetilde V_1(x,y)\cbz, \widetilde W_1(x,y)\sbz)^T$ (spanwise periodic) induced by the
$(a)$~optimal destabilizing,
$(b)$~optimal stabilizing and
$(c)$~first sub-optimal stabilizing wall control,
at $\Rey=50$, $\beta=1$.
Left: vector fields $(U_1,V_1)^T$ at $z=0$ and contours of $W_1$ at $z=\pi/(2\beta)$.
Middle: vector field $(V_1,W_1)^T$ and contours of $U_1$ at $x=2$.
Right: 
induced mean flow correction $\UU_2^{2D} = ( U_2^{2D}(x,y), V_2^{2D}(x,y), 0)^T$ (spanwise invariant), shown with vector field $(U_2^{2D},V_2^{2D})^T$ and contours of velocity magnitude. 
}  
\label{fig:Q1_from_opt_Uw_stab}
\end{figure}

Flow modifications induced by the optimal wall control are shown in Fig.~\ref{fig:Q1_from_opt_Uw_stab}, with the first-order (spanwise-periodic) modification $\QQ_1$ in the left and middle panels ($x-y$ and $z-y$ planes respectively) and the spanwise-invariant component $\QQ_2^{2D}$ of the second-order modification in the right panel. 
The optimal destabilizing wall control [Fig.~\ref{fig:Q1_from_opt_Uw_stab}$(a)$] induces a double-streak pattern in the cylinder wake: moving along $z$, the streamwise velocity takes alternatively positive and negative values in the upper half-domain and the opposite sign in the lower half-domain. 
The mean flow correction has two regions of negative streamwise velocity concentrated in the shear layers on both sides of the recirculation region.
By contrast, the optimal stabilizing wall control [Fig.~\ref{fig:Q1_from_opt_Uw_stab}$(b)$] induces a simple-streak pattern, with high- and low-velocity streaks extending over the whole height of the wake. 
The mean flow correction has again two regions of negative streamwise velocity in the shear layers, and a region of positive streamwise velocity along the centerline $y=0$. The latter is expected to reduce the length of the recirculation region~\cite{Bou14} and to have a stabilizing effect~\cite{Marquet08cyl}.
Note that both first and second-order flow modifications have much larger amplitudes in the stabilizing case than in the destabilizing case.
Finally, the first sub-optimal stabilizing wall control [Fig.~\ref{fig:Q1_from_opt_Uw_stab}$(c)$] induces double streaks which are qualitatively similar to those of the optimal destabilizing case, although stronger and extending farther downstream. The mean flow correction consists primarily of a strong region of positive streamwise velocity along the centerline. 
The different symmetries correspond to  varicose and sinuous streaks, respectively \cite{Kim05, Choi08, Hwang2013, DelGuercio2014parallel}.

%--- calc_Ex_withtext.m
\begin{figure}
\centerline{  
   \begin{overpic}[width=6.6cm, trim=2mm 64mm 22mm 62mm, clip=true, tics=10]{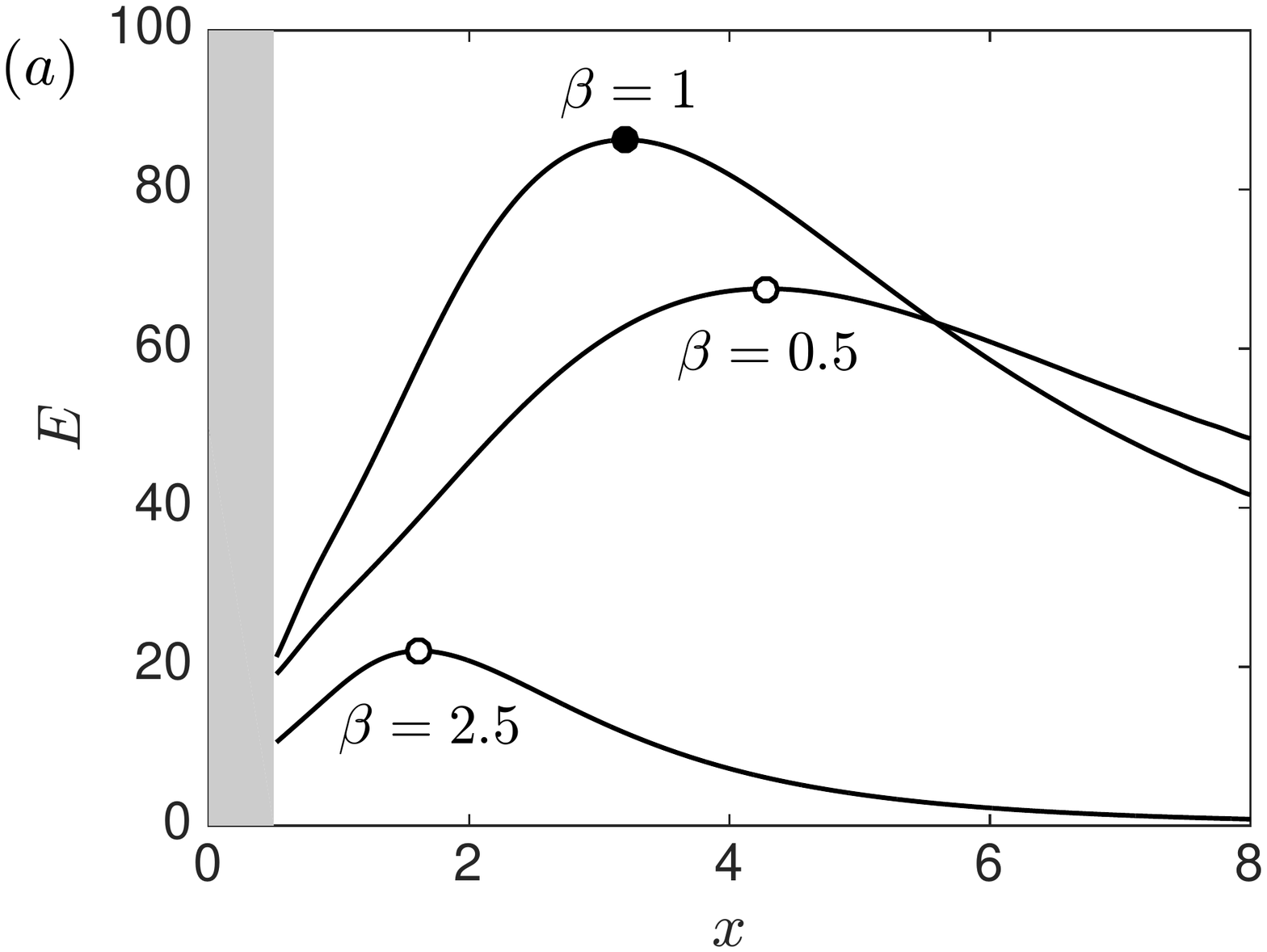} 
   \end{overpic}  
   \hspace{0.5cm} 
   \begin{overpic}[width=6.5cm, trim=0mm 59mm 20mm 60mm, clip=true, tics=10]{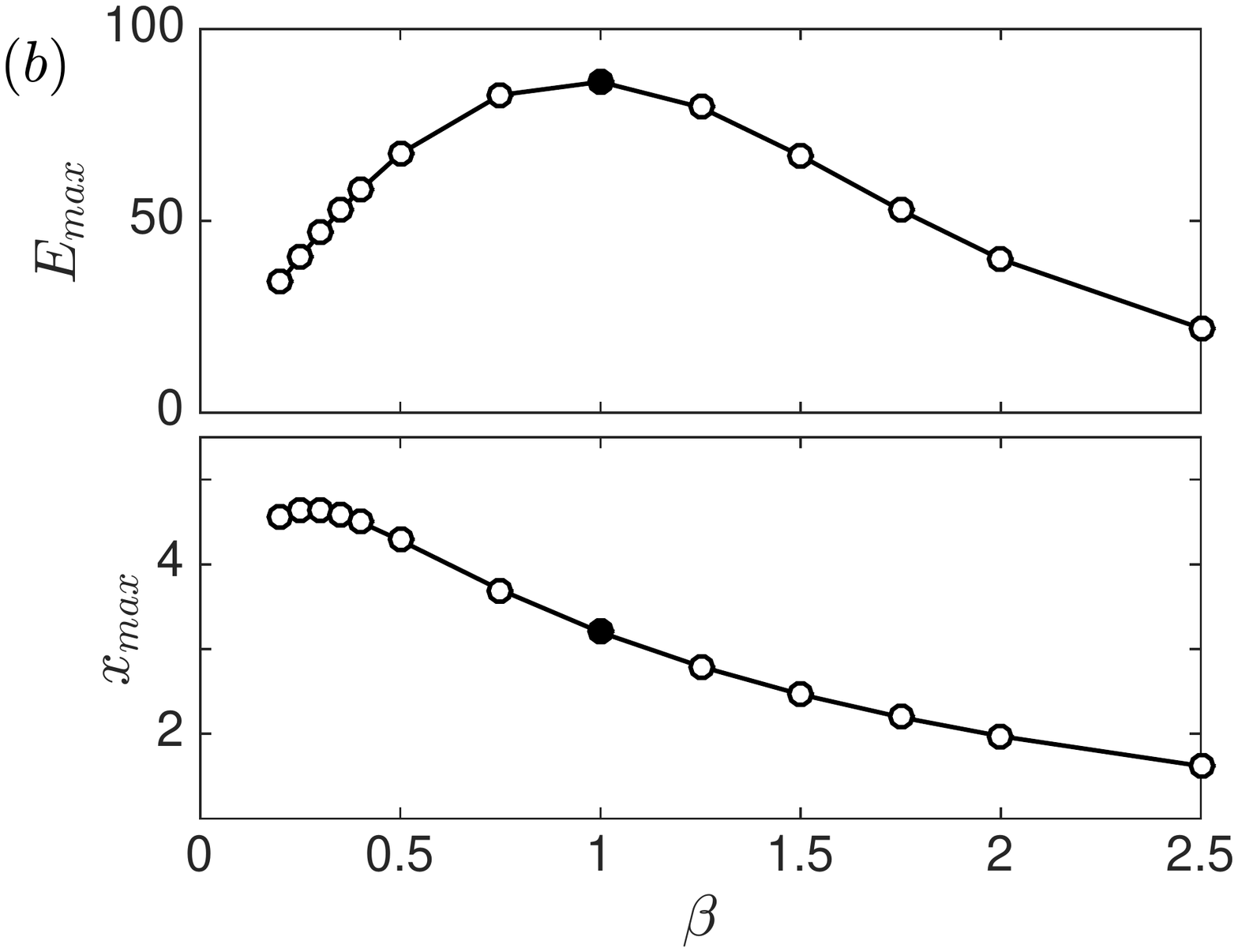} 
   \end{overpic}  
}
\caption{
$(a)$~Energy density $E(x)$ of the flow modification $\QQ_1$ induced by the optimal stabilizing wall control, $\beta=0.5$, 1 and 2.
$(b)$~Maximum energy density and location of the maximum as functions of spanwise wavenumber.
$\Rey=50.$
}  
\label{fig:Ex_vs_x_and_beta}
\end{figure}

The streaks induced by the optimal stabilizing wall control are strongly reminiscent of those induced by the steady, spanwise-periodic wall control optimized for maximal energy amplification, as computed in~\cite{DelGuercio2014globalcyl}.
The streamwise evolution of the streaks, measured by the energy density $E(x)=\int (\widetilde U_1^2+\widetilde V_1^2+\widetilde W_1^2) \,\mathrm{d}y$ shown in Fig.~\ref{fig:Ex_vs_x_and_beta}$(a)$, has a bell shape typical of spatial transient growth: streamwise vortices generated near the cylinder are amplified via the lift-up effect into streamwise streaks, which then decay smoothly by diffusion.
As shown in Fig.~\ref{fig:Ex_vs_x_and_beta}$(b)$, while the location $x_{max}$ of maximal energy density decreases monotonously with $\beta$, the maximum $E_{max}$ itself is largest for $\beta=1$.
This maximum is also strongly increasing with $\Rey$, as shown in Fig.~\ref{fig:Ex_vs_Re}$(a)$, a trend followed closely by the eigenvalue variation.
This increase of $|\ev_{2r}(\Rey)|$ is quicker than that of the linear growth rate $\ev_{0r}(\Rey)$ (approximately exponential and linear close to $\Rey_c$, respectively), which results in the control amplitude $\epsilon_s=\sqrt{\ev_{0r}/ |\ev_{2r}|}$
needed to fully restabilize the flow\footnote{Recall $\ev = \ev_0+\epsilon^2\ev_2$ at second order.}
exhibiting a maximum (for $\Rey\simeq 58$) as shown in Fig.~\ref{fig:Ex_vs_Re}$(b)$.
The decrease in $\epsilon_s$ at larger $\Rey$ seems to suggests that it becomes increasingly easier to stabilize the flow; this point  deserves further investigation because a second eigenmode becomes unstable in the uncontrolled flow, and because the range of validity in $\epsilon$ of the sensitivity prediction may decrease with $\Rey$.

These observations agree in all aspects with those about the optimal streaks of \cite{DelGuercio2014globalcyl}, suggesting that, when using spanwise-periodic wall control, similar mechanisms are at play in optimal spatial growth and optimal stabilization. In other words, our results confirm that maximizing spatial growth is the optimal strategy for stabilizing the flow.

%--- calc_Ex_vs_Re_wall_blowing_withtext.m
\begin{figure}
\centerline{   
   \begin{overpic}[width=6.5cm, trim=0mm 60mm 20mm 60mm, clip=true, tics=10]{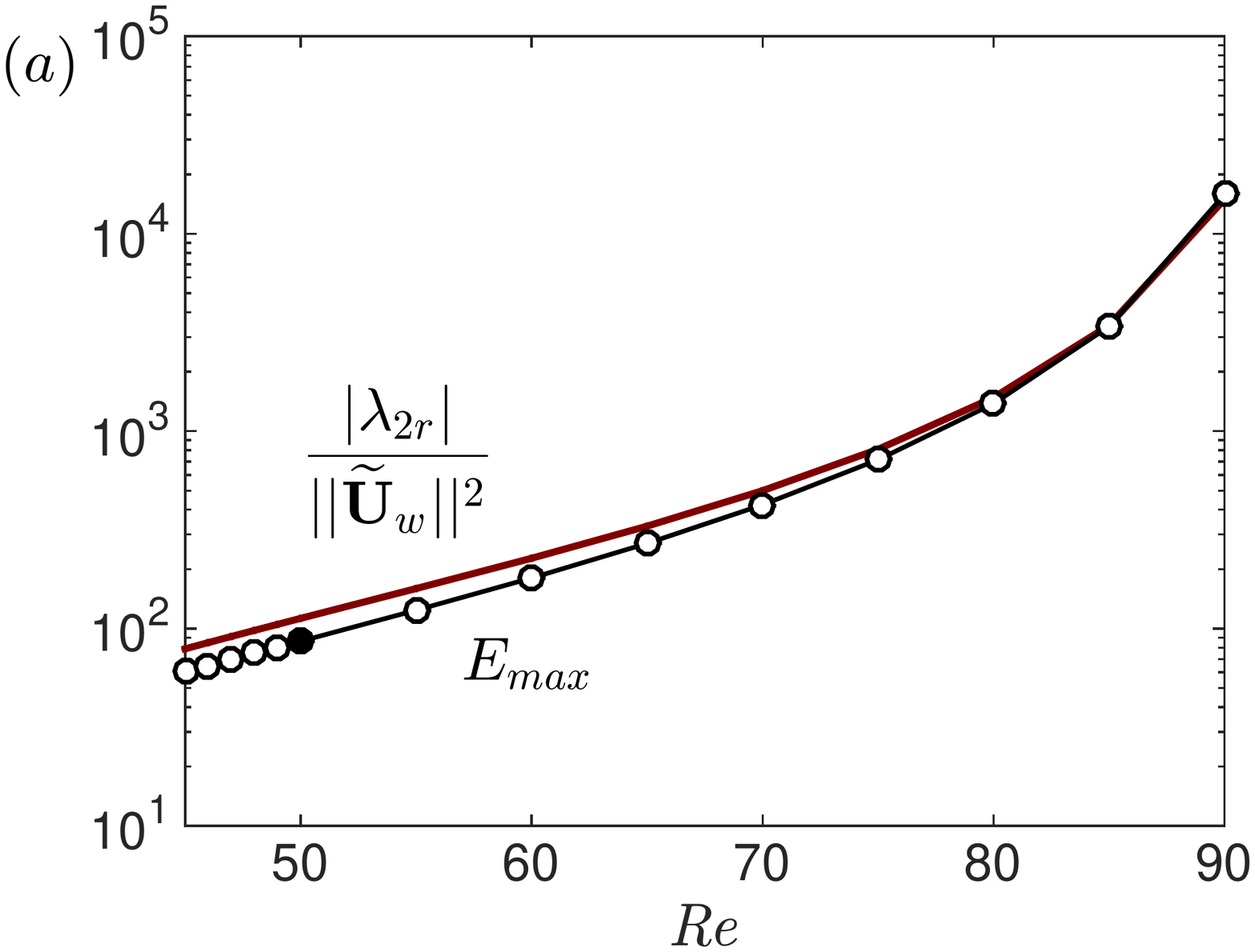} 
   \end{overpic}  
   \hspace{0.5cm} 
   \begin{overpic}[width=6.5cm, trim=0mm 60mm 20mm 60mm, clip=true, tics=10]{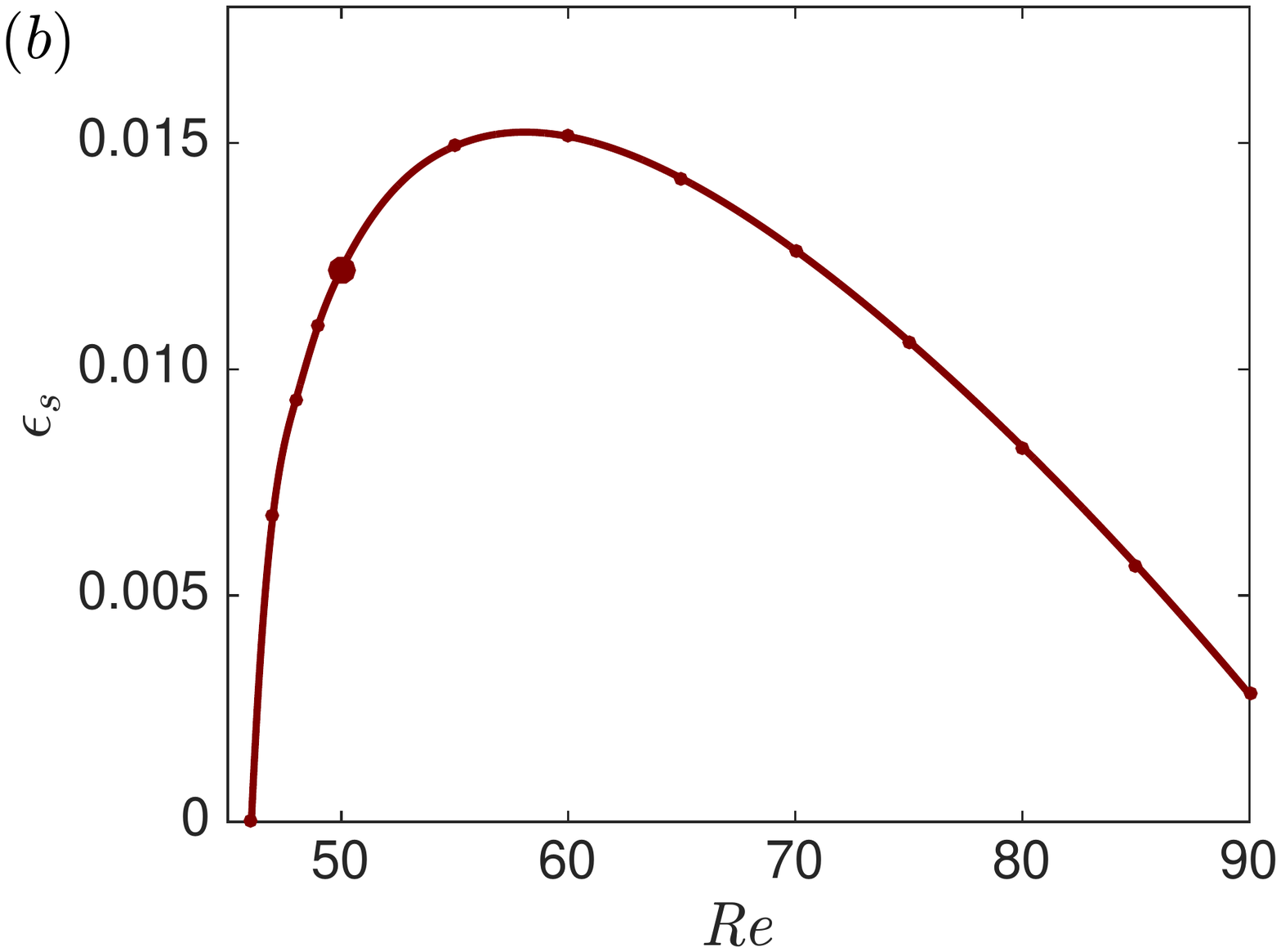} 
   \end{overpic}  
}
\caption{
Variation with Reynolds number.
$(a)$~Normalized growth rate variation (in absolute value), and maximum energy density, both for the flow modification induced by the optimal wall control for stabilization. $(b)$~Wall control amplitude $\epsilon_s=\sqrt{\ev_{0r}/ (|\ev_{2r}|/||\widetilde\UU_w||^2) }$ necessary to fully stabilize the leading eigenvalue.
$\beta=1$.
}  
\label{fig:Ex_vs_Re}
\end{figure}

%-----------------------------------------------
%-----------------------------------------------
\subsection{Competition between amplification and stabilization}

\label{sec:compet_ampli_stab}

We analyze in more detail how much stabilization is due to an efficient \textit{amplification} of the wall control 
into the induced flow modification,
and how much is due to an efficient \textit{stabilization} of this induced flow.
To this aim, we separate the two effects by rewriting the eigenvalue variation, normalized for unit-norm wall control, as
\be 
\dfrac{\lambda_{2r}}{||\widetilde \UU_w||^2}
=
\dfrac{||\widetilde \QQ_1||^2}{||\widetilde \UU_w||^2}
\dfrac{\lambda_{2r}}{||\widetilde \QQ_1||^2}.
\ee
Here
$G^2=||\widetilde \QQ_1||^2/||\widetilde \UU_w||^2$ is the amplification from wall control $\UU_w$ to flow modification $\QQ_1$,
and $\lambda_{2r}/||\widetilde \QQ_1||^2$ is the eigenvalue variation induced by a unit-norm flow modification $\QQ_1/||\widetilde \QQ_1||$.

%--- plot_fig3.m
\begin{figure}
\centerline{     
   \begin{overpic}[width=6.5cm, trim=0mm 60mm 20mm 60mm, clip=true, tics=10]{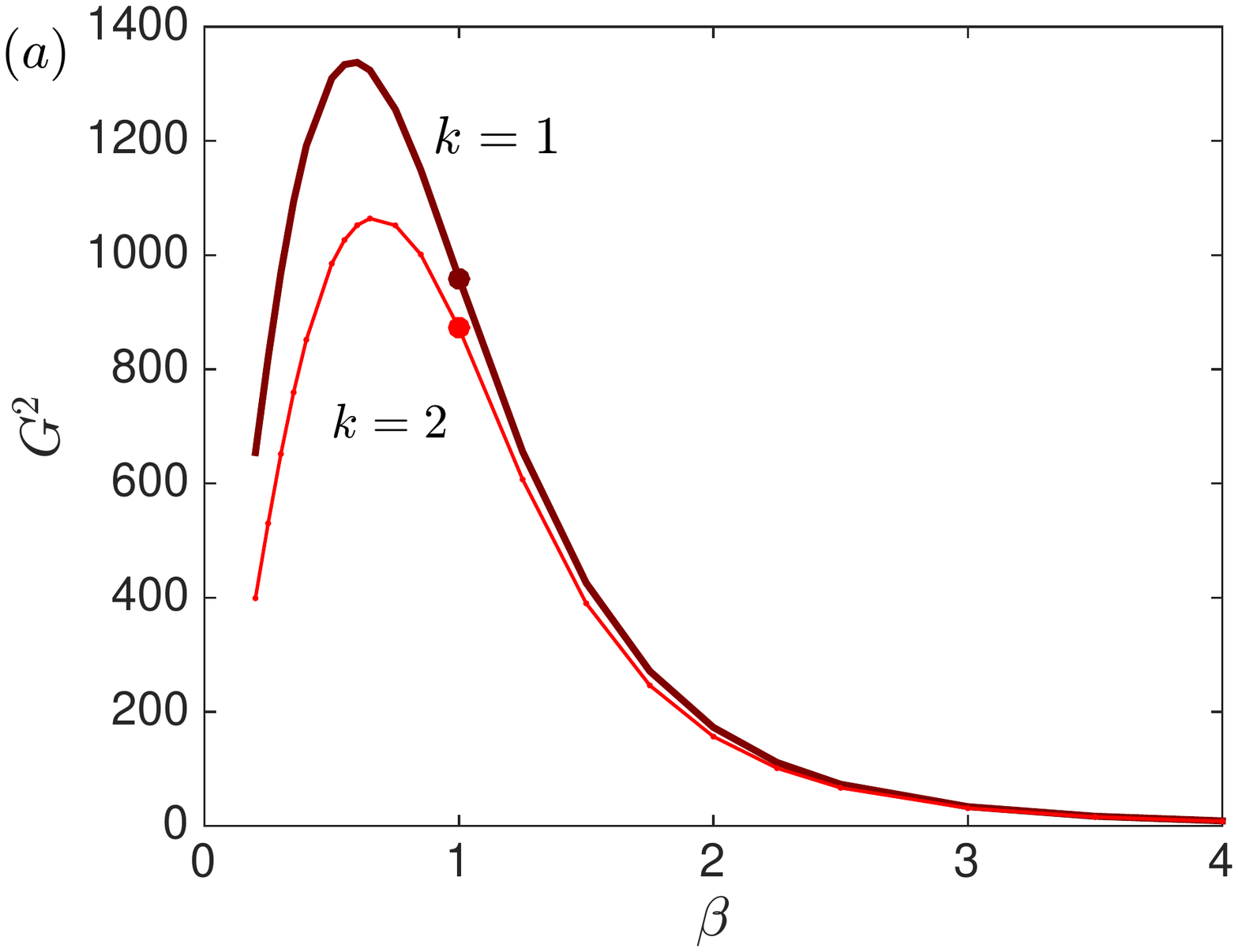}         
   \end{overpic}        
   \hspace{0.5cm}  
   \begin{overpic}[width=6.5cm, trim=0mm 60mm 20mm 60mm, clip=true, tics=10]{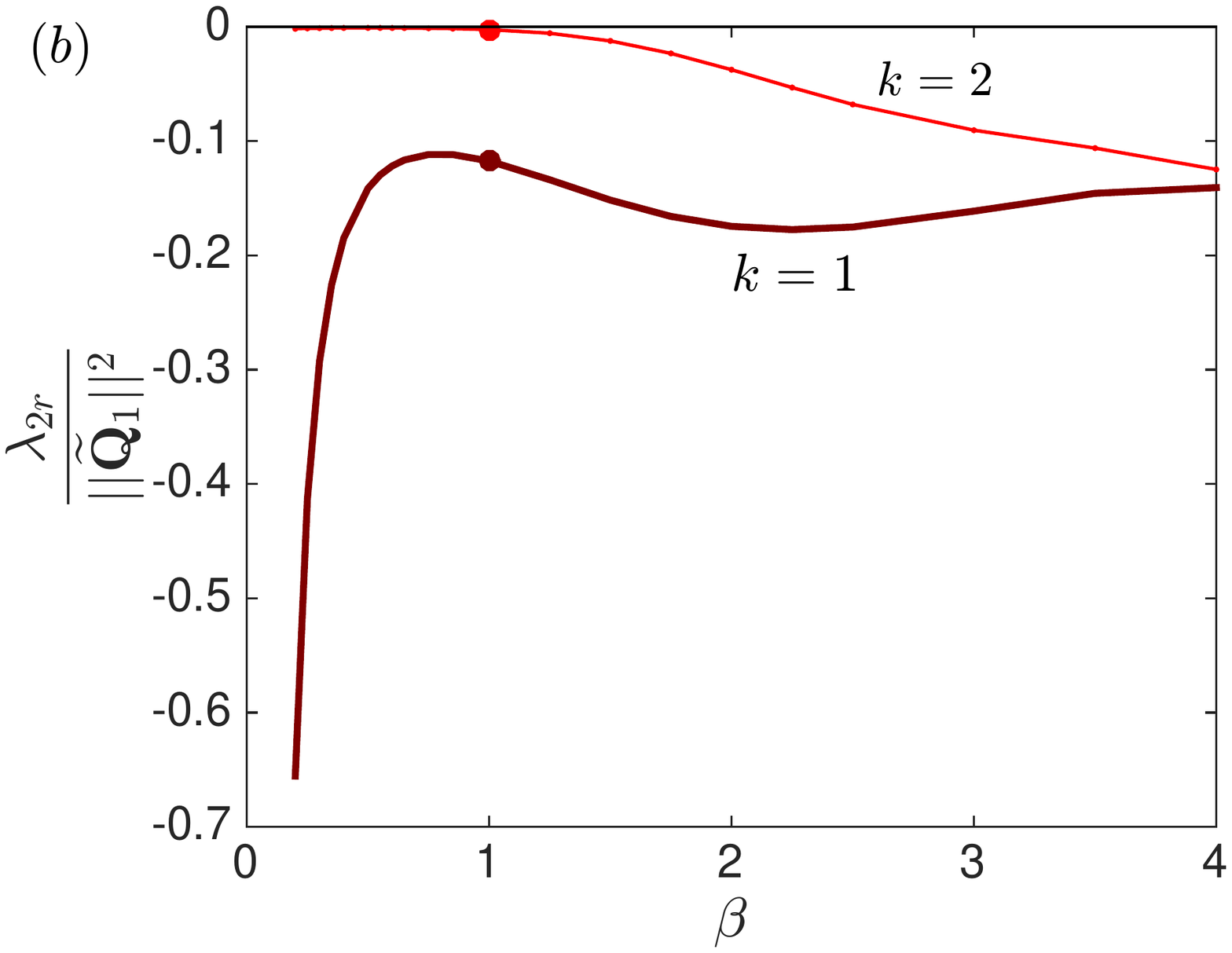}        
   \end{overpic} 
}
\caption{
Decomposition into amplification and stabilization:
$(a)$~gain $G^2=||\QQ_1||^2/||\UU_w||^2$ from wall control to resulting flow modification;
$(b)$~growth rate reduction induced by unit flow modification (same data as Fig.~\ref{fig:lam2r_opt_Uw_r}, normalized by the response norm $||\QQ_1||^2$ rather than the control norm $||\UU_w||^2$).
Optimal ($k=1$) and first sub-optimal ($k=2$) stabilizing wall control. 
$\Rey=50.$
}  
\label{fig:lam2r_opt_Uw_r_normQ1}
\end{figure}

Figure~\ref{fig:lam2r_opt_Uw_r_normQ1} shows that the optimal wall control ($k=1$) is only slightly more amplified into varicose streaks than the first sub-optimal wall control ($k=2$) is amplified into sinuous streaks (e.g. at $\beta=1$, the ratio of gains $G^2$ is $1.1$); 
however, varicose streaks have a much more stabilizing structure than their sinuous counterparts (ratio of $\lambda_{2r}/||\widetilde\QQ_1||^2$ values $\simeq 50$ for $\beta=1$).
Conversely, the second sub-optimal wall control ($k=3$, not shown) induces a flow structure that is actually more stabilizing than the varicose streaks, but experiences such a poor amplification that the net effect is smaller. 
This suggests that optimizing for stabilization only might yield flow modifications requiring impractically large control amplitudes,
and that optimizing for amplification and stabilization simultaneously should be preferred.

%-----------------------------------------------
%-----------------------------------------------
\subsection{3D and 2D contributions}

\label{sec:compet_3D_2D}

As mentioned in section \ref{sec:22}, the second-order eigenvalue variation resulting from a spanwise-periodic control is the sum of two effects: from
(i)~the first-order flow modification $\QQ_1$, and 
(ii)~the second-order flow modification $\QQ_2$ 
(see also (\ref{eq:evp2}) and (\ref{eq:ev2})). 
Specifically, effect (i)~is an interaction between $\QQ_1$ and the 
first-order eigenmode modification $\qq_1$, both of which are spanwise-periodic; this contribution is therefore denoted \textit{3D contribution}.
Conversely, effect (ii)~is an interaction between the spanwise-invariant component of $\QQ_2$ (mean flow correction $\QQ_2^{2D}$) and the original eigenmode $\qq_0$; this contribution is therefore denoted \textit{2D contribution} (not to be confused with the eigenvalue variation that would be induced by a spanwise-uniform control).

Figure~\ref{fig:lam2r_opt_Uw_r_3D_2D} shows these two contributions for optimal stabilizing controls.
The optimal wall control ($k=1$, panel $a$) mainly has a stabilizing effect via its 3D contribution for small $\beta$. 
This contrasts with the observations of Marant and Cossu \cite{Marant2018} on the time-evolving parallel shear layer flow.
As $\beta$ increases and the amplitude of the optimal streaks decreases (Fig.~\ref{fig:Ex_vs_x_and_beta}), the relative 2D contribution increases and reaches $50\%$ for $\beta=4$.
Surprisingly, the first sub-optimal wall control ($k=2$, panel $b$) mainly has a stabilizing effect via its 2D contribution, while the 3D contribution is actually \textit{destabilizing} 
up to $\beta \lesssim 2.5$. 
Note how this destabilizing effect shifts the optimal wavenumber from $\beta=1.5$ (2D contribution only) to $\beta=2$ (net effect).
Two very different mechanisms can therefore be distinguished:
(i)~varicose streaks stabilizing via their direct spanwise-periodic effect on $\QQ_1$and $\qq_1$, 
and
(ii)~sinuous streaks stabilizing via their mean flow correction $\QQ_2^{2D}$.

%--- plot_fig3.m
\begin{figure}
\def \thiswidth {4.5cm}
\centerline{    
   \begin{overpic}[width=6.5cm, trim=0mm 60mm 20mm 60mm, clip=true, tics=10]{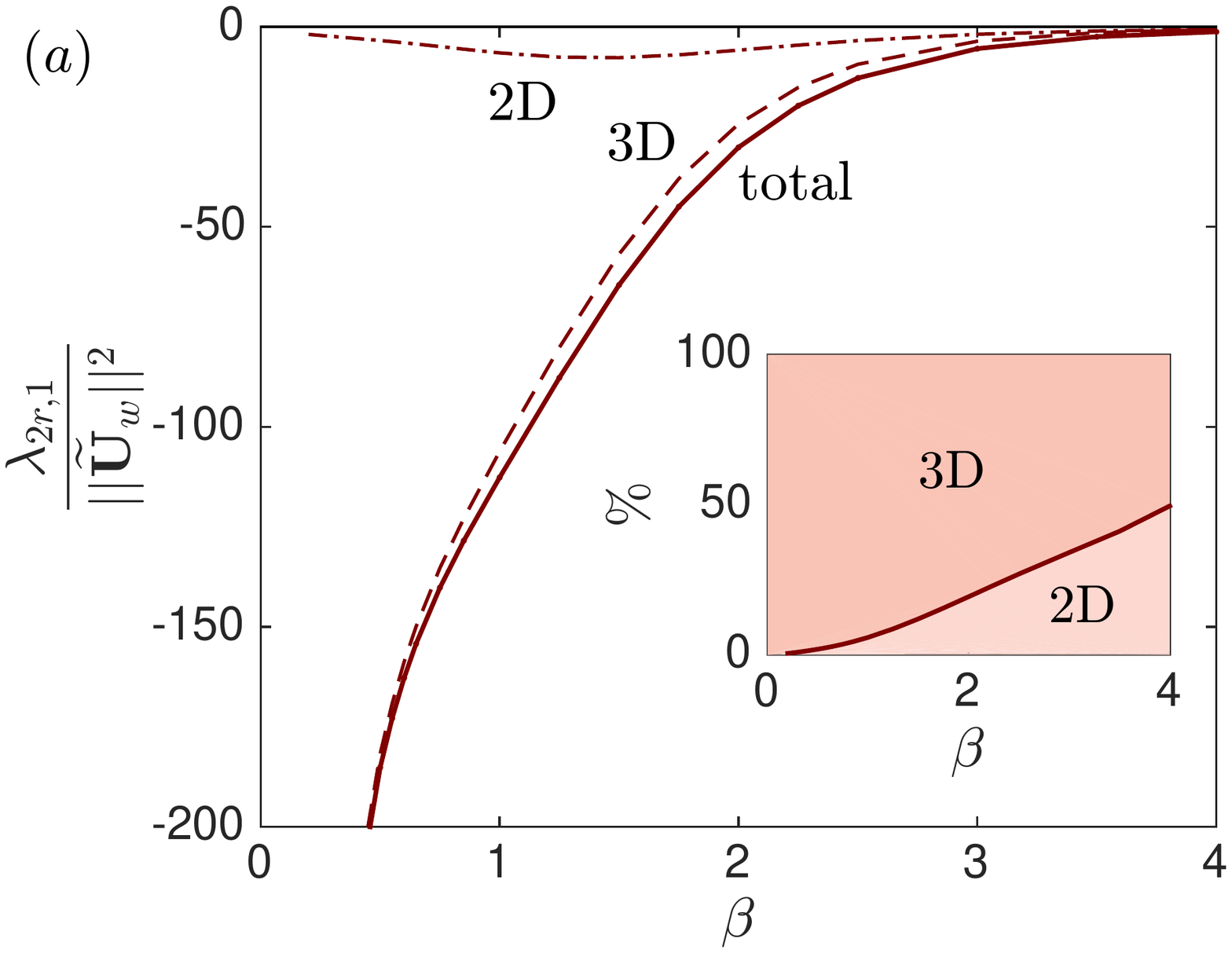}
   \end{overpic}  
   \hspace{0.5cm}
   \begin{overpic}[width=6.5cm, trim=0mm 60mm 20mm 60mm, clip=true, tics=10]{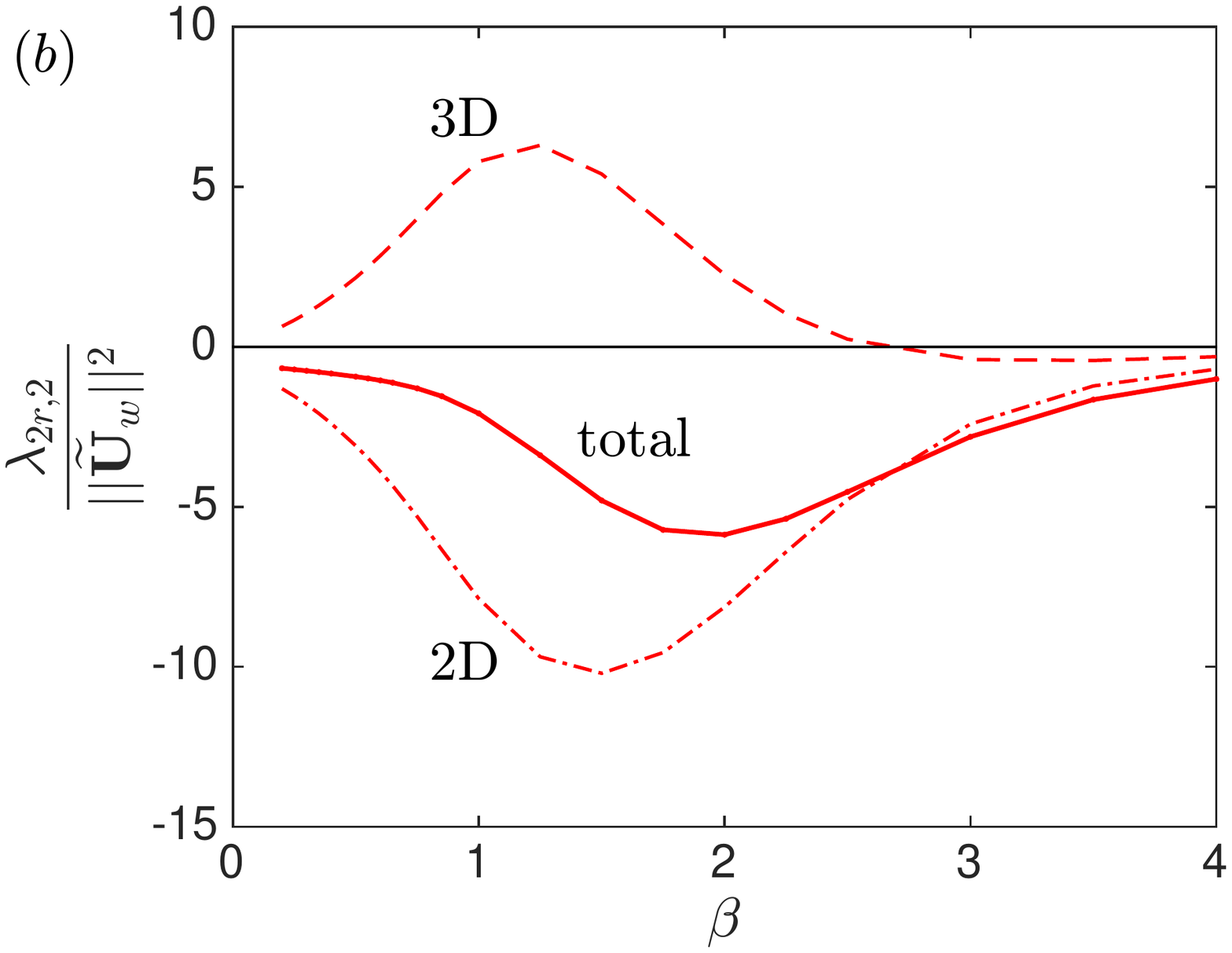} 
   \end{overpic}  
}
\caption{
3D contribution ($- -$, interaction between 3D fields $\QQ_1$ and $\qq_1$) and 2D contribution ($-\cdot-$, interaction between 2D fields $\QQ_2^{2D}$ and $\qq_0$) 
to the total growth rate variation (solid line) induced by 
$(a)$ the optimal ($k=1$) and 
$(b)$ first sub-optimal ($k=2$) stabilizing wall control $\UU_w$.
$\Rey=50.$
See text for details.
}  
\label{fig:lam2r_opt_Uw_r_3D_2D}
\end{figure}

%-----------------------------------------------
%-----------------------------------------------
\subsection{Effect on frequency}

\label{sec:effect_on_freq}

It is important, in some applications, to know the overall effect of the optimal control.
Figure~\ref{fig:lam2i_opt_Uw_r}
shows the effect on frequency ($\ev_{2i}$) of the control optimized for stabilization ($\ev_{2r}<0$)  discussed so far in this section~\ref{sec:growth-wallactu}.
For all $\beta$ values, the optimal stabilizing control is seen to induce a frequency \textit{increase} ($\ev_{2i}>0$),  smaller that the growth rate decrease.
Interestingly, this frequency increase is close to the optimal frequency increase, and the associated controls are similar, as will be seen in section~\ref{sec:freq}.

We note that the effect in Fig.~\ref{fig:lam2i_opt_Uw_r} is mainly due to the 3D contribution (not shown), and that 
the first stabilizing sub-optimal $k=2$ induces a frequency \textit{decrease} (not shown).

%--- plot_fig3.m
\begin{figure}
\centerline{   
   \begin{overpic}[width=6.5cm, trim=0mm 60mm 20mm 60mm, clip=true, tics=10]{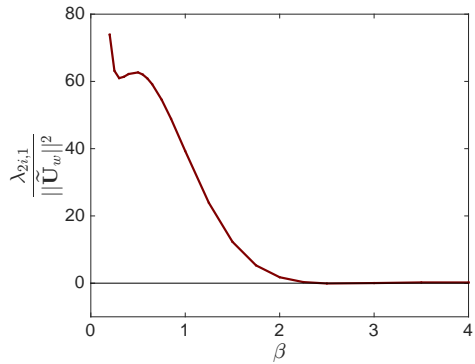}      
   \end{overpic} 
}
\caption{
Normalized effect on frequency of the optimal stabilizing wall control $\UU_w$.
$\Rey=50.$
}  
\label{fig:lam2i_opt_Uw_r}
\end{figure}

%-----------------------------------------------
%-----------------------------------------------
\subsection{Simplified wall actuation}
\label{sec:simplified_wall}

Figure~\ref{fig:opt_Uw_dest_stab} shows that the spanwise component of the optimal control is generally small compared to the normal and tangential components.
We now assess quantitatively the contributions of the different velocity components by optimizing for simplified wall controls: namely in-plane (normal and tangential components only $\UU_w=(U_n,U_t,0)^T$, no spanwise component $W$), normal ($\UU_w=(U_n,0,0)^T$), or tangential ($\UU_w=(0,U_t,0)^T$).
This is implemented by restricting the prolongation operator $\PP$ to the velocity components of interest (see sections \ref{sec:sensit_general}-\ref{sec:sensit_spanwise_periodic}).

Figures~\ref{fig:compare_UnUtUw_UnUt_Un_Ut}-\ref{fig:opt_Uw_stab_simple} show that the
in-plane optimal control and 
normal optimal control
are very similar to the full 3D optimal control, and lead to very similar eigenvalue variations.
This implies that $U_n$ is by far the most effective component, while $W$ has a negligible effect.

Interestingly, purely tangential wall control is much less effective (Fig.~\ref{fig:compare_UnUtUw_UnUt_Un_Ut}), but the  flow modification induced downstream of the cylinder is qualitatively similar (not shown) to the that in Fig.~\ref{fig:Q1_from_opt_Uw_stab}$(b)$, indicating that  the  optimal stabilizing mechanism is the creation of streamwise streaks, irrespective of the velocity component(s) used as wall control.

Note that we have also evaluated $\ev_2$ by simply setting $W=0$ or $U_t=W=0$ \textit{a posteriori} in the full 3D optimal control $(U_n,U_t,W)^T$, and noticed only minor changes compared to the  
in-plane optimal control and  
normal optimal control.

%--- plot_comparison_UnUtW_UnUt_Un_Ut_R1.m
\begin{figure}
\centerline{   
   \begin{overpic}[width=7.cm, trim=0mm 60mm 20mm 60mm, clip=true, tics=10]{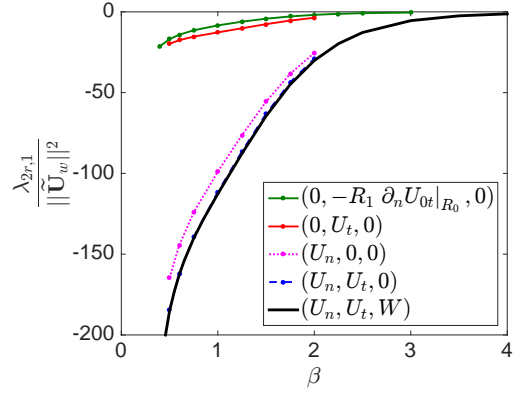}     
   \end{overpic}  
 }  
\caption{
Normalized growth rate variation for several optimal stabilizing wall controls $\UU_w$:
 full 3D   optimal, % $(U_n,U_t,W)$, 
in-plane   optimal, % $(U_n,U_t,0)$,
normal     optimal, % $(U_n,0,  0)$,
tangential optimal. % $(0,U_t,  0)$,
Also shown is the variation for the tangential wall control equivalent to the optimal wall deformation $R_1$
(see section~\ref{sec:growth-walldef}). 
$\Rey=50$, $\beta=1$.
}  
\label{fig:compare_UnUtUw_UnUt_Un_Ut}
\end{figure}

%--- plot_fig13.m
\begin{figure}
\centerline{   
   \begin{overpic}[width=5.7cm, trim=0mm 75mm 40mm 65mm, clip=true, tics=10]{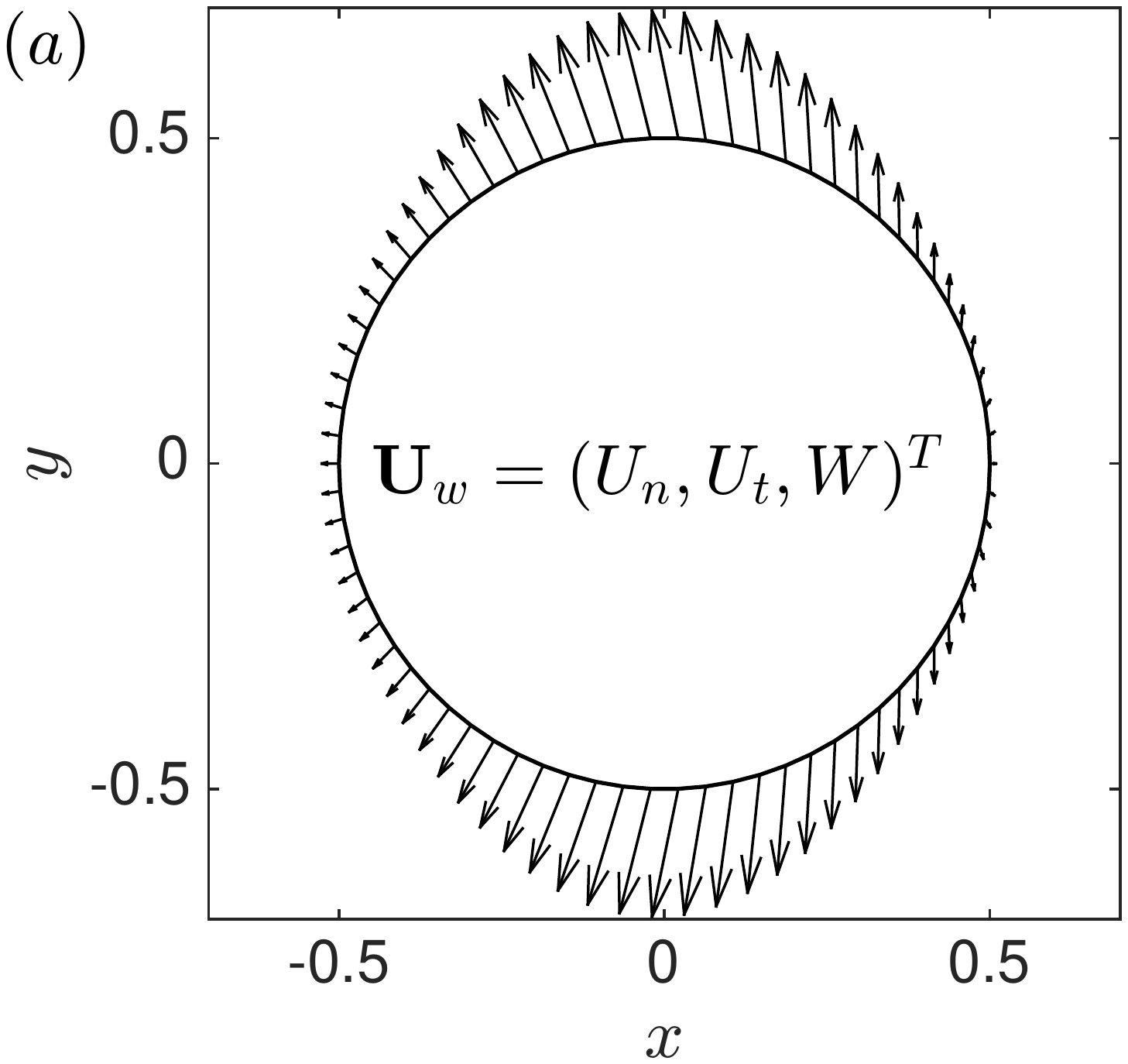} 
   \end{overpic} 
   \begin{overpic}[width=6.cm, trim=15mm 75mm 15mm 65mm, clip=true, tics=10]{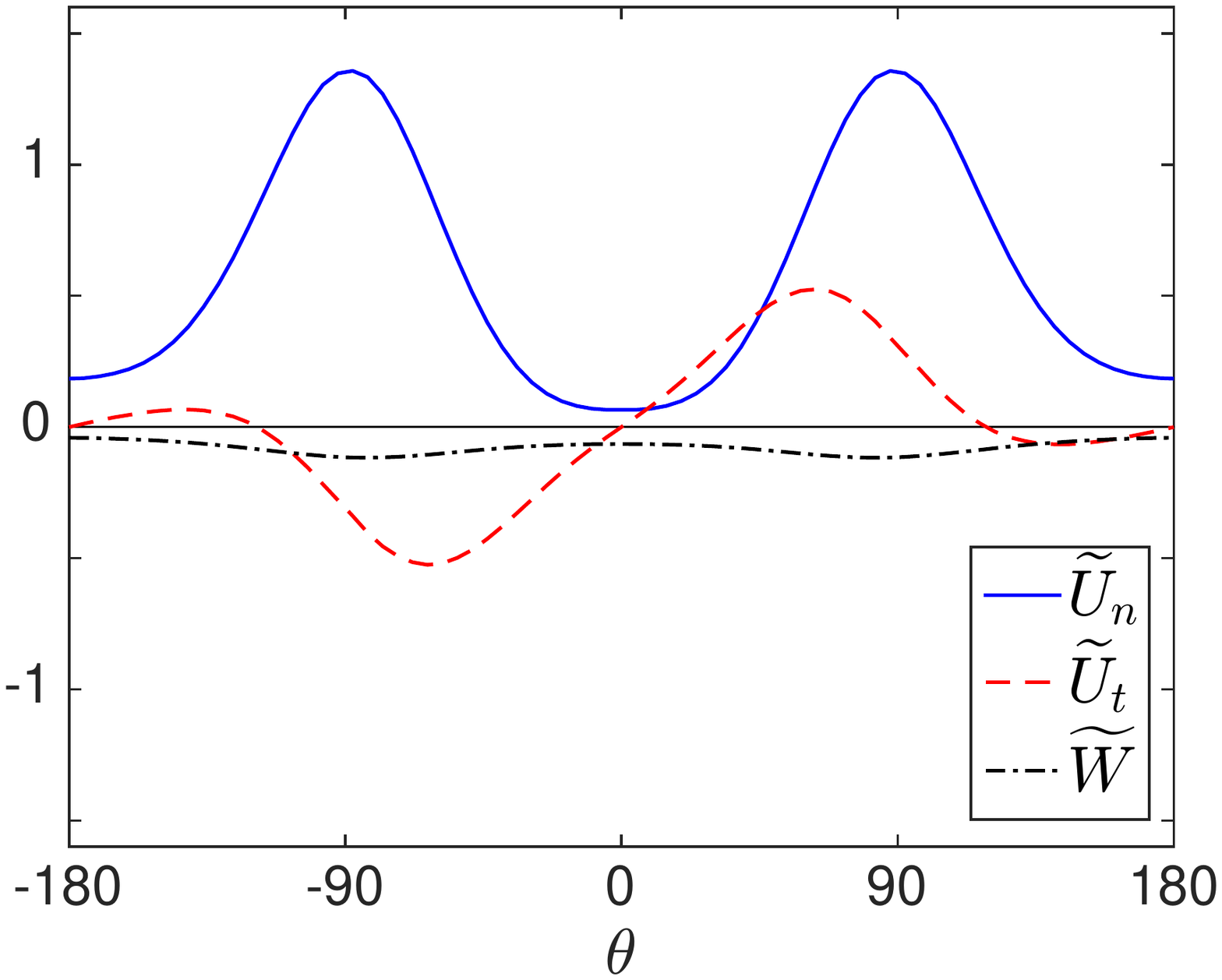}   
   \end{overpic}  
}
\centerline{   
   \begin{overpic}[width=5.7cm, trim=0mm 75mm 40mm 65mm, clip=true, tics=10]{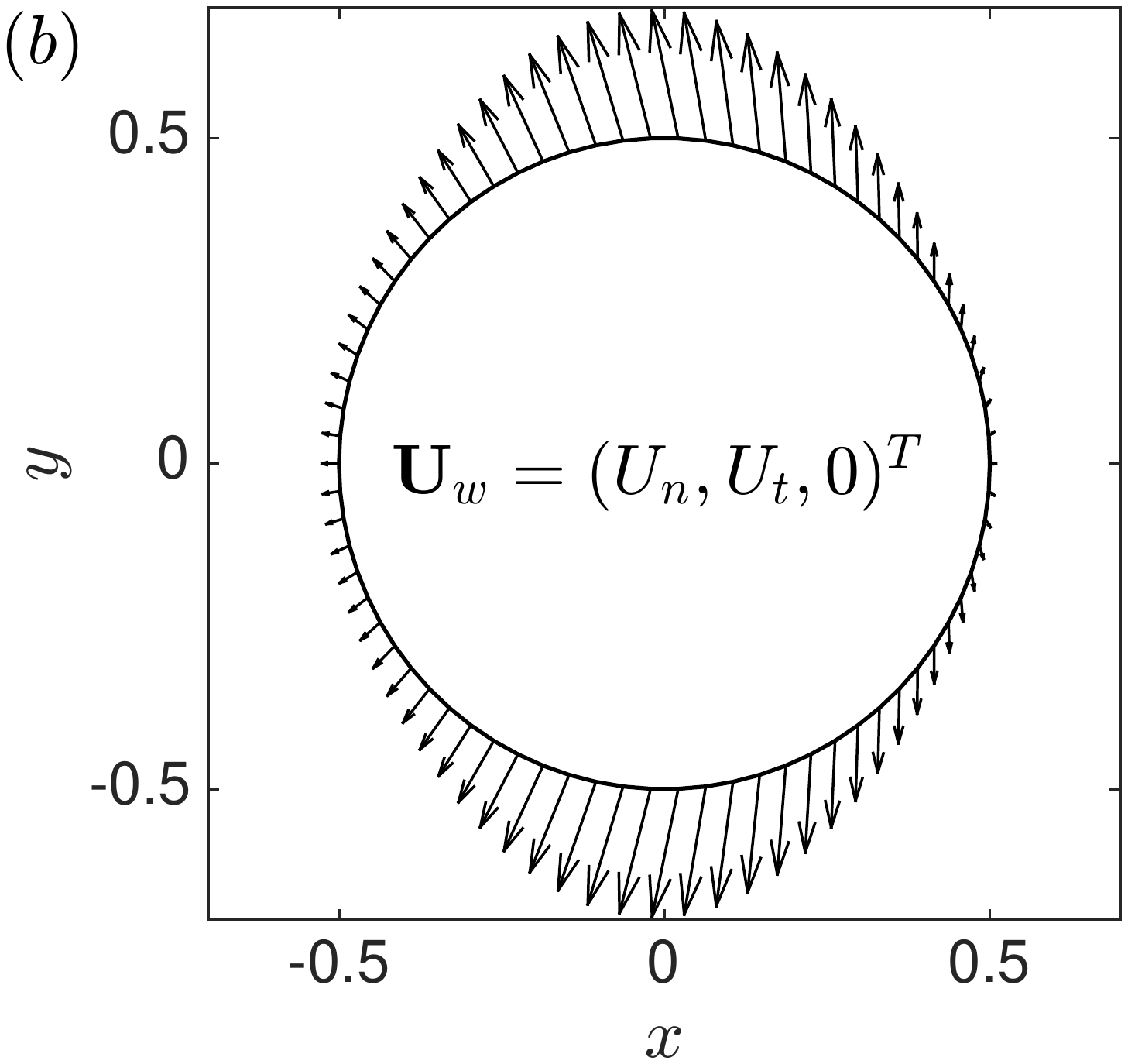} 
   \end{overpic} 
   \begin{overpic}[width=6.cm, trim=15mm 75mm 15mm 65mm, clip=true, tics=10]{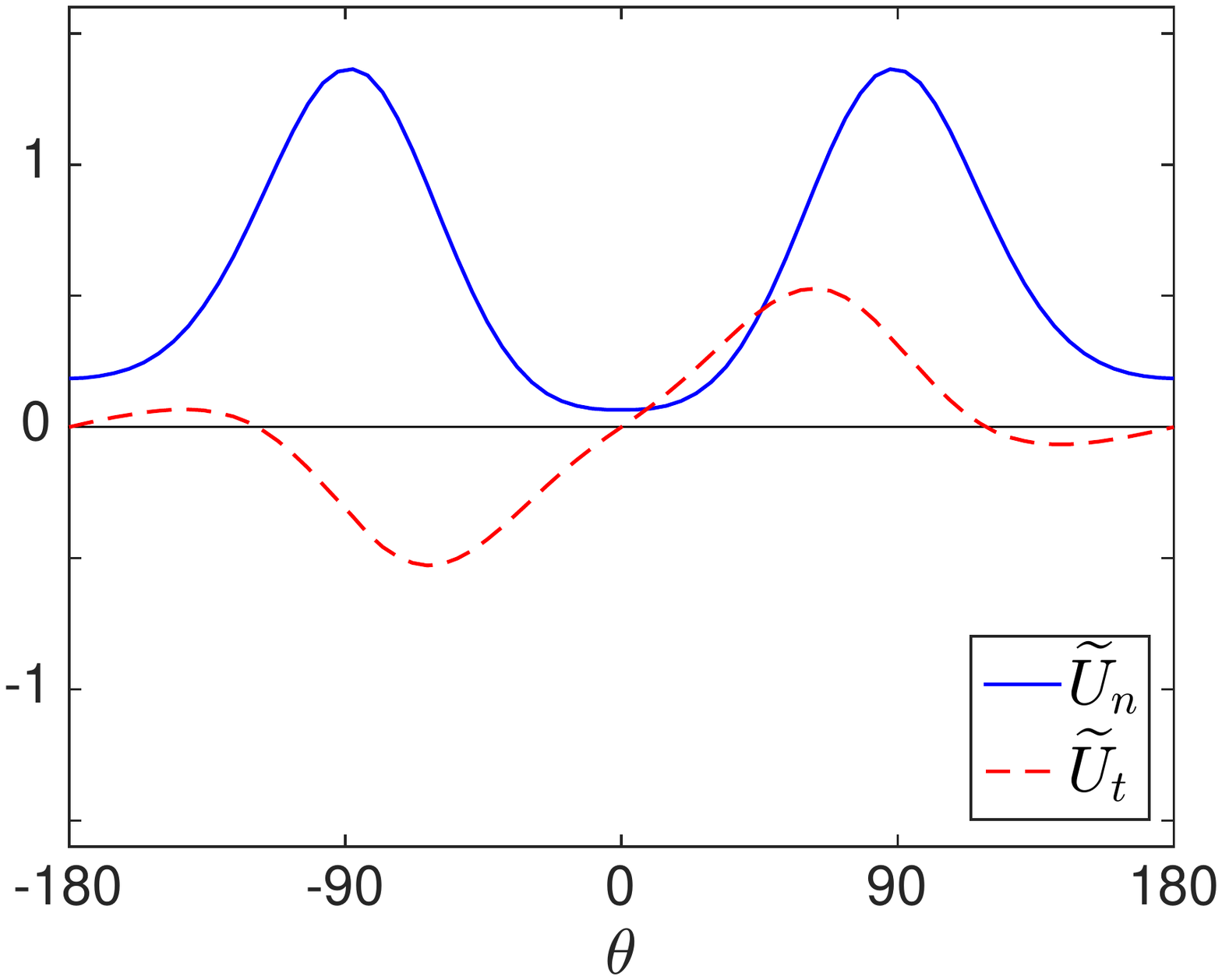}   
   \end{overpic}  
}
\centerline{   
   \begin{overpic}[width=5.7cm, trim=0mm 65mm 40mm 65mm, clip=true, tics=10]{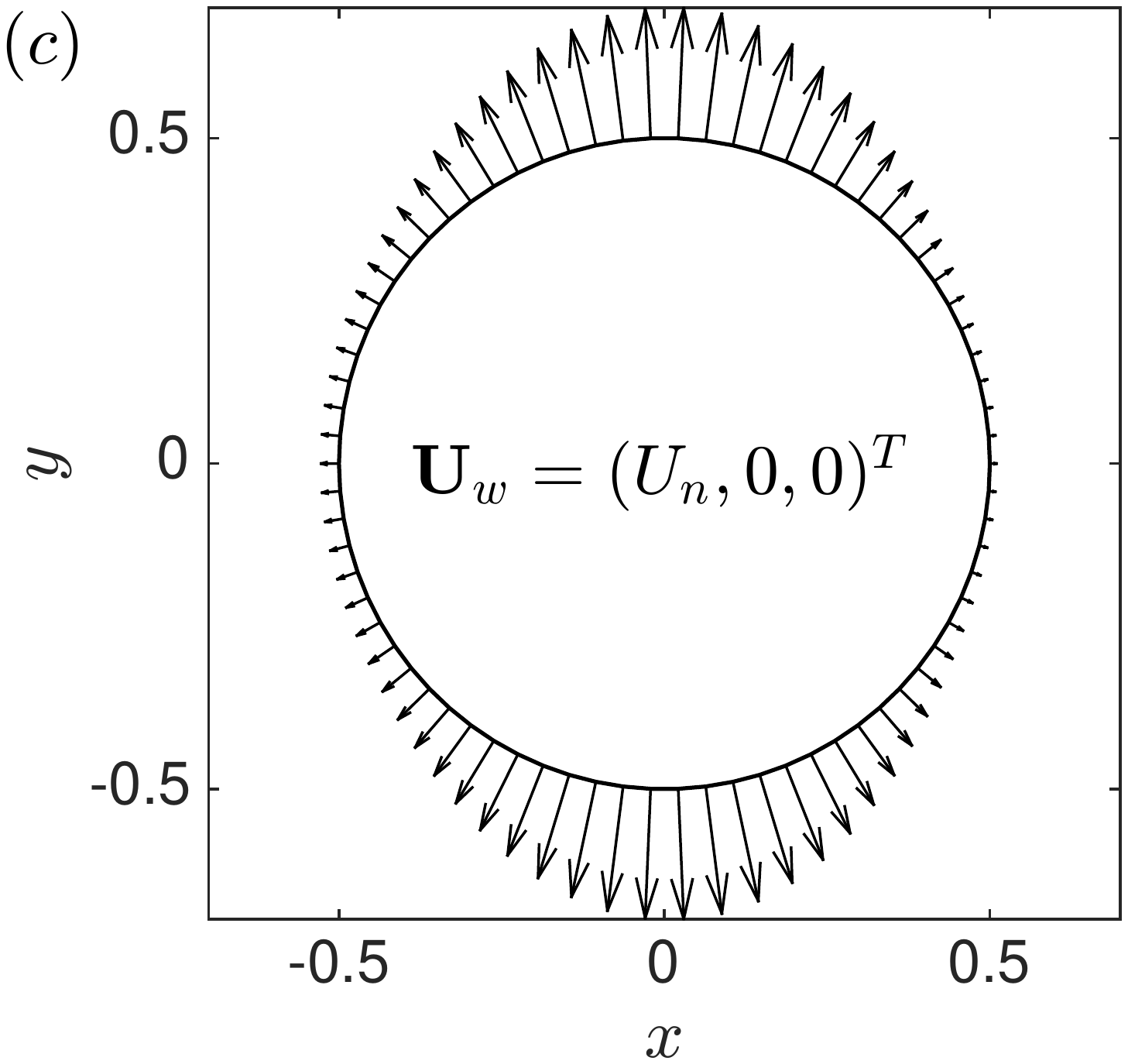} 
   \end{overpic} 
   \begin{overpic}[width=6.cm, trim=15mm 65mm 15mm 65mm, clip=true, tics=10]{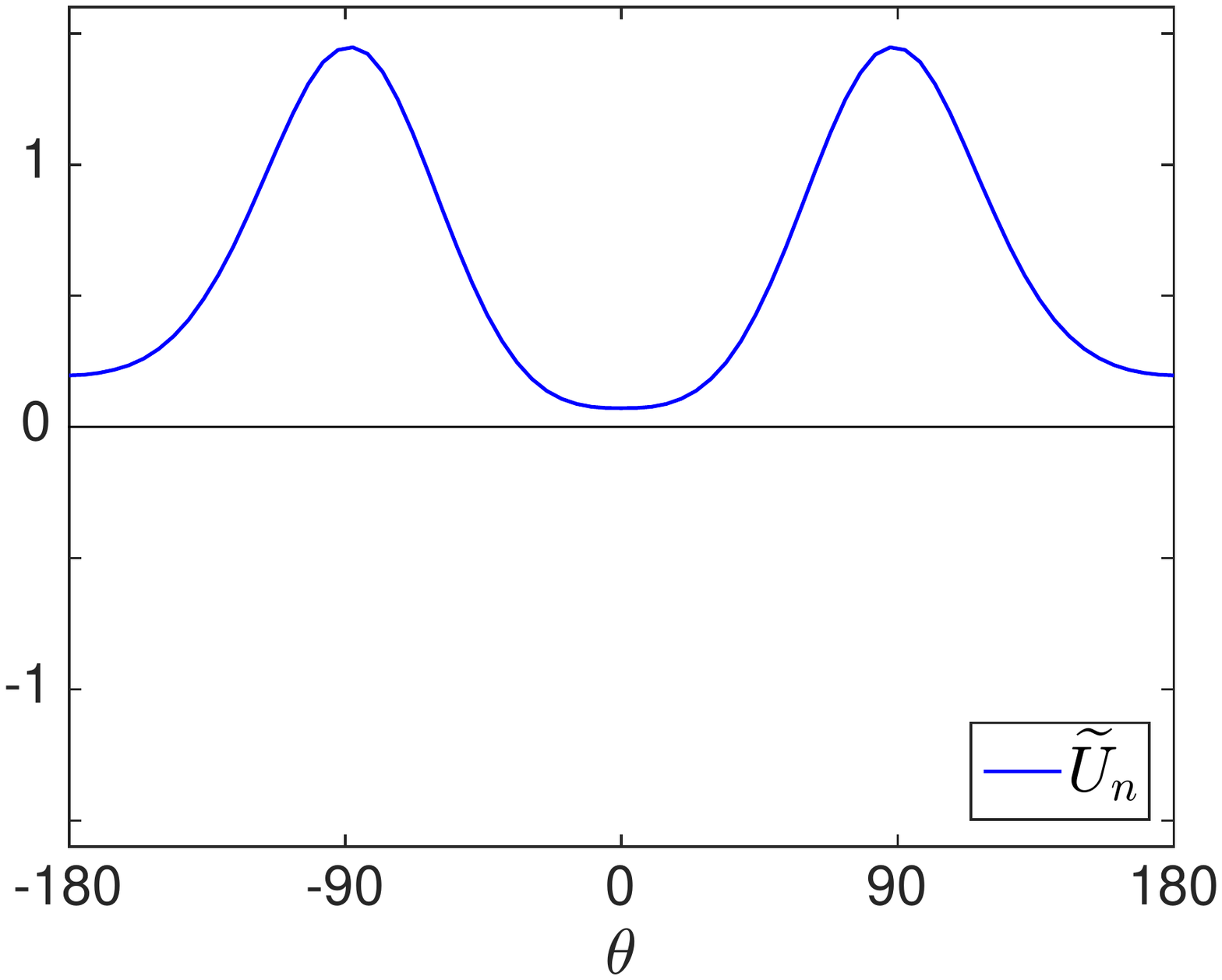}   
   \end{overpic}  
}
\caption{
Optimal stabilizing $\widetilde \UU_w$ normalized to 1:
$(a)$~full optimal 3D control,
%$\UU_w=(U_n,U_t,W)^T$, 
$(b)$~in-plane optimal control, 
%$(U_n,U_t,0)^T$,
$(c)$~normal optimal control.
%$(U_n,0,0)^T$.
$\Rey=50$, $\beta=1$.
}  
\label{fig:opt_Uw_stab_simple}
\end{figure}

%----------------------------------------
%-----------------------------------------------
%-----------------------------------------------
\section{Optimal wall deformation for stabilization}
\label{sec:growth-walldef}

Wall deformation constitutes an interesting alternative to wall blowing/suction, in particular thanks to a relatively easier implementation. 
We therefore consider spanwise-periodic wall deformation $R(\theta,z) = R_0+\epsilon R_1(\theta,z)
= R_0+\epsilon \widetilde R_1(\theta) \cbz$, 
as described in section \ref{sec:opt_wall_def},
and compute the wall deformations that optimally stabilize or destabilize the leading eigenmode.

\begin{figure}
\centerline{    
   \begin{overpic}[width=6.5cm, trim=5mm 60mm 20mm 65mm, clip=true, tics=10]{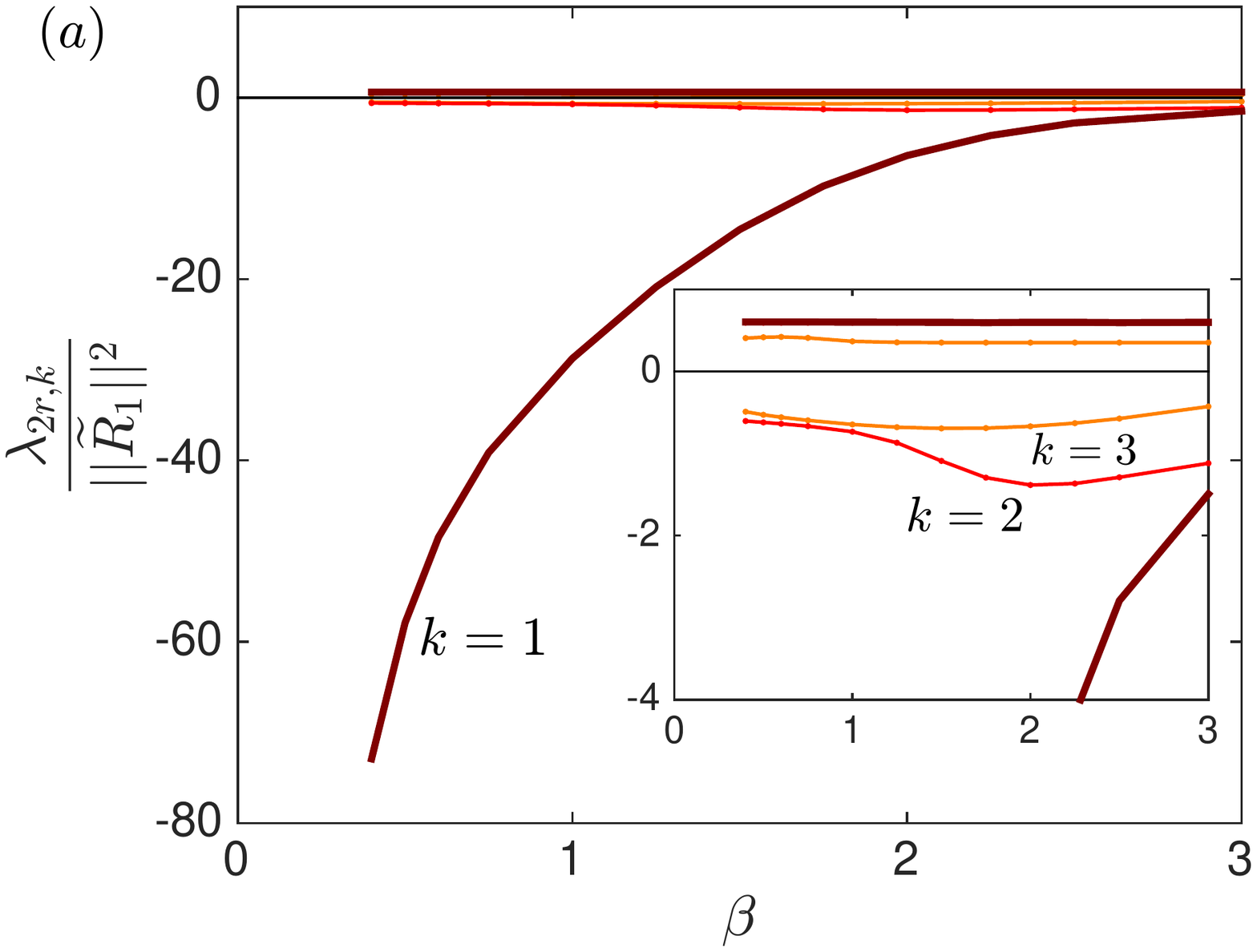} 
   \end{overpic}  
   \hspace{1cm}
   \begin{overpic}[width=6.5cm, trim=5mm 60mm 20mm 65mm, clip=true, tics=10]{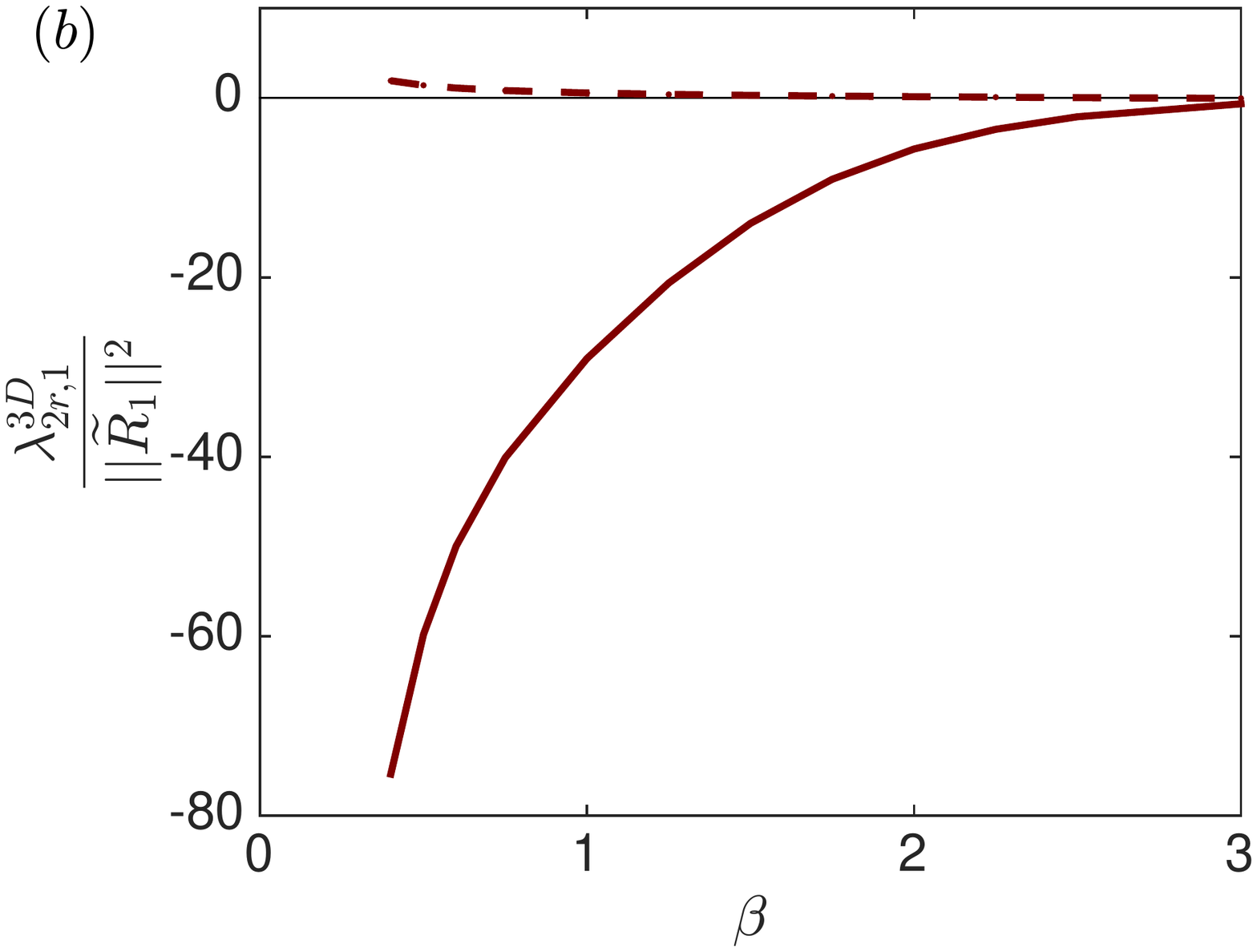} 
   \end{overpic}  }
\caption{
$(a)$~Normalized growth rate variation induced by the  optimal wall deformation $R_1$ for stabilization ($\lambda_{2r}<0$) and destabilization ($\lambda_{2r}>0$).
Optimal ($k=1$) and first sub-optimals ($k=2$, $3$).
$\Rey=50.$
$(b)$~The effect of wall deformation via the eigenmode's boundary condition is much smaller than via  the base flow modification: 
evaluation of (\ref{eq:ev2q1explicit}) with $\widetilde\qq_{1}$ (solid line), and difference between the evaluations with $\widetilde\qq_{1,def}$ and  $\widetilde\qq_{1}$ (dashed line). 
See text for details.
} 
\label{fig:lam2r_opt_WALLDEF_r} 
\end{figure}

Focusing first on the optimal stabilizing wall deformation, we come back to Fig.~\ref{fig:compare_UnUtUw_UnUt_Un_Ut} and observe that the growth rate variation (green line)
follows with $\beta$ a trend qualitatively similar to wall actuation.
In quantitative terms, the variation (normalized with respect to the equivalent wall velocity) is close to that of the optimal tangential actuation (red line).

We now move to Fig.~\ref{fig:lam2r_opt_WALLDEF_r}$(a)$, where the optimal and first sub-optimal growth rate variations at $\Rey=50$ are shown as function of the spanwise wavenumber $\beta$
(now normalized with respect to wall deformation). 
Similar to wall actuation (section \ref{sec:growth-wallactu}), several observations can be made:
(i) the effect of the optimal stabilizing wall deformation is decreasing with $\beta$,
(ii) sub-optimal stabilizing wall deformations are much less efficient than the optimal one,
(ii) the potential for destabilization is much smaller than that for stabilization.

%--- plot_opt_R1_and_equivUt_withtext.m
\begin{figure}[]
\centerline{   
   \begin{overpic}[height=5.cm, trim=5mm 63mm 20mm 65mm, clip=true, tics=10]{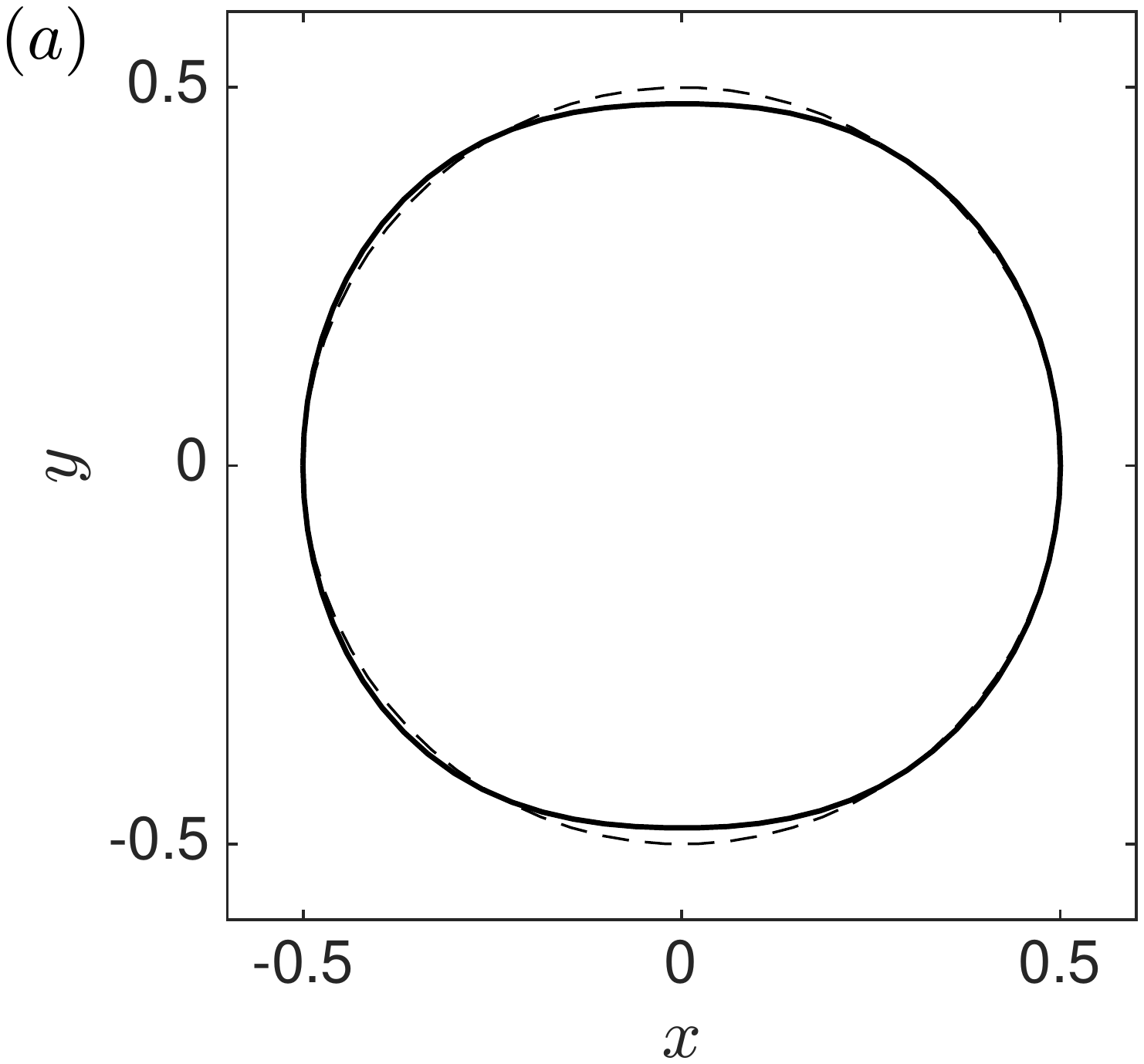} 
   \end{overpic}
   \hspace{1cm}
   \begin{overpic}[height=5cm, trim=5mm 58mm 15mm 65mm, clip=true, tics=10]{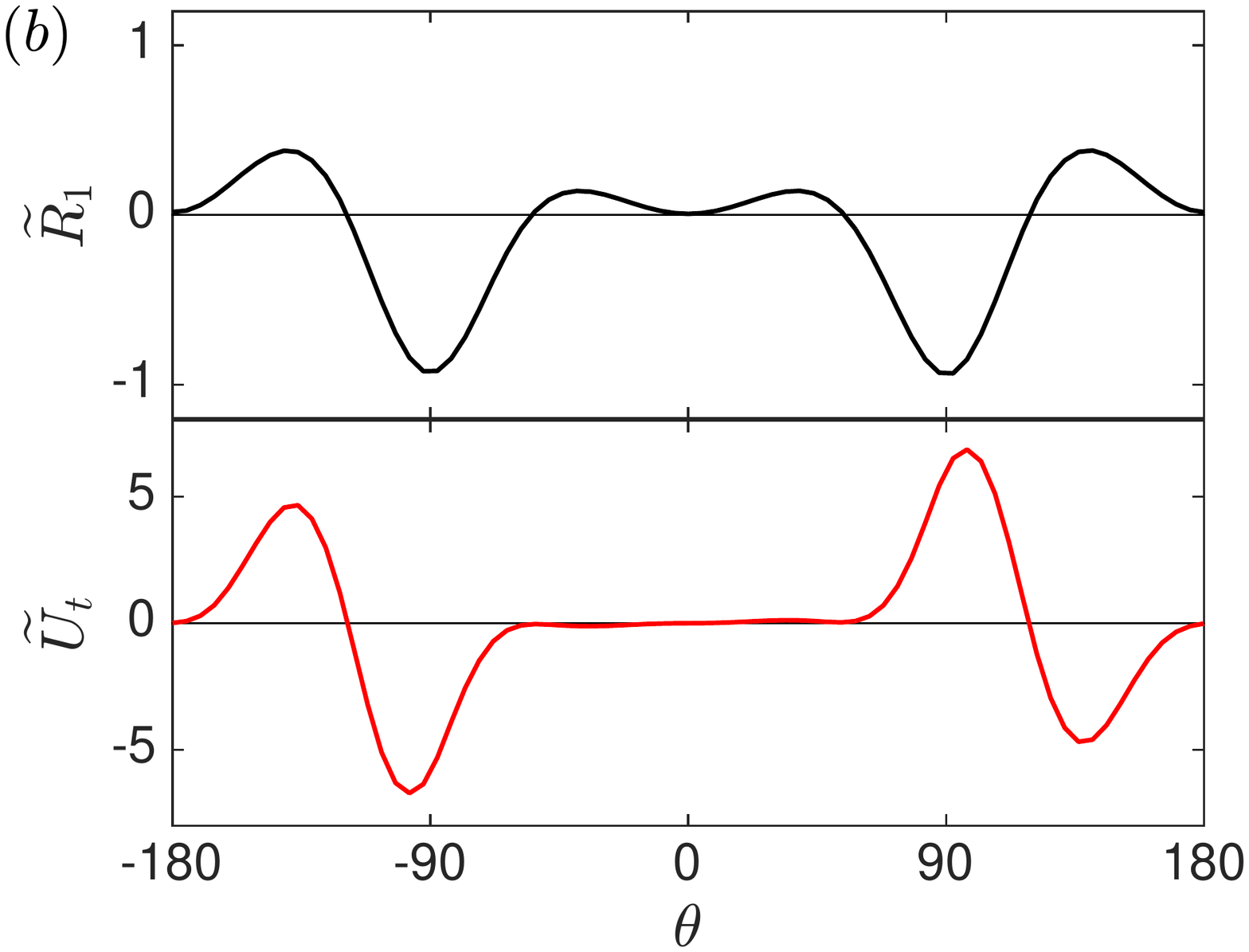} 
   \end{overpic}
}
\caption{
$(a)$ Optimal wavy cylinder for stabilization. Radius $R(\theta)=R_0+\epsilon \widetilde R_1(\theta)\cbz$ shown here at $z=0$, and with the amplitude $\epsilon=0.023$ that just brings the flow back to marginal stability.
(Dashed line: straight cylinder $R=R_0$.)
$(b)$ Optimal wall deformation  for stabilization ($||\widetilde R_1||=1$), and equivalent tangential actuation.
$\Rey=50$, $\beta=1$.
}  
\label{fig:optWALLDEF}
\end{figure}

As mentioned in section~\ref{sec:opt_wall_def}, wall deformation affects the eigenmode  (i) because it modifies the base flow and (ii) because it displaces the  no-slip boundary for the eigenmode. 
In order to compare these two effects, we 
evaluate in (\ref{eq:ev2}) the relevant 3D contribution to $\ev_2$, namely
\be 
\ev_2^{3D} 
= \ps{ \widetilde\qq_0^\dag }{\widetilde\AAA_1 \widetilde\qq_1},
\label{eq:ev2q1explicit}
\ee
in two different ways: 
first, using the eigenmode modification $\widetilde\qq_{1}$ given by (\ref{eq:q1})  with a no-slip boundary condition on the undeformed cylinder;
second, using $\widetilde\qq_{1,def}$
calculated with the flattened version of the no-slip boundary condition on the deformed cylinder, i.e.
\be 
 (\ev_0\EE+\widetilde\AAA_0) \widetilde\qq_{1,def} = \widetilde\AAA_{1,def} \widetilde\qq_0
\ee 
with a modified operator $\widetilde\AAA_{1,def}$ enforcing (\ref{eq:flattenedBC_eigmode}).
Figure~\ref{fig:lam2r_opt_WALLDEF_r}$(b)$ shows the results for the optimal stabilizing deformation $R_1$: the solid line corresponds to the evaluation of (\ref{eq:ev2q1explicit}) with $\widetilde\qq_{1}$, and the dashed line is the difference between the evaluations with $\widetilde\qq_{1,def}$ and  $\widetilde\qq_{1}$. 
Since this difference is negligible, one concludes that  wall deformation has a much larger effect on $\ev_2$ via the base flow modification than 
via the eigenmode's boundary condition.

The optimal stabilizing wavy cylinder is shown in Fig.~\ref{fig:optWALLDEF}$(a)$ at $z=0$, together with the straight circular cylinder for reference (dashed line). 
The geometry is top-down symmetric, and the  wall deformation is concentrated around $|\theta|=\pi/2$, corresponding to successive thinning and thickening of the vertical extent of the cylinder (recall that the deformation has opposite signs at $z=0$ and $z=\pi/\beta$). 
This optimal wavy cylinder is in good agreement with the shape found by \cite{Tammisola2017} with a different method, and without including the effect of the mean flow correction.
Figure~\ref{fig:optWALLDEF}$(b)$ shows a ``developed'' view of the wall deformation as function of $\theta$, as well as the leading-order equivalent tangential wall actuation (\ref{eq:U1t}) which, at $z=0$, consists of
upstream blowing on either side of the cylinder (around $|\theta|=\pi/2$) and
downstream blowing on the front region (around $|\theta|=3\pi/4$).
This contrasts with the tangential component of the optimal stabilizing wall actuation (Fig. \ref{fig:opt_Uw_stab_simple}), which is oriented upstream over the whole cylinder, because in this latter case flow stabilization is also (and prominently) achieved with the normal component.

%--- plot_variation_vs_epsilon.m
\begin{figure}
\centerline{   
   \begin{overpic}[width=7cm, trim=5mm 65mm 20mm 60mm, clip=true, tics=10]{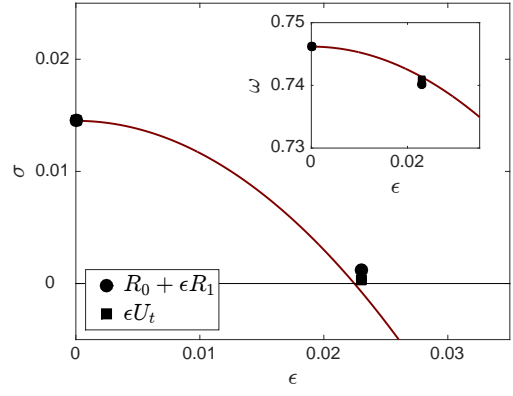} 
   \end{overpic}
}
\caption{
Effect of the optimal stabilizing wall deformation $R_1$ on growth rate and frequency.
Line: sensitivity prediction, symbols: 3D stability analysis (circles {\Large $\bullet$}: wavy cylinder, 
squares {\scriptsize $\blacksquare$}: equivalent tangential blowing/suction).
$\Rey=50$, $\beta=1$.
}  
\label{fig:valid3}
\end{figure}

The amplitude $\epsilon=0.023$ used in Fig.~\ref{fig:optWALLDEF} is the amplitude needed to bring the flow back to marginal stability, as illustrated in Fig.~\ref{fig:valid3}.
Validations against 3D linear stability analysis show a good agreement for both wall deformation and equivalent tangential actuation.

%--- plot_mono_C1_Q1_Q2_real.m
%--- plot_mono_C1_Q1_Q2_real_arrows_xy.m
%--- plot_mono_C1_Q1_Q2_real_arrows_yz.m
\begin{figure}
\centerline{   
   \begin{overpic}[trim=5mm 14mm 0mm 17mm, clip, height=4.1cm,tics=10]{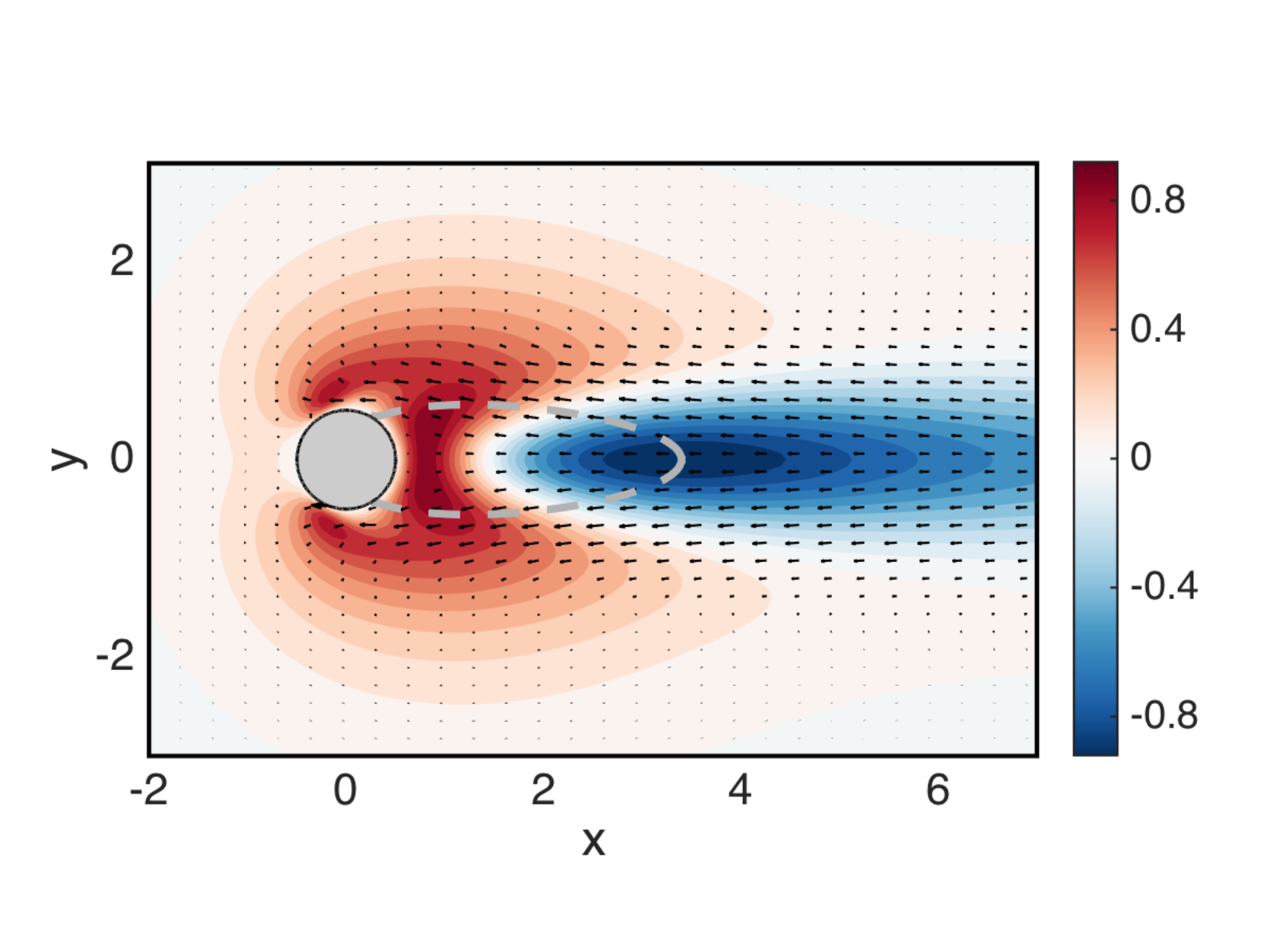}
   \end{overpic}      
   \begin{overpic}[trim=23mm 0 5mm 0, clip, height=4.05cm,tics=10]{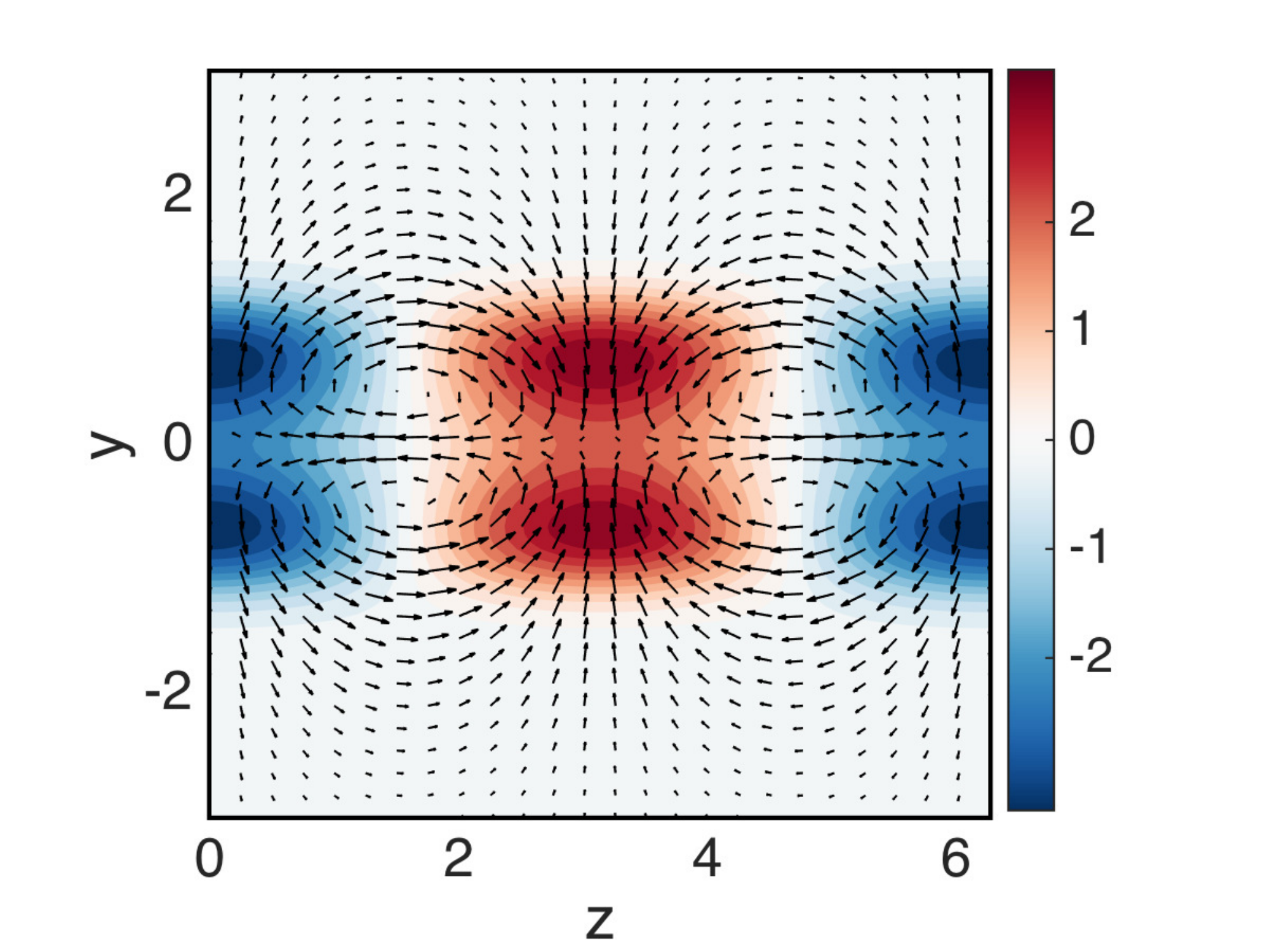}
   \end{overpic} 
   \begin{overpic}[trim=15mm 14mm 5mm 17mm, clip, height=4.1cm,tics=10]{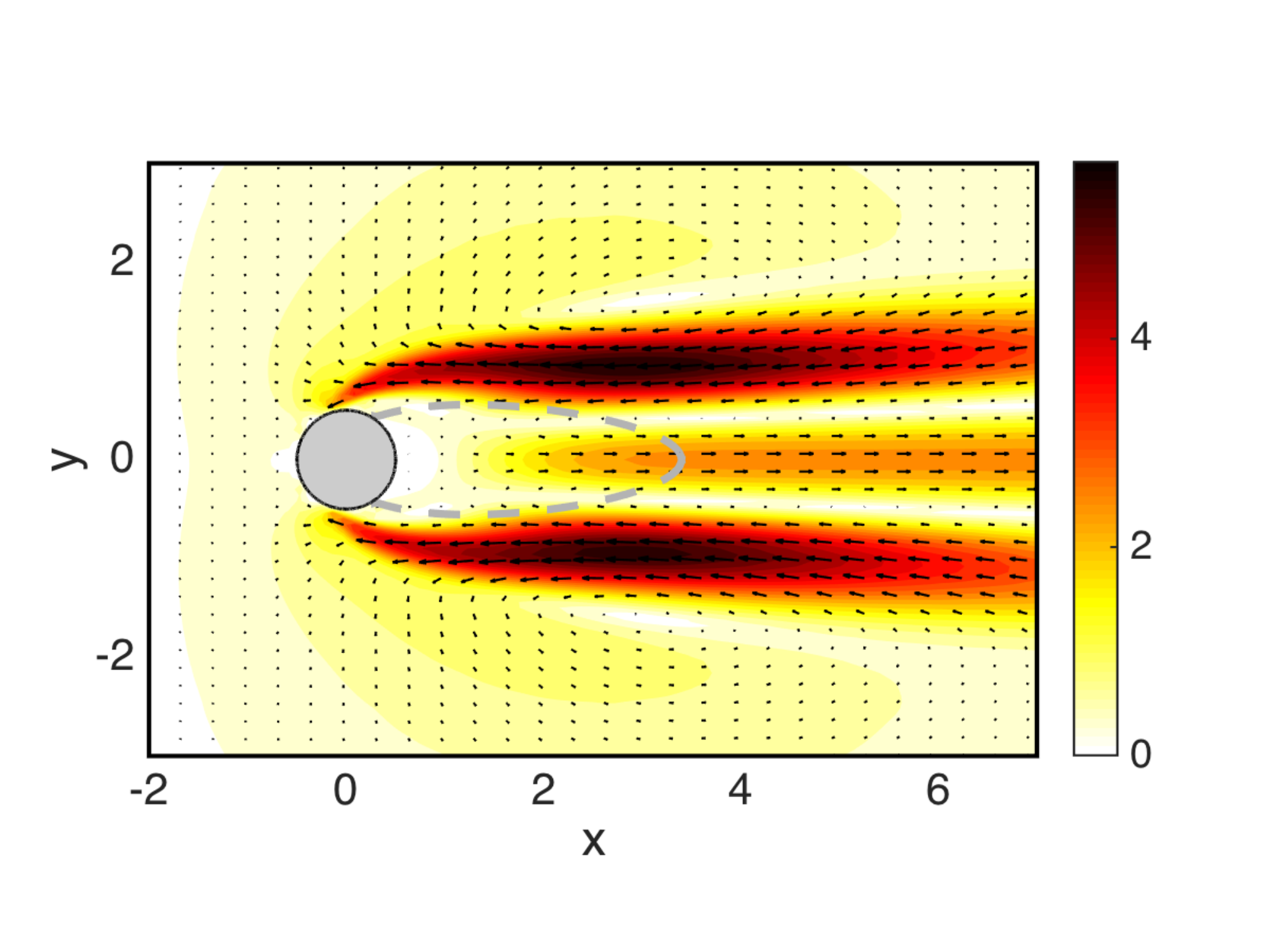}     
   \end{overpic}   
}
\caption{
First-order flow modification $\UU_1 = ( \widetilde U_1(x,y)\cbz, \widetilde V_1(x,y)\cbz, \widetilde W_1(x,y)\sbz)^T$ (spanwise periodic) induced by the optimal stabilizing wall deformation (or equivalent tangential wall control) at 
$\Rey=50$, $\beta=1$.
Left: vector fields $(U_1,V_1)^T$ at $z=0$ and contours of $W_1$ at $z=\pi/(2\beta)$.
Middle: vector field $(V_1,W_1)^T$ and contours of $U_1$, at $x=2$.
Right: 
induced mean flow correction $\UU_2^{2D} = ( U_2^{2D}(x,y), V_2^{2D}(x,y), 0)^T$ (spanwise invariant), shown with vector field $(U_2^{2D},V_2^{2D})^T$ and contours of velocity magnitude. 
}  
\label{fig:Q1_from_opt_WALLDEF_stab}
\end{figure}

The first- and second-order flow modifications induced by the optimal wall deformation, shown in Fig.~\ref{fig:Q1_from_opt_WALLDEF_stab},
are sensibly similar to those induced by the optimal wall actuation [Fig.~\ref{fig:Q1_from_opt_Uw_stab}$(b)$], except for a weaker (resp. stronger) mean flow correction on the symmetry axis $y=0$ in the wake (resp. on either side of the cylinder in the immediate vicinity of the walls).

%--- calc_Ex_vs_Re_wall_deform.m
\begin{figure}
\centerline{
   \begin{overpic}[width=6.5cm, trim=0mm 60mm 20mm 60mm, clip=true, tics=10]{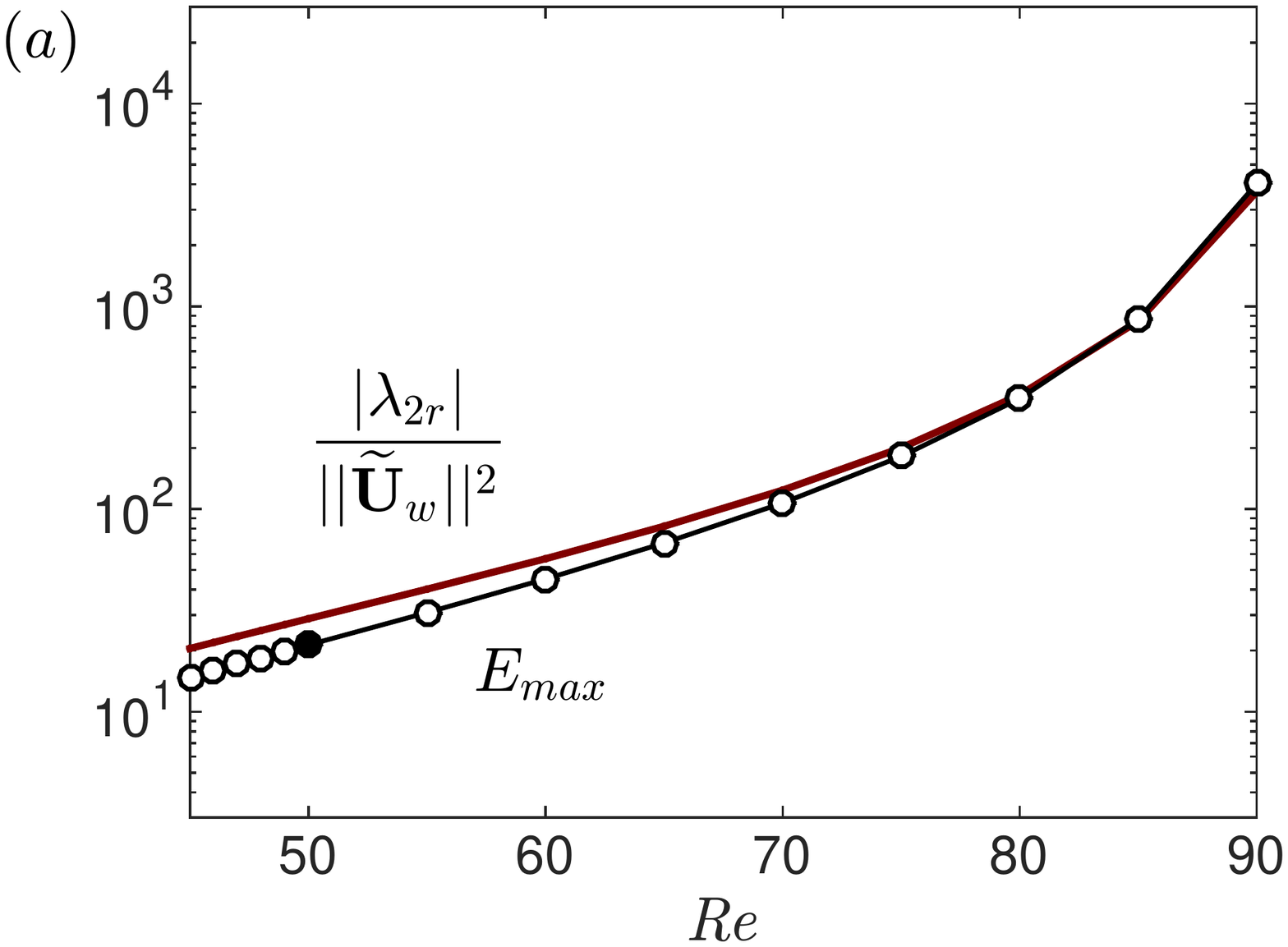} 
   \end{overpic}  
   \hspace{0.5cm} 
   \begin{overpic}[width=6.5cm, trim=0mm 60mm 20mm 60mm, clip=true, tics=10]{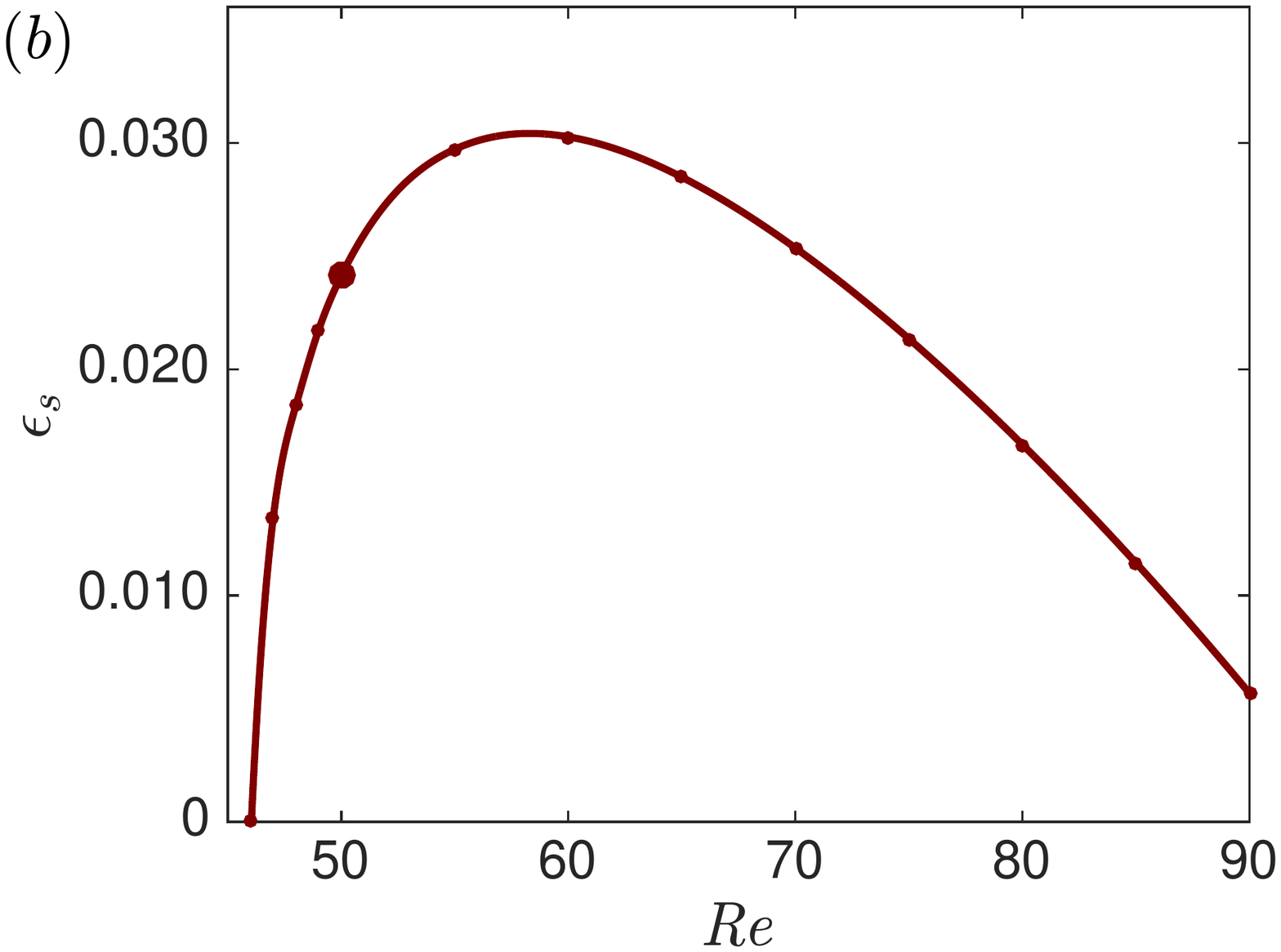} 
   \end{overpic}  
}
\caption{
Variation with Reynolds number.
$(a)$~Normalized growth rate variation (in absolute value), and maximum energy density, both for the flow modification induced by the optimal wall deformation for stabilization. $(b)$~Wall deformation amplitude $\epsilon_s=\sqrt{\ev_{0r}/ (|\ev_{2r}|/||\UU_w||^2) }$ necessary to fully stabilize the leading eigenmode.
$\beta=1$.
}  
\label{fig:Ex_vs_Re-Wall_def}
\end{figure}

Figure~\ref{fig:Ex_vs_Re-Wall_def} shows that, similar to wall actuation (Fig.~\ref{fig:Ex_vs_Re}), the effect of wall deformation on both the induced flow modification (as measured by the maximum energy density of $\QQ_1$) and the leading growth rate increases with $\Rey$, and the amplitude needed to stabilize the leading eigenvalue has a maximum around $\Rey \simeq 58$.

%----------------------------------------
%----------------------------------------
%----------------------------------------
\section{Optimal wall actuation for frequency modification}
\label{sec:freq}

While stabilization is a major objective of flow control, the ability to alter vortex shedding frequency is appealing in some applications too. 
Here we use our optimization method to control the frequency of the leading eigenvalue (obtained in the present case from linear stability about the base flow), which is  close to the actual vortex shedding frequency at and slightly above the onset of instability. 
Note that controlling the nonlinear frequency would require targeting the eigenvalue obtained from linear stability about the mean flow \cite{Barkley06, Sipp2007}, for instance in the spirit of \cite{Meliga2016}.

%--- plot_fig19.m
\begin{figure}
%\label{fig:}
\def \thiswidth {4.5cm}
\centerline{   
     \begin{overpic}[width=7cm, trim=5mm 65mm 20mm 60mm, clip=true, tics=10]{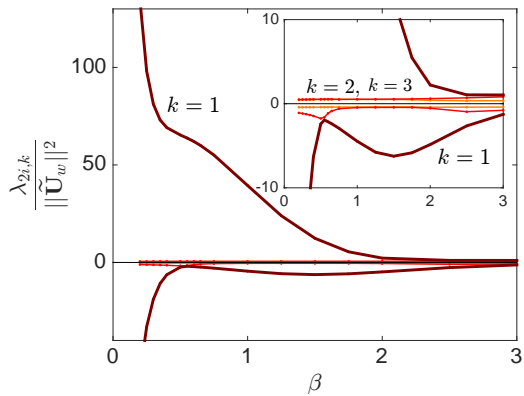}       
   \end{overpic}   
}
\caption{
Normalized frequency variation induced by the optimal wall actuation $\UU_w$ for frequency increase ($\lambda_{2i}>0$) and frequency reduction ($\lambda_{2i}<0$).
Optimal ($k=1$) and first sub-optimals ($k=2, 3$). 
$\Rey=50.$
} 
\label{fig:lam2i_opt_Uw_i} 
\end{figure}

The frequency variation induced the by the optimal wall actuation is shown in Fig.~\ref{fig:lam2i_opt_Uw_i}.
Over a wide range of wavenumber $\beta$,
it is easier to increase than to reduce the frequency.
The optimal frequency variation (in absolute value) is generally smaller than the optimal growth rate variation (in absolute value) for $\beta \lesssim 2$.

In the range of relevant $\beta$ values, 
the frequency variation (both positive and negative) is mainly due to the 3D contribution for the optimal control $k=1$.
There is a significant 2D contribution for the sub-optimal $k=2$, however the total effect is much smaller than for $k=1$ (not shown). 
This is similar to the results for optimal control for stabilization/destabilization (section~\ref{sec:compet_3D_2D}).

Regarding the competition between amplification (from wall control to flow modification) and normalized effect (frequency variation induced by a unit-norm flow modification),
the optimal control $k=1$ is much more amplified than the first sub-optimal $k=2$ (up to two orders of magnitude at $\beta=1.2$),
whereas the normalized effect is comparable for $k \leq 3$ (not shown).
This is fundamentally different from the results for optimal control for stabilization/destabilization (section~\ref{sec:compet_ampli_stab}).

%%--- plot_fig19.m
\begin{figure}
\def \thiswidth {4.5cm}
\centerline{   
   \begin{overpic}[width=6.5cm, trim=5mm 65mm 20mm 60mm, clip=true, tics=10]{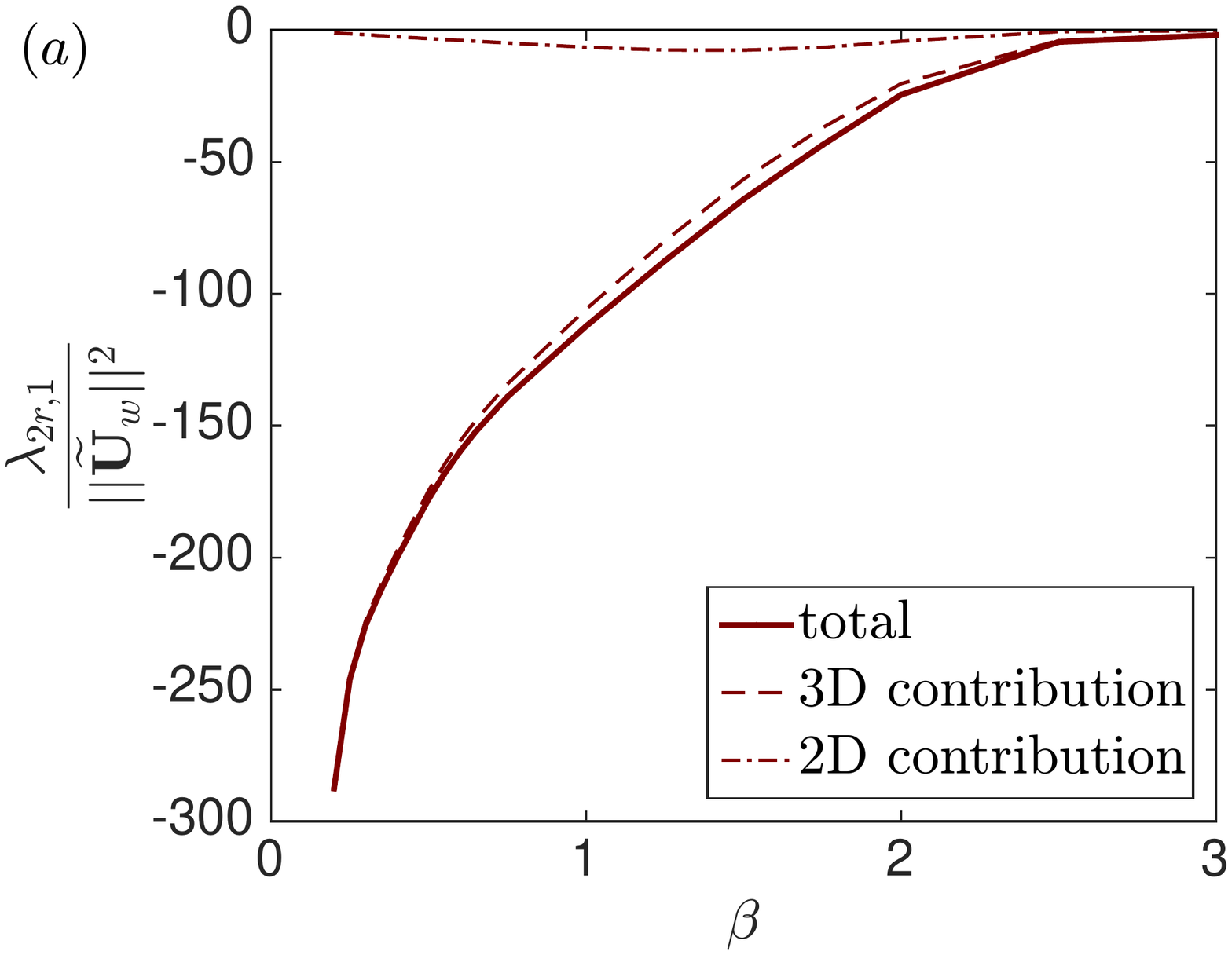} 
   \end{overpic}     
\hspace{1.5cm}
   \begin{overpic}[width=6.5cm, trim=5mm 65mm 20mm 60mm, clip=true, tics=10]{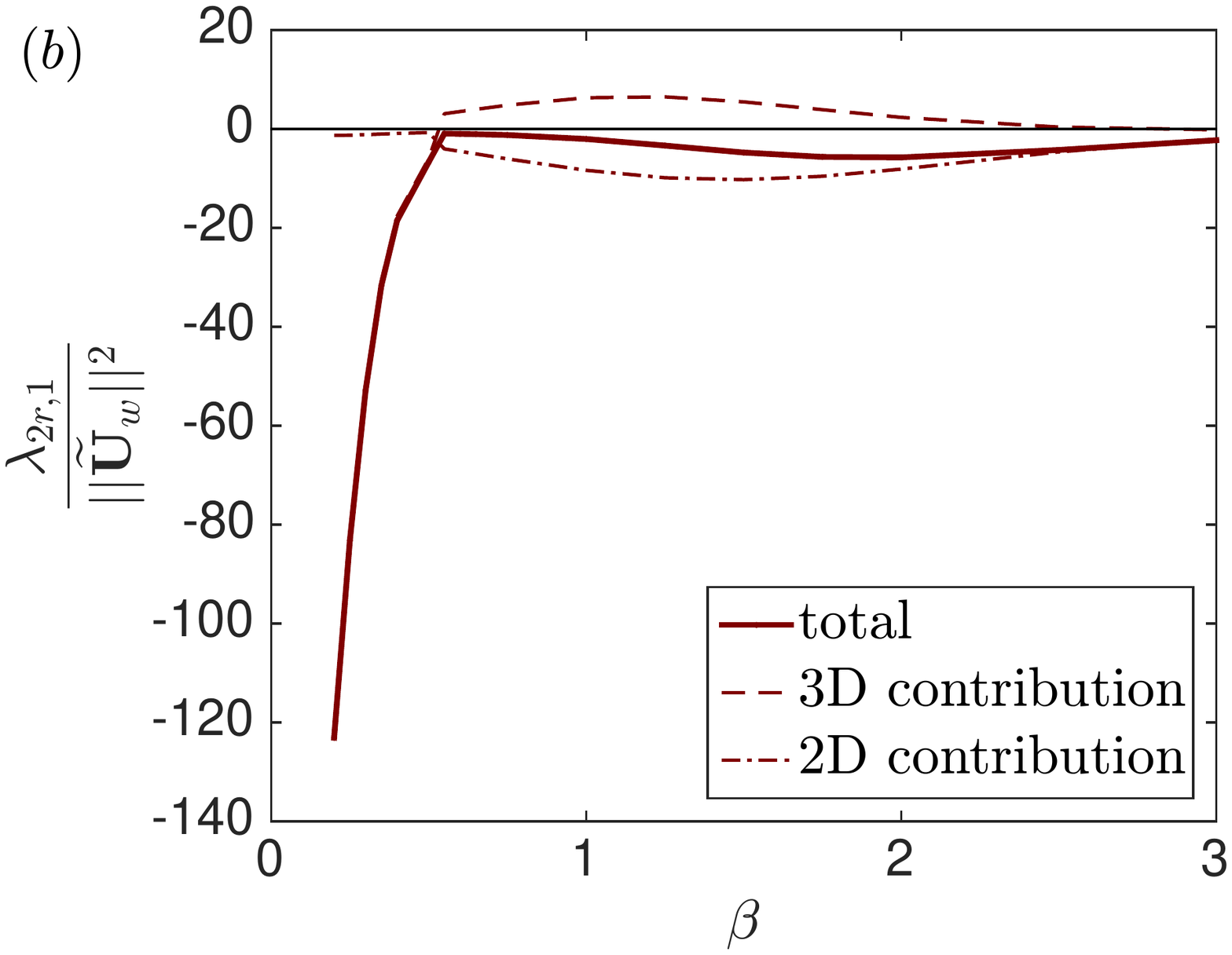} 
   \end{overpic}     
}
\caption{
Normalized effect on growth rate of the optimal $(a)$~frequency-increasing and $(b)$~frequency-reducing wall control $\UU_w$.
3D ($- -$) and 2D ($-\cdot-$) contributions.
$\Rey=50.$
}  
\label{fig:lam2r_opt_Uw_i}
\end{figure}

%--- plot_opt_Uwall_cplx.m
\begin{figure}
\centerline{   
   \begin{overpic}[width=5.7cm, trim=0mm 75mm 40mm 65mm, clip=true, tics=10]{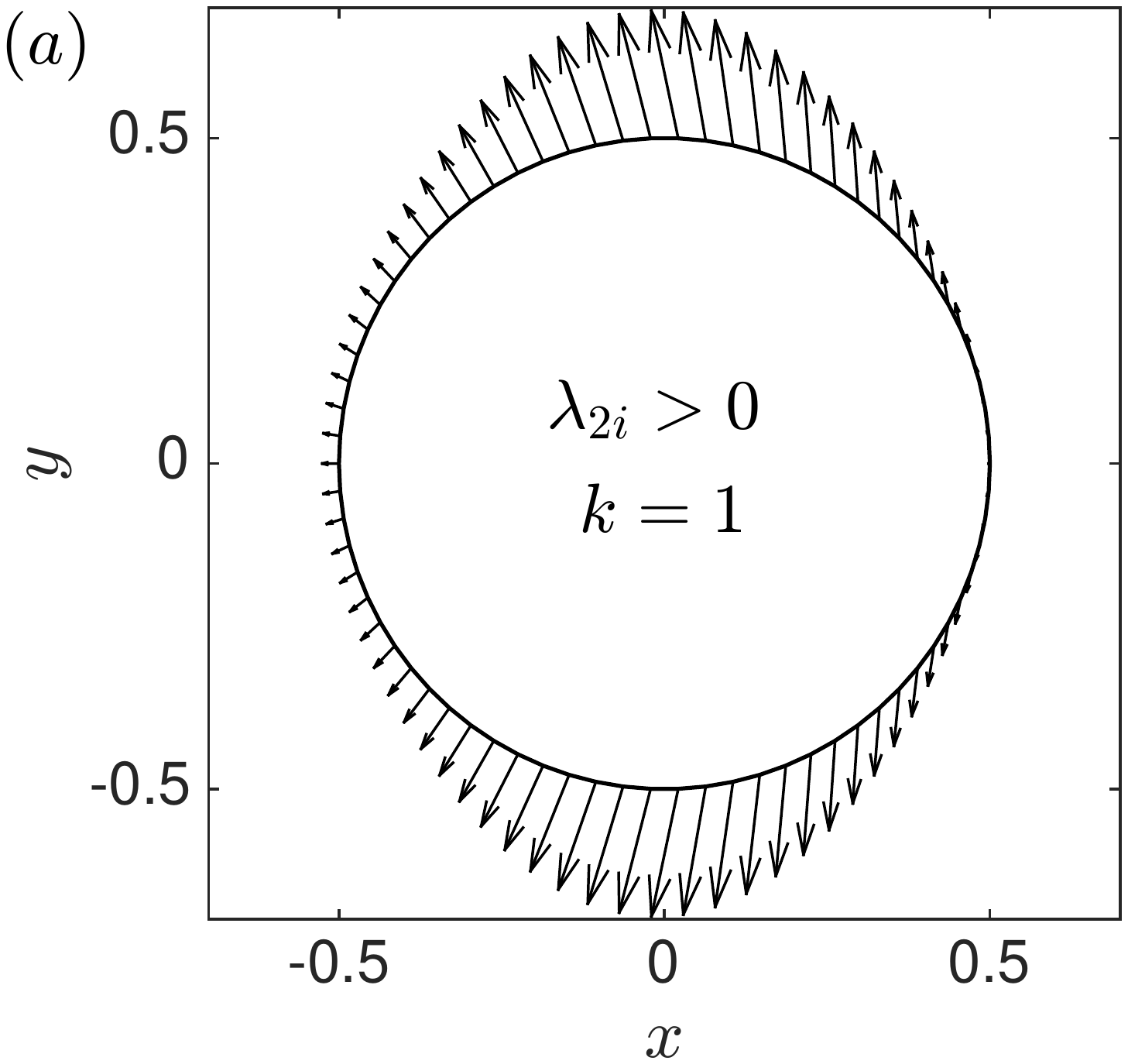} 
   \end{overpic} 
   \begin{overpic}[width=6.cm, trim=15mm 75mm 15mm 65mm, clip=true, tics=10]{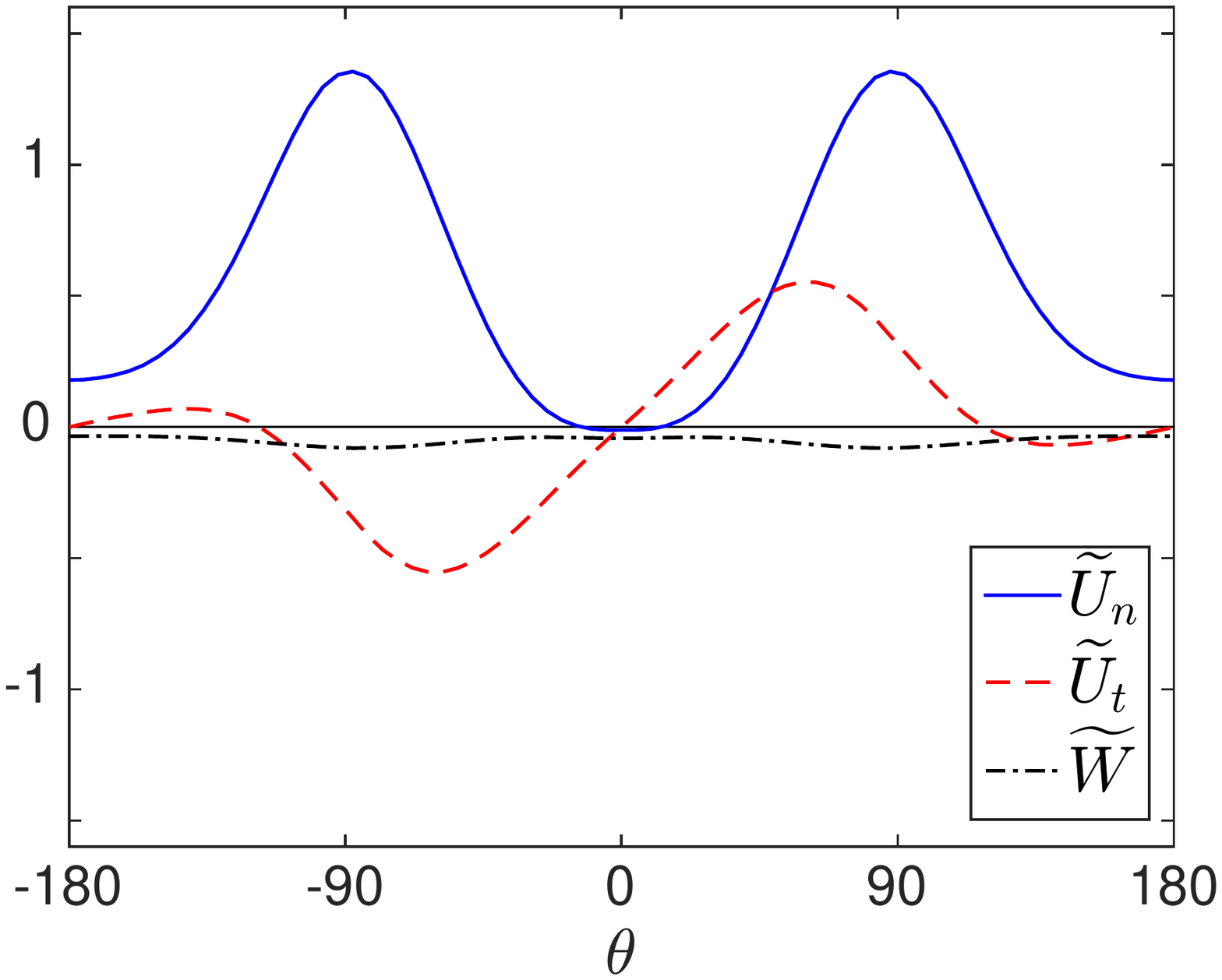}   
   \end{overpic}  
}
\centerline{   
   \begin{overpic}[width=5.7cm, trim=0mm 75mm 40mm 65mm, clip=true, tics=10]{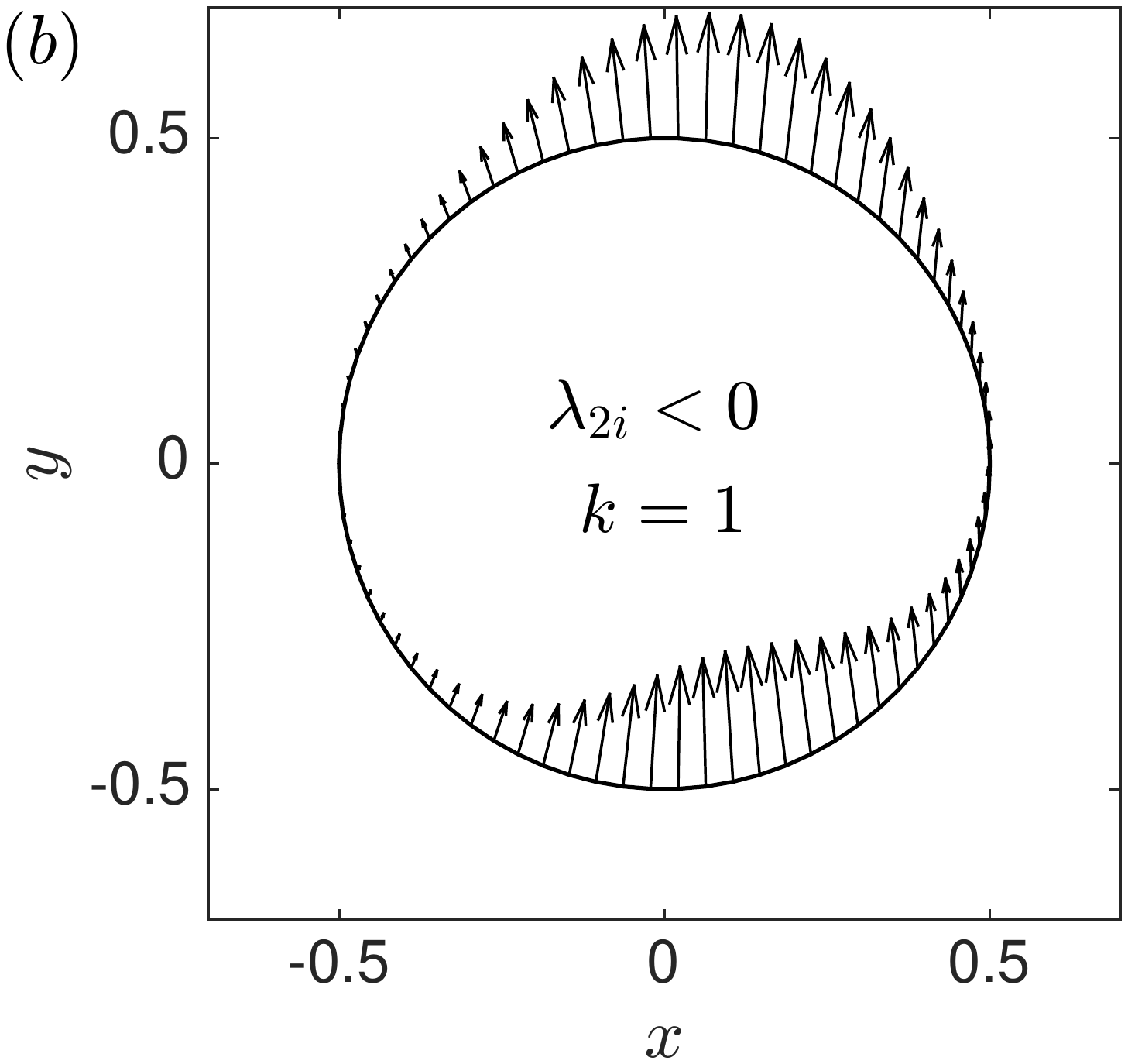} 
   \end{overpic} 
   \begin{overpic}[width=6.cm, trim=15mm 75mm 15mm 65mm, clip=true, tics=10]{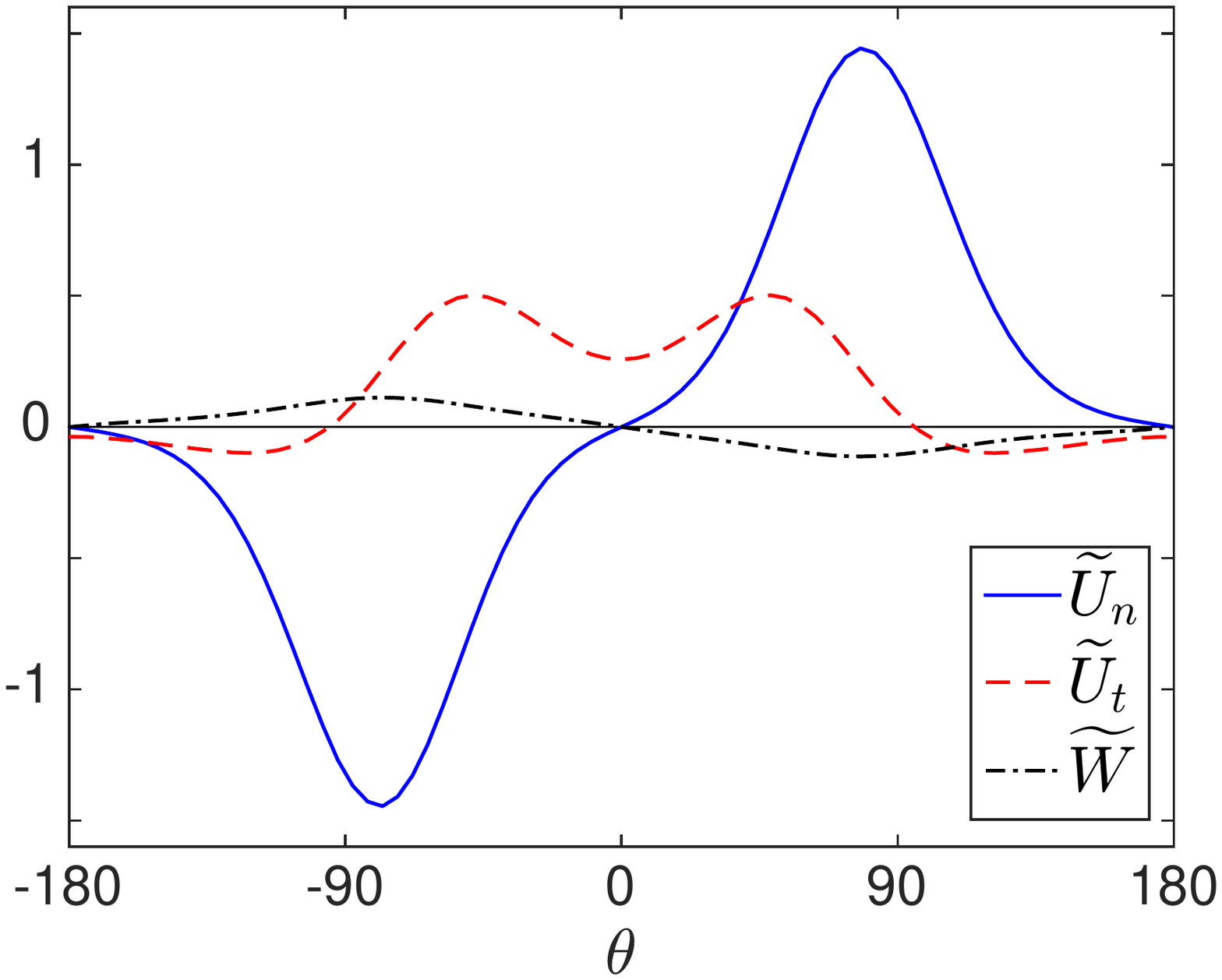}   
   \end{overpic}  
}
\caption{
Optimal $(a)$ frequency-increasing and $(b)$ frequency-reducing wall control $\UU_w$.
$\Rey=50$, $\beta=1$.
}  
\label{fig:opt_Uw_freqplus}
\end{figure}

Figure~\ref{fig:lam2r_opt_Uw_i} shows the effect on the growth rate of the  wall actuation optimized for frequency increase or reduction.
In both cases, this effect is stabilizing.
Interestingly, the wall actuation optimized for frequency increase (panel $a$) has an effect almost as large as the optimal stabilizing wall actuation (Fig.~\ref{fig:lam2r_opt_Uw_r}).
The wall actuation optimized for frequency decrease (panel $b$) has a substantially smaller effect.
This can be explained by the distributions shown in Fig.~\ref{fig:opt_Uw_freqplus}: in the former case, all velocity components are very similar to those of the optimal stabilizing actuation [Fig.~\ref{fig:opt_Uw_dest_stab}$(b)$], therefore achieving quasi-optimal stabilization. This is consistent with the observations of section~\ref{sec:effect_on_freq}.
In the latter case, the normal velocity component, and to a lesser extent the tangential and spanwise components, are similar to those of the first sub-optimal stabilizing actuation [Fig.~\ref{fig:opt_Uw_dest_stab}$(c)$]. The actuation is top/bottom symmetric and antisymmetric, respectively, and smoothly varying around the separation point in both cases.
Finally, and as expected, the induced flow modifications shown in Fig.~\ref{fig:Q1_from_opt_Uw_freqplus} are very similar to those  induced by the optimal and first sub-optimal stabilizing wall actuations [Fig.~\ref{fig:Q1_from_opt_Uw_stab}$(b-c)$].

%--- plot_mono_C1_Q1.m
%--- plot_mono_C1_Q1_Q2_real_arrows_xy.m
%--- plot_mono_C1_Q1_Q2_real_arrows_yz.m
\begin{figure}
\centerline{   
   \begin{overpic}[trim=5mm 14mm 0mm 17mm, clip, height=4.1cm,tics=10]{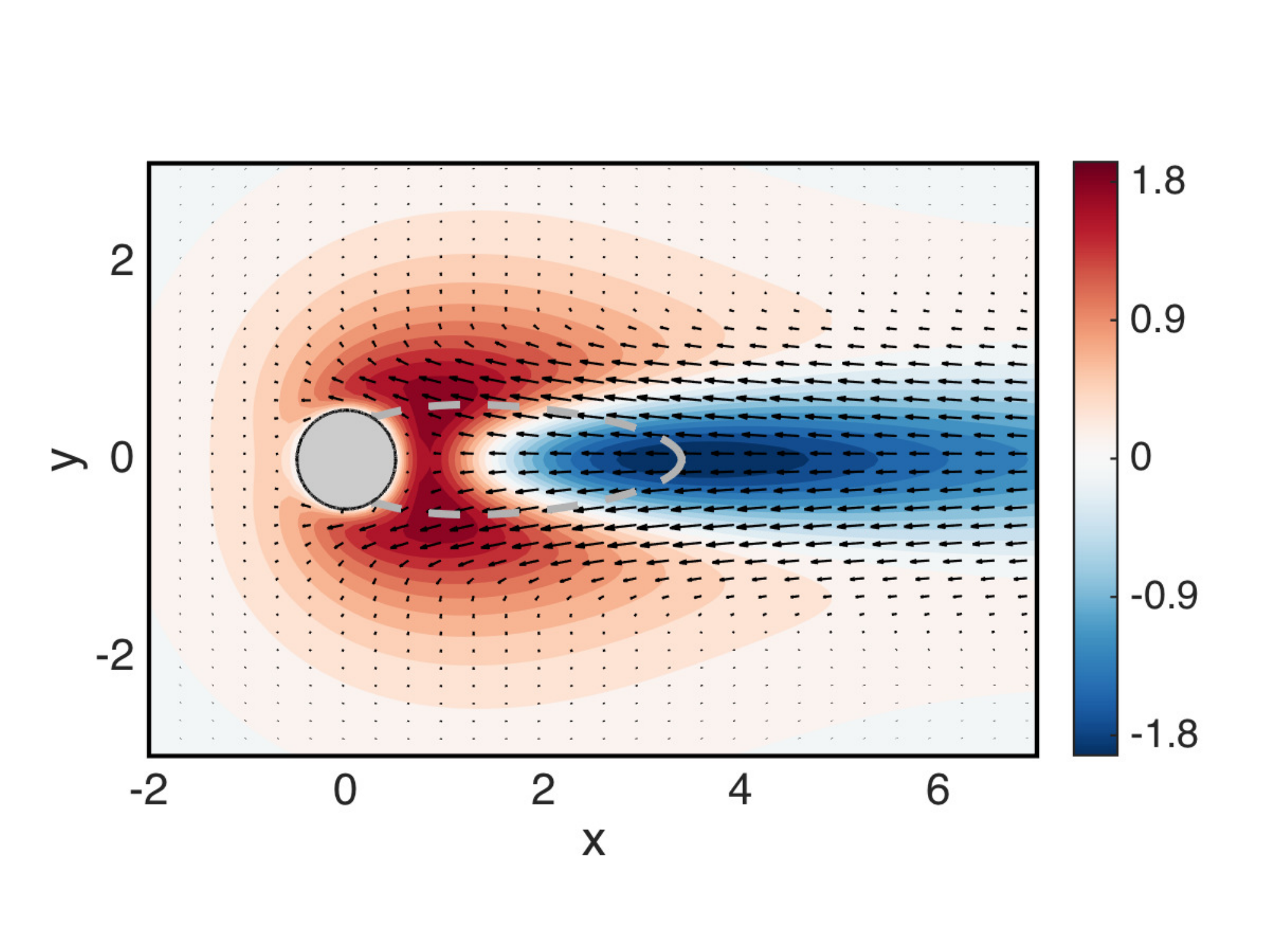}     
      \put(-2,54){$(a)$} 
   \end{overpic}      
   \begin{overpic}[trim=23mm 0 5mm 0, clip, height=4.05cm,tics=10]{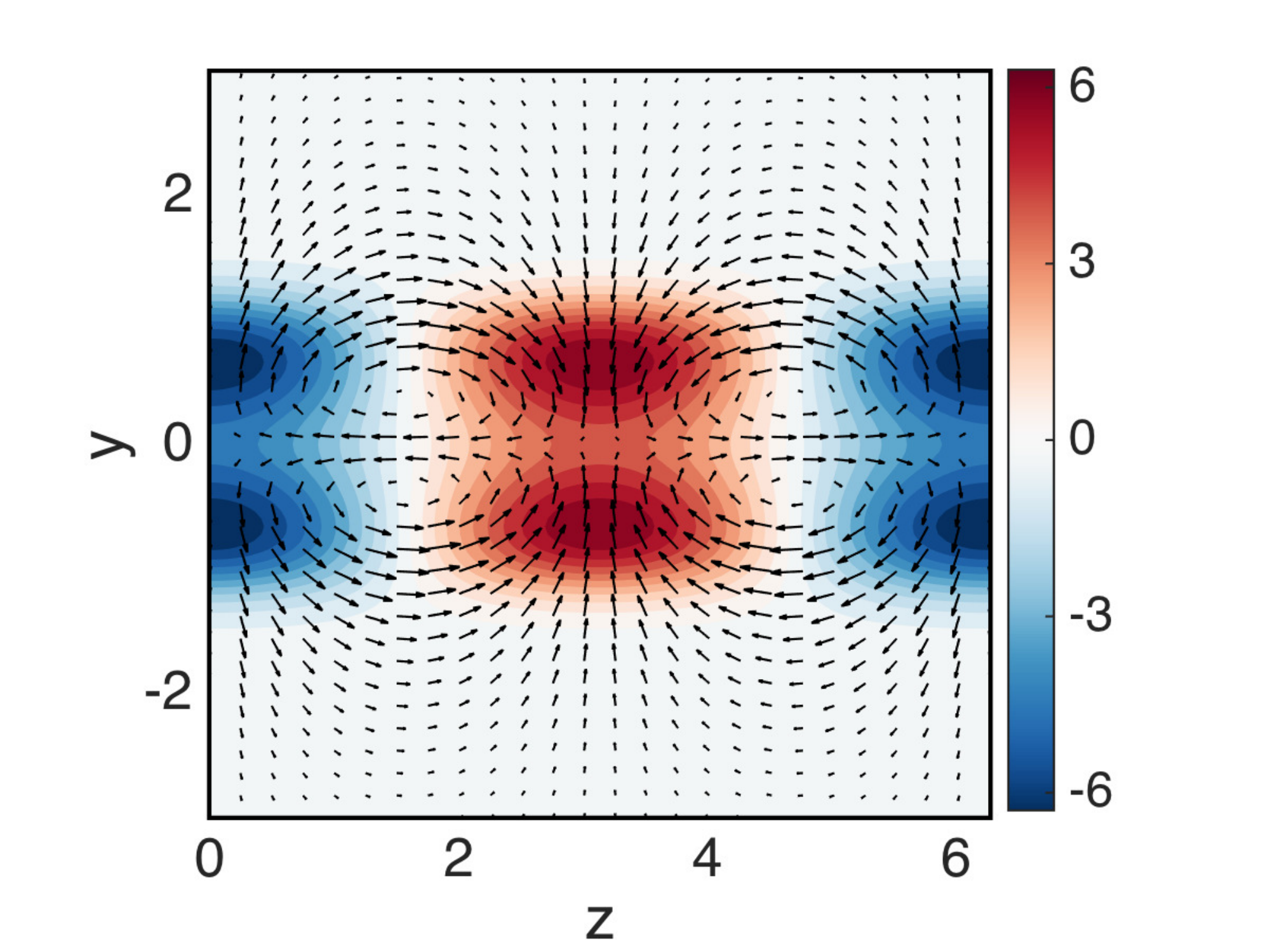}
   \end{overpic} 
   \begin{overpic}[trim=15mm 14mm 5mm 17mm, clip, height=4.1cm,tics=10]{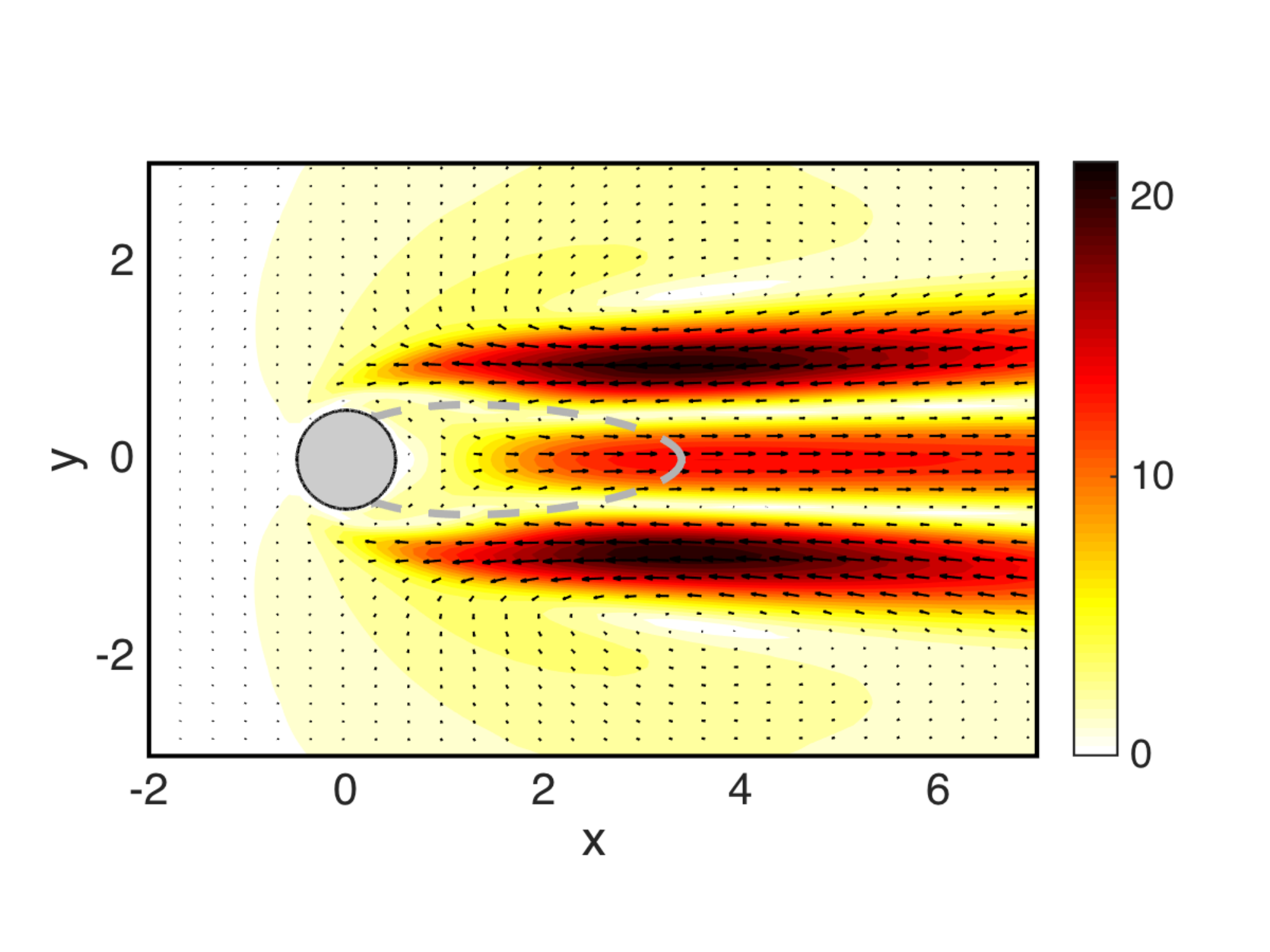}     
   \end{overpic}   
}
\centerline{   
   \begin{overpic}[trim=5mm 14mm 0mm 17mm, clip, height=4.1cm,tics=10]{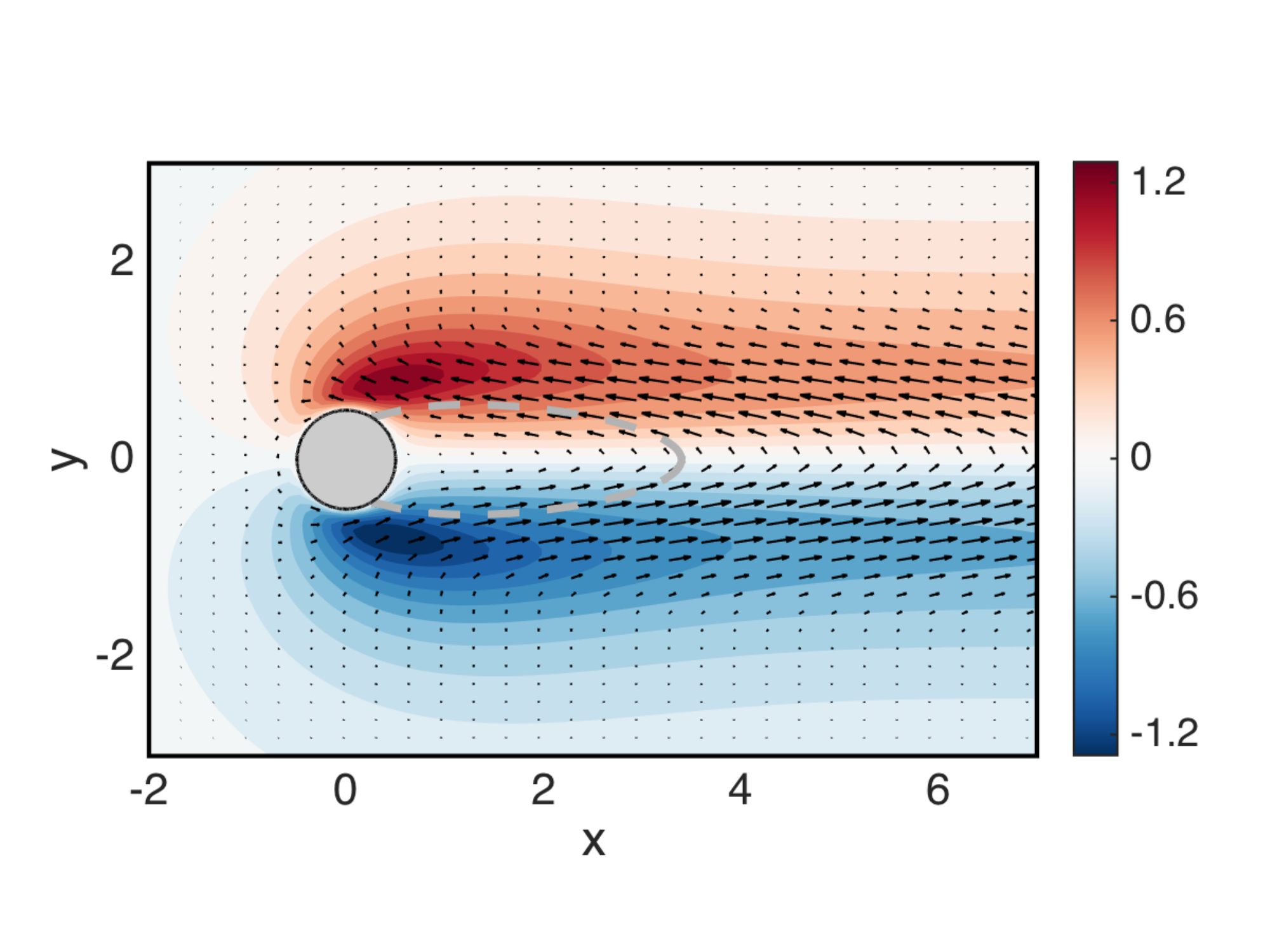}  
      \put(-2,54){$(b)$} 
   \end{overpic}      
   \begin{overpic}[trim=23mm 0 5mm 0, clip, height=4.05cm,tics=10]{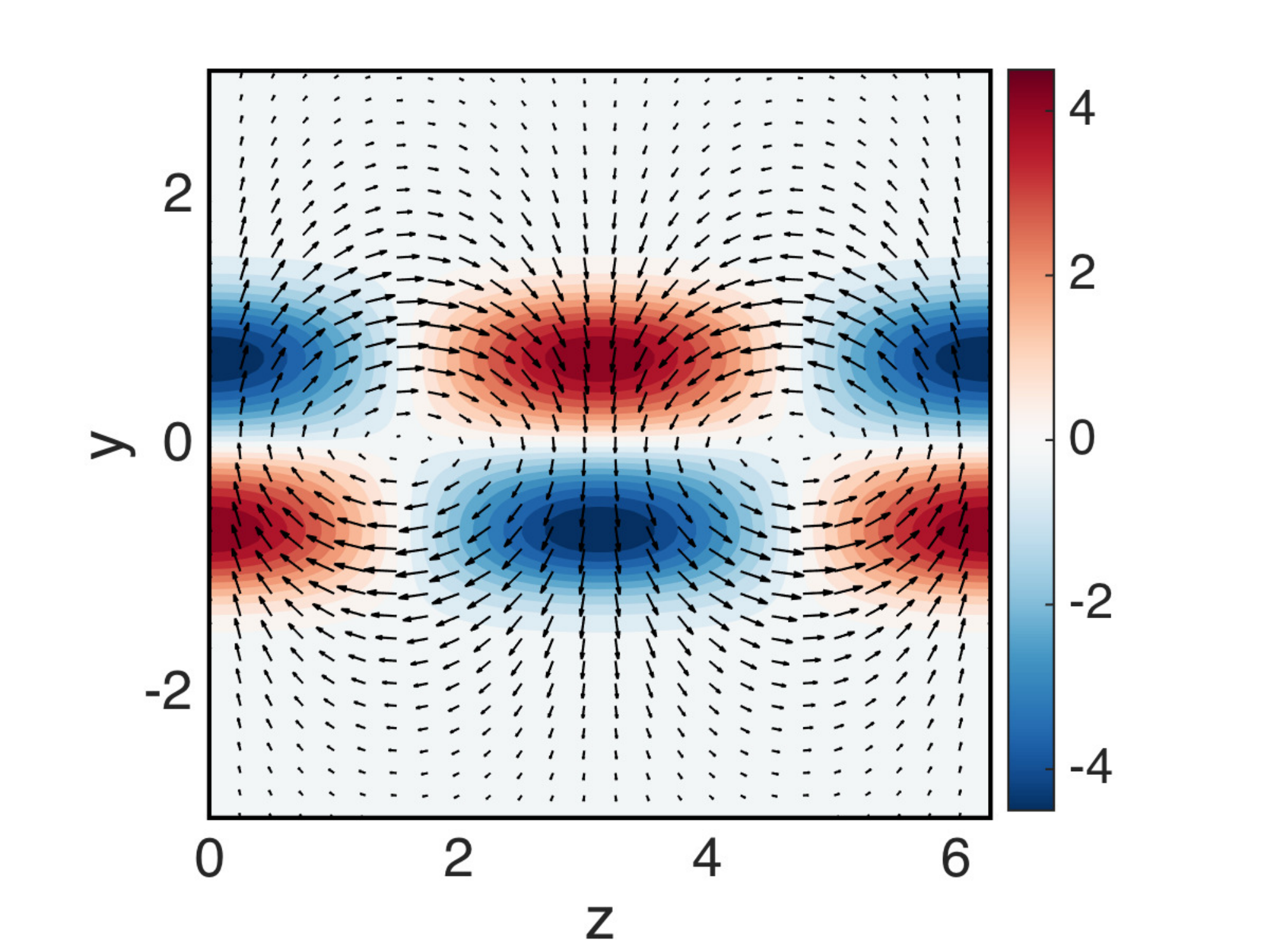}
   \end{overpic} 
   \begin{overpic}[trim=15mm 14mm 5mm 17mm, clip, height=4.1cm,tics=10]{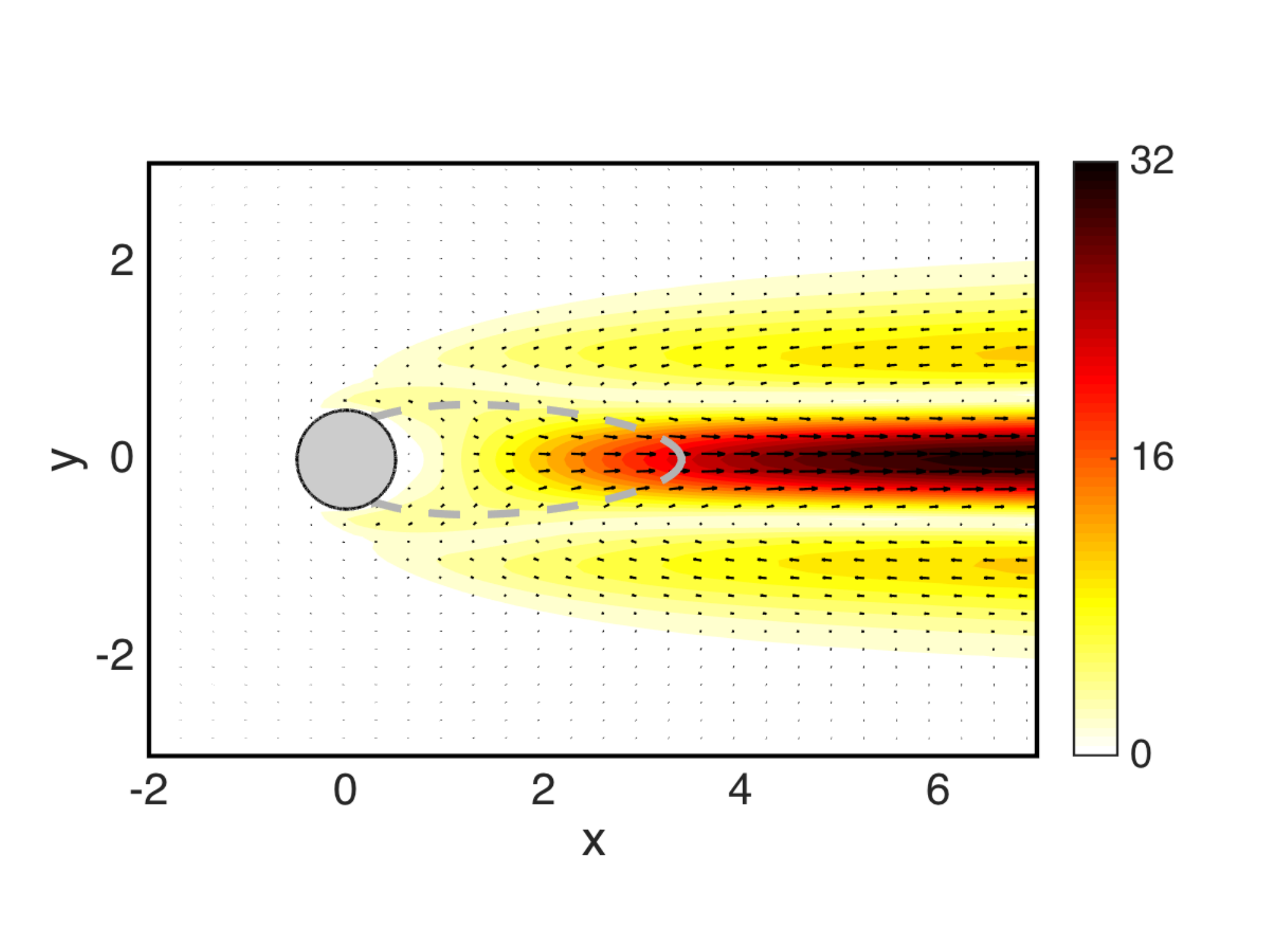}     
   \end{overpic}   
}
\caption{
First-order flow modification $\UU_1 = ( \widetilde U_1(x,y)\cbz, \widetilde V_1(x,y)\cbz, \widetilde W_1(x,y)\sbz)^T$ (spanwise periodic) induced by the optimal
$(a)$~frequency-increasing and 
$(b)$~frequency-reducing wall control,
at $\Rey=50$, $\beta=1$.
Left: vector fields $(U_1,V_1)^T$ at $z=0$ and contours of $W_1$ at $z=\pi/(2\beta)$.
Middle: vector field $(V_1,W_1)^T$ and contours of $U_1$, at $x=2$.
Right: 
induced mean flow correction $\UU_2^{2D} = ( U_2^{2D}(x,y), V_2^{2D}(x,y), 0)^T$ (spanwise invariant), shown with vector field $(U_2^{2D},V_2^{2D})^T$ and contours of velocity magnitude.  
}  
\label{fig:Q1_from_opt_Uw_freqplus}
\end{figure}

%----------------------------------------
%----------------------------------------
%----------------------------------------
\section{Conclusion}
\label{sec:conclusion}

We use an adjoint method to compute the second-order sensitivity to small-amplitude control of eigenvalues encountered in global linear stability analysis (i.e. solutions to the eigenvalue problem resulting from linearization of the Navier--Stokes equations).
In 2D flows, spanwise-periodic control has a zero net first-order (linear) effect, therefore the second-order (quadratic) effect is the leading effect.
The sensitivity operator allows one to predict the effect of any small-amplitude control on an eigenvalue, without actually computing the controlled flow.
Further, we compute the optimal control (the most effective control) for a variety of objectives: stabilization, destabilization, frequency modification. 
Apart from the quadratic approximation, our method is exact in that it does not rely on a projection of the optimal control onto basis functions to keep the problem tractable.
Instead, we take advantage of the very spanwise-periodic nature of the control and reduce tremendously the computational complexity from that of a fully 3D problem to that of a 2D problem. 
As a result, the operator inversion involved when computing the sensitivity or the optimal control is easily performed given the size of the 2D problem.

We apply the approach to the leading eigenvalue of the incompressible,  laminar flow around a circular cylinder. 
We consider three kinds of spanwise-periodic control: volume control (via a body force), wall actuation (via blowing/suction), and wall deformation; we focus on the latter two and give illustrative results.
We optimize alternatively for the linear growth rate and for the linear frequency, motivated by two different issues: stability of the leading eigenmode, and linear frequency of this mode when unstable, respectively. 
Both issues are of interest in vortex-induced vibrations, for instance, when the integrity of mechanical structures must be guaranteed, or conversely when oscillations should be promoted for energy extraction. Applications in aeroacoustics such as tonal noise are also relevant.

We find that sensitivity results are in good agreement with 3D validations (3D nonlinear controlled base flow and its 3D linear stability analysis), within the range of control amplitudes where quadratic effects are dominant.
We also observe that among the two second-order effects at play, the 3D contribution (related to the spanwise-periodic first-order flow modification $\QQ_1$) is generally larger than the 2D contribution (related to the mean flow correction, i.e. the spanwise-invariant component $\QQ_2^{2D}$ of the second-order flow modification $\QQ_2$).

We show that, over a wide range of control spanwise wavenumber $\beta$, the optimal control for flow stabilization is top-down symmetric and leads to varicose streaks in the cylinder wake, consistent with previous observations.
Conversely, the optimal control for flow destabilization is antisymmetric and leads to sinuous streaks. 
Symmetry alone cannot explain stabilization/destabilization: for instance, the first sub-optimal stabilizing control is antisymmetric.
However, a detailed analysis of the competition between amplification (from the cylinder to the wake) and stabilizing effect (of the flow modification) provides more insight: in the stabilizing case, the optimal varicose streaks, generated through an amplification of the same order as the optimal sinuous streaks, stabilize the flow more efficiently.

Regarding wall blowing/suction, spanwise actuation has a negligible contribution to the optimal control; therefore, in-plane  actuation $(U_n,U_t,0)^T$ is practically optimal.
Tangential actuation has a non-negligible but significantly smaller contribution too; therefore, normal-only actuation $(U_n,0,0)^T$ is a good trade-off between simplicity and effectiveness.
 
Regarding wall deformation, the optimal stabilizing deformation (and the equivalent tangential blowing/suction)  induces a flow modification very similar to that induced by the optimal wall actuation.

Our method is applicable to any other 2D flow, and can easily be extended to axisymmetric flows. We expect it to produce interesting results in flows where spanwise-periodic (or azimuthal-periodic) control has received less attention so far than bluff-body wakes.
The extension of this approach to a variety of other control objectives (e.g. aerodynamic forces, non-normal amplification/non-modal stability, flow geometry, etc.) is worth investigating and bears great promise for the systematic design of efficient control techniques.

%---------------------------------------
%---------------------------------------
%---------------------------------------
%\begin{acknowledgments}
%...
%\end{acknowledgments}

%---------------------------------------
%---------------------------------------
%---------------------------------------
\appendix

%----------------------------------------
%----------------------------------------
%----------------------------------------
\section{General second-order sensitivity}
\label{sec:apdx_general_sensitivity}

The second-order eigenvalue variation $\ev_2$ can be evaluated from (\ref{eq:ev2}) for a given control (wall deformation $R_1$, wall actuation $\UU_c$, or volume control $\CC$), provided the induced flow modifications $\QQ_1$ and $\QQ_2$ are computed with~(\ref{eq:Q0}).

Interestingly, $\ev_2$ can also be evaluated directly as a simple scalar product between the control and a second-order sensitivity operator.
This sensitivity operator is independent from the control (and thus needs only be computed once, irrespective of the number of specific control configurations considered); furthermore, it does not require computing the induced flow modifications $\QQ_1$ and $\QQ_2$.

Recall the expression (\ref{eq:ev2}) of the second-order eigenvalue variation:
\begin{align}
\ev_2 
&= \pps{ \qq_0^\dag }{-\AAA_2\qq_0} 
+ \pps{ \qq_0^\dag }{\AAA_1(\ev_0\EE+\AAA_0)^{-1} \AAA_1 \qq_0}
\\
&= \pps{ \qq_0^\dag }{-\AAA_2\qq_0} 
+ \pps{\AAA_1^\dag \qq_0^\dag }{(\ev_0\EE+\AAA_0)^{-1} \AAA_1 \qq_0},
\label{eq:ev22}
\end{align}
where $\AAA_1^\dag$ is the adjoint of $\AAA_1$ (recall the general definition (\ref{eq:adjoint})). 
We introduce linear operators $\LL$ and $\MM$ that depend only on $\qq_0$ and $\qq_0^\dag$ such that: 
\begin{align}
\AAA_1 \qq_0      
&= 
\left[\begin{array}{cc}
\UU_1 \bcdot\bnabla\uu_0 + \uu_0 \bcdot \bnabla\UU_1 & 0 \\
0 &  0
\end{array}\right]
= \LL\QQ_1 ,
\label{eq:L}
\\
\AAA_2 \qq_0      
&=
\left[\begin{array}{cc}
\UU_2 \bcdot \bnabla\uu_0  + \uu_0 \bcdot \bnabla\UU_2 & 0 \\
0 &  0
\end{array}\right]
= \LL\QQ_2,
\label{eq:N}
\\
\AAA_1^\dag \qq_0^\dag 
&= 
\left[\begin{array}{cc}
-\UU_1 \bcdot \bnabla\uu_0^\dag   + \uu_0^\dag \bcdot \bnabla\UU_1^T & 0 \\
0 &  0
\end{array}\right]
=\MM\QQ_1.
\label{eq:M}
\end{align}
%------------------------------------
Substituting into (\ref{eq:ev22}) yields
\begin{align}
 \lambda_2 
 &=  \pps{ \qq_0^\dag }{-\LL\QQ_2} 
 + \pps{\MM\QQ_1 }{(\ev_0\EE+\AAA_0)^{-1} \LL\QQ_1}
 \\
  &=  \pps{ \LL^\dag\qq_0^\dag }{-\QQ_2} 
 + \pps{\QQ_1 }{\MM^\dag(\ev_0\EE+\AAA_0)^{-1} \LL\QQ_1},
 \label{eq:tmp1}
\end{align}
where we have introduced the adjoint operators 
\begin{align}
\LL^\dag=
\left[\begin{array}{cc}
()\bcdot \bnabla\uu_0^H   - \overline\uu_0 \bcdot \bnabla()   & 0 \\
0 &  0
\end{array}\right],
\quad
\MM^\dag=
\left[\begin{array}{cc}
 -() \bcdot \bnabla\overline\uua_0   
 -() \bcdot \bnabla\overline\uua_0^T   & 0 \\
0 &  0
\end{array}\right]
\end{align}
(recall that the overbar stands for the complex conjugate).
The first term is rearranged by making use of (\ref{eq:Q2}):
\begin{align}
\ev_2 
&= \pps{ \LL^\dag\qq_0^\dag }{\AAA_0^{-1} (\UU_1 \bcdot  \bnabla \UU_1,0)^T} 
 + \pps{\QQ_1 }{\MM^\dag (\ev_0\EE+\AAA_0)^{-1} \LL\QQ_1} 
 \\
%&= \pps{{\AAA_0^\dag}^{-1} \NN^\dag\qq_0^\dag }{(\UU_1 \bcdot  \bnabla\UU_1,0)^T } 
% + \pps{\QQ_1 }{\MM^\dag (\ev_0\EE+\AAA_0)^{-1} \LL\QQ_1} 
%% \\ 
%% &=  -\pps{\QQa }{(- \UU_1 \bcdot \bnabla \UU_1,0)^T } + \pps{\QQ_1 }{\MM^\dag (\ev_0\EE+\AAA_0)^{-1} \LL\QQ_1}
%\\ 
 &= \pps{\QQa }{( \UU_1 \bcdot \bnabla \UU_1,0)^T } + \pps{\QQ_1 }{\MM^\dag (\ev_0\EE+\AAA_0)^{-1} \LL\QQ_1},
 \label{eq:tmp1}
 \end{align}
where the field $\QQa=(\UUa,\Pa)^T$ is a solution of 
\be
\AAA_0^\dag \QQa = \LL^\dag  \qq_0^\dag. 
\ee 
Since the first term is linear in $\QQa$ and quadratic in $\QQ_1$, we define another linear operator
\begin{align}
& \qquad \KK
%= \left[\begin{array}{cccc}
%\overline \Ua \partial_x & \overline\Va \partial_x & \overline\Wa \partial_x & 0 \\
%\overline\Ua \partial_y & \overline\Va \partial_y & \overline\Wa \partial_y & 0 \\
%\overline\Ua \partial_z & \overline\Va \partial_z & \overline\Wa \partial_z & 0 \\
%0 & 0 & 0 & 0
%\end{array}\right] 
= \left[\begin{array}{cc}
\overline\UUa \bcdot \bnabla()^T & 0 \\
0 &  0
\end{array}\right]
\\
\mbox{such that} \quad
&\pps{\QQa }{( \UU_1 \bcdot \bnabla \UU_1,0)^T }
= \pps{\QQ_1} {\KK \QQ_1},
\end{align}
which allows us to obtain the following expression for the second-order eigenvalue variation
\begin{align}
 \lambda_2 
 &=  \pps{\QQ_1}{ \KK \QQ_1 } 
 + \pps{\QQ_1 }{\MM^\dag (\ev_0\EE+\AAA_0)^{-1} \LL\QQ_1} .
\end{align}
Finally, the total second-order sensitivity to flow modification is:
\begin{align}
%\SSS_{2,\QQ_1} &= 
%\underbrace{\KK(\QQa)}_{\mbox{2D}} 
%+
%\underbrace{\MM^\dag (\ev_0\EE+\AAA_0)^{-1} \LL}_{\mbox{3D}}.
%% \\
%%       &=  \KK(\QQa) + \MM^\dag (\ev_0\EE+\AAA_0)^{-1} \DD(\uu_0).
%\\
\SSS_{2,\QQ_1} &= 
\KK
+
\MM^\dag (\ev_0\EE+\AAA_0)^{-1} \LL,
\quad \mbox{ such that }
\ev_2 = \pps{\QQ_1}{\SSS_{2,\QQ_1} \QQ_1}.
\label{eq:S2U}
\end{align}

From the sensitivity to flow modification (\ref{eq:S2U}), one can derive the sensitivity to control.
Let us define the prolongation operator $\PP$ from velocity-only space to velocity-pressure space such that
$\PP\UU = (\UU,0)^T$ and 
$\UU=\PP^T(\UU,0)^T$.
The second-order sensitivity to volume control reads 
\begin{align}
\SSS_{2,\CC} &=  \PP^T {\AAA_{0,\CC}^\dag}^{-1} \SSS_{2,\QQ_1} {\AAA_{0,\CC}}^{-1} \PP,
%\nonumber
%\\
%&= \PP^T {\AAA_{0,\CC}^\dag}^{-1} \left[ \KK(\QQa) + \MM^\dag (\ev_0\EE+\AAA_{0,\CC})^{-1} \LL \right] {\AAA_0}^{-1} \PP,
\quad \mbox{ such that }
\ev_2 = \pps{\CC}{\SSS_{2,\CC} \CC}.
\label{eq:S2C}
\end{align}
where $\AAA_{0,\CC}$ is defined by the volume-control-only (no wall control, $\UU_c=\00$) version of (\ref{eq:Q1}):
\be
\AAA_{0,\CC} \QQ_1 = (\CC,0)^T = \PP\CC
 \, \mbox{ in } \Omega,
\qquad
 \UU_1=\00 \, \mbox{ on } \Gamma.
\ee
Likewise, the second-order sensitivity to wall actuation reads 
\begin{align}
\SSS_{2,\UU_c} &=  \PP^T {\AAA_{0,\UU_c}^\dag}^{-1} \SSS_{2,\QQ_1} {\AAA_{0,\UU_c}}^{-1} \PP,
%\nonumber
%\\
%&= \PP^T {\AAA_{0,\UU_c}^\dag}^{-1} \left[ \KK(\QQa) + \MM^\dag (\ev_0\EE+\AAA_0)^{-1} \LL \right] {\AAA_{0,\UU_c}}^{-1} \PP,
\quad \mbox{ such that }
\ev_2 = \ppsw{\UU_c}{\SSS_{2,\UU_c} \UU_c}.
\label{eq:S2Uc}
\end{align}
where $\AAA_{0,\UU_c}$ is defined in this case by the wall-actuation-only (no volume control, $\CC=\00$) problem:
\begin{align}
& \AAA_{0,\UU_c} \QQ_1 = \00 \, \mbox{ in } \Omega,
\qquad
\UU_1=\UU_c \, \mbox{ on } \Gamma.
\end{align}

%----------------------------------------
%----------------------------------------
%----------------------------------------
\section{Spanwise-periodic sensitivity, simplification to a 2D problem}
\label{sec:apdx_spanwise_sensitivity}

The second-order sensitivity operators (\ref{eq:S2C}) and (\ref{eq:S2Uc}) depend on the spanwise coordinate $z$ via the spanwise-periodic flow modification $\QQ_1$. 
We now derive \textit{reduced} $z$-independent, yet exact, expressions of the sensitivity operators. 
As detailed below, this expression makes it possible to  evaluate the eigenvalue variation $\ev_2$ and determine the optimal spanwise-periodic controls $\CC$ and $\UU_c$ using  only 2D fields, making these operations significantly more computationally affordable than with 3D fields.

We consider, without loss of generality, the following harmonic wall forcing on $\Gamma$ and harmonic volume forcing in $\Omega$:
\be 
\UU_c(x,y,z) =  \left( \begin{array}{c}
\widetilde U_{c}(x,y) \cbz \\ \widetilde V_{c}(x,y) \cbz \\ \widetilde W_{c}(x,y) \sbz
\end{array} \right),
\quad
\CC(x,y,z) =  \left( \begin{array}{c}
\widetilde C_x(x,y) \cbz \\ \widetilde C_y(x,y) \cbz \\ \widetilde C_z(x,y) \sbz 
\end{array} \right).
\label{eq:harm_U_C}
\ee
This might seem more restrictive than 
\begin{align}
\UU_c(x,y,z) &= \widetilde\UU_{c}^{c}(x,y) \cbz + \widetilde\UU_{c}^{s}(x,y) \sbz,
\label{eq:Uc_full}
\\
\CC(x,y,z) &= \widetilde\CC^{c}(x,y) \cbz + \widetilde\CC^{s}(x,y) \sbz,
\label{eq:C_full}
\end{align}
(or an equivalent complex formulation),
but this is actually not the case, as will be touched upon later in this section.
With the spanwise-harmonic control (\ref{eq:harm_U_C}), the flow response at first order $\epsilon^1$ is 
\begin{align}
\QQ_1=
\left(\begin{array}{c}
\widetilde U_1(x,y) \cbz \\
\widetilde V_1(x,y) \cbz \\
\widetilde W_1(x,y) \sbz \\
\widetilde P_1(x,y) \cbz
\end{array}\right),
\end{align}
and the first-order problem 
$\AAA_0 \QQ_1 = (\CC,0)^T$ 
can be rewritten in the \textit{reduced} form
$\widetilde \AAA_0 \widetilde\QQ_1 = (\widetilde\CC,0)^T,$
where 
\begin{align}
{\widetilde \AAA_0} 
= \left[\begin{array}{cccc}
 U_0\ddd_x +V_0\ddd_y+\ddd_x U_0 - \widetilde D  & \ddd_y U_0 & 0 &   \ddd_x \\
\ddd_x V_0 & U_0\ddd_x +V_0\ddd_y+\ddd_y V_0 - \widetilde D & 0 &   \ddd_y \\
0 & 0 &  U_0\ddd_x +V_0\ddd_y  - \widetilde D & -\beta \\
\ddd_x & \ddd_y & \beta & 0 
\end{array}\right],
\end{align}
 \begin{align}
\widetilde D =  \nnu (\ddd_{xx}+\ddd_{yy}-\beta^2).
\end{align}

Next, we note that the  forcing term of the problem 
$\AAA_0 \QQ_2 = (- \UU_1 \bcdot  \bnabla\UU_1,0)^T$ 
is the sum of a 2D term (wavenumber 0) and a 3D term (wavenumber $2\beta$): 
\begin{align}
-\UU_1 \bcdot  \bnabla\UU_1
&=
\ff^{2D}(x,y) + \ff^{3D}(x,y,z),
\\
\ff^{2D} 
&= 
-\frac{1}{2} 
(\widetilde U_1 \ddd_x + \widetilde V_1 \ddd_y - \beta \widetilde W_1 )
\left(\begin{array}{c}
 \widetilde U_1 
\\  \widetilde V_1
\\ 
0
\end{array}\right),
\\
\ff^{3D}
&=
-\frac{1}{2} 
(\widetilde U_1 \ddd_x + \widetilde V_1 \ddd_y + \beta \widetilde W_1 )
\left(\begin{array}{c}
 \widetilde U_1 \cos(2\beta z)
\\ 
 \widetilde V_1 \cos(2\beta z)
\\
 \widetilde W_1 \sin(2\beta z)
\end{array}\right).
\end{align}
At second order $\epsilon^2$, 
the flow response can therefore be decomposed into
the response to each of the above two forcing terms:
\begin{align}
\QQ_2 &= \QQ_2^{2D}(x,y) + \QQ_2^{3D}(x,y,z),
\\
\AAA_0 \QQ_2^{2D} &= (\ff^{2D},0)^T,
\quad
\AAA_0 \QQ_2^{3D}  = (\ff^{3D},0)^T.
\end{align}
The 2D response 
$\QQ_2^{2D}(x,y) = (U_2^{2D}, V_2^{2D}, 0, P_2^{2D})^T$
can also be written as a solution of the \textit{reduced} equation
$\widehat \AAA_0^{2D} \QQ_2^{2D} = (\ff^{2D},0)^T$, 
with purely 2D operators:
\begin{align}
{\widehat \AAA_0^{2D}} = \left[\begin{array}{cccc}
 U_0\ddd_x +V_0\ddd_y+\ddd_x U_0 - \widehat D  & \ddd_y U_0 & 0 &   \ddd_x \\
\ddd_x V_0 & U_0\ddd_x +V_0\ddd_y+\ddd_y V_0 - \widehat D & 0 &   \ddd_y \\
0 & 0 &  0 & 0 \\
\ddd_x & \ddd_y & 0 & 0 
\end{array}\right],
\label{eq:A02D}
\end{align}
\begin{align}
\widehat D =  \nnu (\ddd_{xx}+\ddd_{yy}).
\end{align}
The 3D forcing term induces a 3D response $\QQ_2^{3D}$ that is a solution of $\AAA_0 \QQ_2^{3D} = (\ff^{3D}, 0)^T$ and that is $z$-periodic of wavelength $\pi/\beta$.
Therefore, its contribution to 
$\ev_2 = \pps{ \qq_0^\dag }{-\AAA_2\qq_0 + \ldots}$ 
in (\ref{eq:ev2}) will average out to zero.
In other words, the contribution of $\QQ_2$ only comes from the mean flow correction $\QQ_2^{2D}$, not from the harmonic field $\QQ_2^{3D}$.

Finally, for spanwise-periodic control,
 the second-order sensitivity operator (\ref{eq:S2U})
reduces to
\begin{align}
\widetilde \SSS_{2,\widetilde\QQ_1} &= 
\widetilde \KK
+
\widetilde{\MM}^\dag (\ev_0\EE+\widetilde\AAA_0)^{-1} \widetilde\LL,
\quad \mbox{ such that }
\ev_2 = \ps{\widetilde\QQ_1}{\widetilde\SSS_{2,\widetilde\QQ_1} \widetilde\QQ_1},
\label{eq:S2U_red}
\end{align}
where 
\begin{align}
\widetilde \LL &= \LL,
%\quad
%\widetilde \LL^\dag = \LL^\dag,
\quad
\widetilde \MM^\dag = \MM^\dag,
%\\
\quad
\widetilde \KK = 
\left[\begin{array}{cccc}
\overline{ \widetilde \Ua} \partial_x & \overline{ \widetilde \Va} \partial_x & 0 & 0 \\
\overline{ \widetilde \Ua} \partial_y & \overline{ \widetilde \Va} \partial_y & 0 & 0 \\
-\beta \overline{ \widetilde \Ua}  & -\beta \overline{ \widetilde \Va}  & 0 & 0 \\
0 & 0 & 0 & 0
\end{array}\right],
\\ 
\widetilde \QQa &= \left( \widetilde \Ua, \widetilde \Va, 0, \widetilde \Pa \right)^T
\mbox{ is a solution of }
\widehat \AAA_0^{2D\dag} \widetilde\QQa = \widetilde\LL^\dag  \qq_0^\dag.
\end{align}
Similarly, the second-order sensitivity operators (\ref{eq:S2C}) and (\ref{eq:S2Uc}) reduce to 
\begin{align}
\widetilde \SSS_{2,\widetilde\CC} &= 
\PP^T 
\left. \widetilde{\AAA}^\dag_{0,\CC} \right.^{-1} 
\widetilde \SSS_{2,\widetilde\QQ_1} 
\left. \widetilde{\AAA}_{0,\CC} \right.^{-1} 
\PP,
\quad \mbox{ such that }
\ev_2 = \ps{\widetilde \CC}{\widetilde\SSS_{2,\widetilde\CC} \widetilde\CC},
\label{eq:S2C_red}
\\
\widetilde \SSS_{2,\widetilde\UU_c} 
&=
\PP^T 
\left. \widetilde{\AAA}_{0,\UU_c}^\dag \right.^{-1} \widetilde \SSS_{2,\widetilde\QQ_1} 
\left. \widetilde{\AAA}_{0,\UU_c} \right.^{-1} 
\PP,
\quad \mbox{ such that }
\ev_2 = \psw{\widetilde \UU_c}{\widetilde \SSS_{2,\widetilde\UU_c} \widetilde \UU_c}.
\label{eq:S2Uc_red}
\end{align}

Coming back to the specific choice of (\ref{eq:harm_U_C}) as a control,
detailed calculations show that using 
(\ref{eq:Uc_full})-(\ref{eq:C_full}) instead does not affect the results, as far as the second-order eigenvalue variation (\ref{eq:ev2}) is concerned.
Indeed, the two control fields 
\be 
\left( \begin{array}{c}
\widetilde U_{c}^c(x,y) \cbz \\ 
\widetilde V_{c}^c(x,y) \cbz \\ 
\widetilde W_{c}^s(x,y) \sbz
\end{array} \right)
\quad\mbox{and}\quad
\left( \begin{array}{c}
\widetilde U_{c}^s(x,y) \sbz \\ 
\widetilde V_{c}^s(x,y) \sbz \\ 
\widetilde W_{c}^c(x,y) \cbz
\end{array} \right)
\label{eq:Uc_full_2}
\ee
do interact in  $\ev_2$ through quadratic terms such as 
$\UU_1 \bcdot \bnabla\UU_1$ and 
$\AAA_1 \qq_1$; 
however, this interaction induces new terms that do not affect $\ev_2$,
(i)~either because they are spanwise periodic and average out,
(ii)~or because they appear on $z$ components and cannot contribute in 
$\pps{ \qq_0^\dag }{\AAA_2 \qq_0}$
and
$\pps{ \qq_0^\dag }{\AAA_1 \qq_1}$
since $\qq_0$ and $\qqa_0$ are 2D (spanwise invariant, no spanwise component).
In other words, the two fields (\ref{eq:Uc_full_2}) contribute independently to $\ev_2$. 
Looking for an optimal control with the additional degrees of freedom 
(\ref{eq:Uc_full})-(\ref{eq:C_full}) returns the same field twice (up to sign differences), therefore optimizing for (\ref{eq:harm_U_C}) is sufficient. 

Of course, it is still possible to actually implement a control of the form (\ref{eq:Uc_full})-(\ref{eq:C_full}), or a similar form for the radius $R_1$ in the case of wall deformation.
In particular, this allows the implementation of a ``helical''  control or  deformation. Indeed,  a ``traveling wave'' in $(z,\theta)$ can be decomposed as the sum of two different ``varicose'' controls/deformations (``standing waves''): 
\begin{align}
\left( \begin{array}{c}
U \cos(\beta z + m\theta) \\ 
V \cos(\beta z + m\theta) \\ 
W \sin(\beta z + m\theta)
\end{array} \right)
&=
\left( \begin{array}{c}
U \cos(m\theta) \cbz \\ 
V \cos(m\theta) \cbz \\ 
W \cos(m\theta) \sbz
\end{array} \right)
+
\left( \begin{array}{c}
-U \sin(m\theta) \sbz \\ 
-V \sin(m\theta) \sbz \\ 
W \sin(m\theta) \cbz
\end{array} \right)
\nonumber
\\
&=
\left( \begin{array}{c}
\widetilde U_c^c(\theta) \cbz \\ 
\widetilde V_c^c(\theta) \cbz \\ 
\widetilde W_c^s(\theta) \sbz
\end{array} \right)
+
\left( \begin{array}{c}
\widetilde U_c^s(\theta) \sbz \\ 
\widetilde V_c^s(\theta) \sbz \\ 
\widetilde W_c^c(\theta) \cbz
\end{array} \right),
\end{align}
\begin{align}
R \cos(\beta z + m\theta) 
&=
R\cos(m\theta) \cbz - R\sin(m\theta) \sbz
\nonumber
\\
&=
\widetilde R_1^c(\theta) \cbz + \widetilde R_1^s(\theta) \sbz.
\end{align}
However, because at least one of the two varicose controls/deformations is not the optimal control/deformation (for instance, $\widetilde R_1^c$ differs necessarily from $\widetilde R_1^s$), the resulting $\ev_2$ will necessarily be sub-optimal, i.e. smaller than the optimal $\ev_2$ obtained with the optimal control/deformation of same norm.

%----------------------------------------
%----------------------------------------
%----------------------------------------
\section{optimization}
\label{sec:apdx_optimization}

On a technical note, reduced 2D second-order sensitivity operators (\ref{eq:S2C_red})-(\ref{eq:S2Uc_red}) involve several inversion operations; therefore, although much smaller that their full 3D counterparts $\SSS_{2,*}$, in practice the operators $\widetilde\SSS_{2,*}$ are never formed explicitly.
Instead, the eigenvalue problems (\ref{eq:maxlam2rC}), (\ref{eq:maxlam2rUc}) etc. are solved iteratively by evaluating  matrix-vector products (i.e. computing the action of operators on vectors) and solving linear systems of equations.
Specifically, operators $\widetilde\AAA_{0}$, $\widetilde\AAA_{0}^\dag$, $\widetilde\KK$, $\widetilde\LL$, $\widetilde\MM^\dag$ etc. are first built with \textit{FreeFem++} and subsequently imported in \textit{Matlab}, where lower-upper (LU) decompositions of the operators to be inverted ($\widetilde\AAA_{0}$, $\ev_0\EE+\widetilde\AAA_0$, etc.) are performed once for all before solving the eigenvalue problems. 
During the eigenvalue problem resolution, whenever evaluating the action of $\widetilde\SSS_{2,*}$ on a vector $\xx$ is needed, the action of inverse operators is computed by substitution using the triangular LU factors (i.e. for instance if $\widetilde\AAA_{0}=\LL_\AAA \UU_\AAA$, then 
the vector 
$\yy = \widetilde\AAA_{0}^{-1} \xx = \UU_\AAA^{-1} \LL_\AAA^{-1} \xx$ 
is computed by solving 
$\LL_\AAA \zz=\xx$ for $\zz$, and 
$\UU_\AAA \yy=\zz$ for $\yy$).

Note, however, that the need to take the real or imaginary part and actually compute
$\,\widetilde\SSS_{2,*,r} \xx\,$ or 
$\,\widetilde\SSS_{2,*,i} \xx\,$ rather than
$\,\widetilde\SSS_{2,*} \xx\,$ 
is not straightforward since the operator 
$\widetilde\SSS_{2,*}$ is not formed explicitly. 
This is circumvented by noting that the following relations hold:
\begin{align} 
&\widetilde \SSS_{2,*,r}\xx 
= \frac{1}{2} \left( \widetilde \SSS_{2,*}\xx+\overline{\widetilde \SSS_{2,*}\overline\xx} \right),
\quad
\widetilde \SSS_{2,*,r}^T\xx 
= \frac{1}{2} \left( \widetilde \SSS_{2,*}^T\xx+\overline{\widetilde \SSS_{2,*}^T\overline\xx} \right),
\\%---------------------------------------
&\widetilde \SSS_{2,*,i}\xx 
= \frac{1}{2i} \left( \widetilde \SSS_{2,*}\xx - \overline{\widetilde \SSS_{2,*}\overline\xx} \right),
\quad
\widetilde \SSS_{2,*,i}^T\xx 
= \frac{1}{2i} \left( \widetilde \SSS_{2,*}^T\xx - \overline{\widetilde \SSS_{2,*}^T\overline\xx} \right).
\end{align} 
Using these relations, one only needs to evaluate matrix-vector \textit{products} (and vector complex conjugates), 
rather than evaluating the real/imaginary part of \textit{operators}.

%----------------------------------------
%----------------------------------------
%----------------------------------------
\section{Optimal volume control for stabilization}
\label{sec:apdx_volume_control}

%--- plot_mono_C1_Q1.m
%--- plot_mono_Uabf.m
\begin{figure}
\centerline{   
 \begin{overpic}[trim=5mm 14mm 0mm 17mm, clip, height=4.1cm,tics=10]{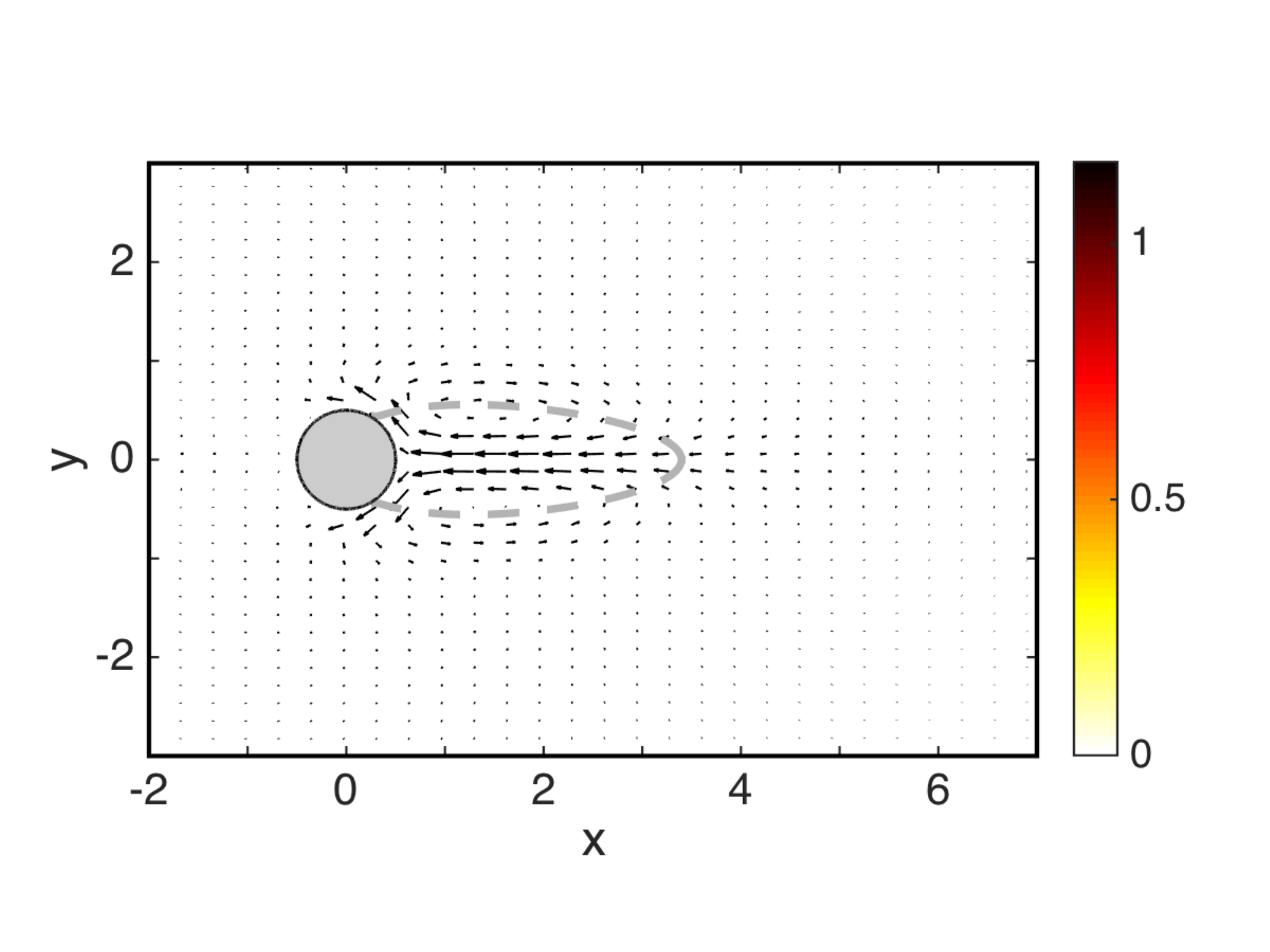}     
      \put(-2,54){$(a)$} 
   \end{overpic}     
   \begin{overpic}[trim=15mm 14mm 5mm 17mm, clip, height=4.1cm,tics=10]{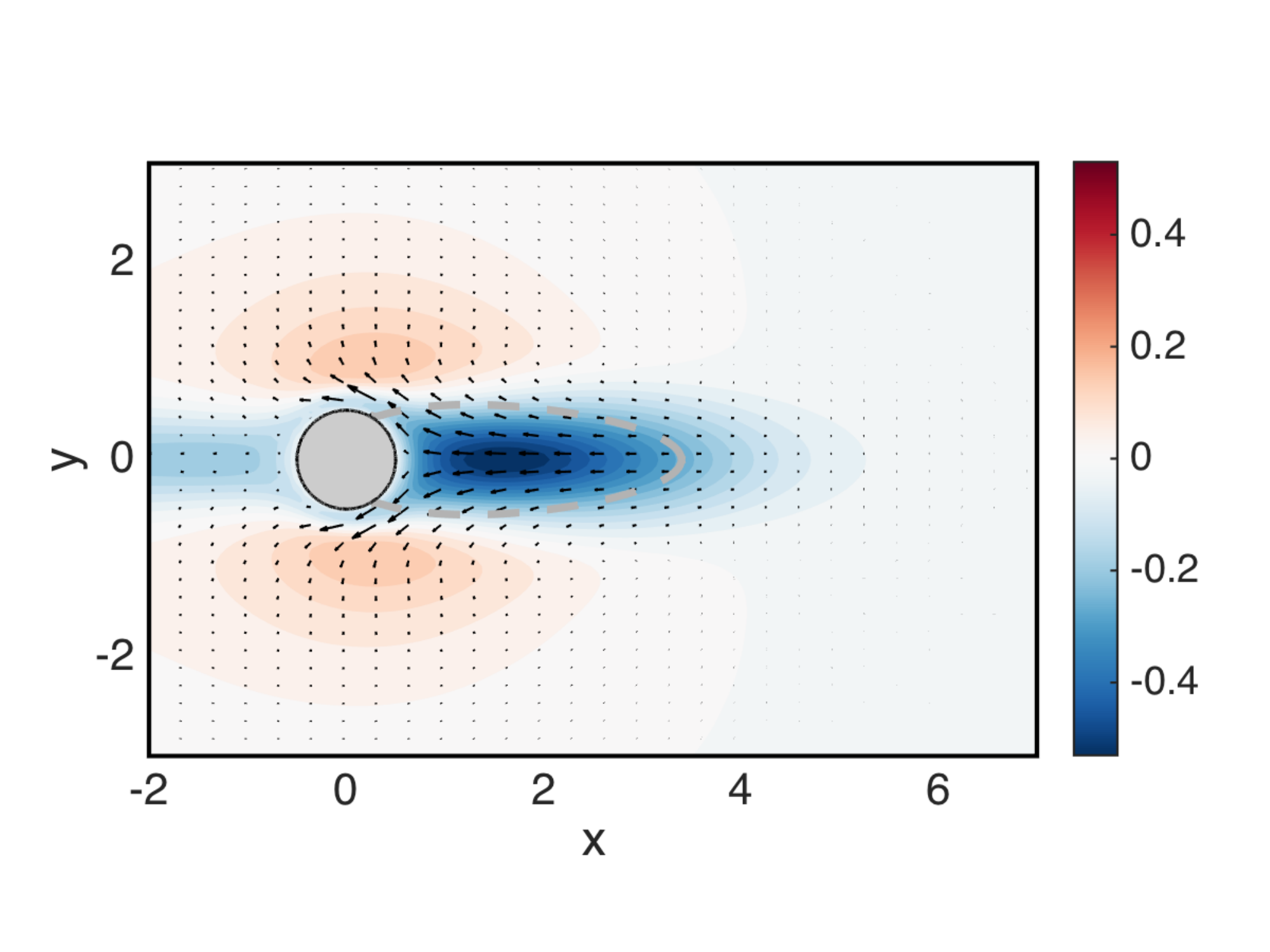}     
      \put(-8,58){$(b)$} 
   \end{overpic}      
   \begin{overpic}[trim=23mm 0 5mm 0, clip, height=4.05cm,tics=10]{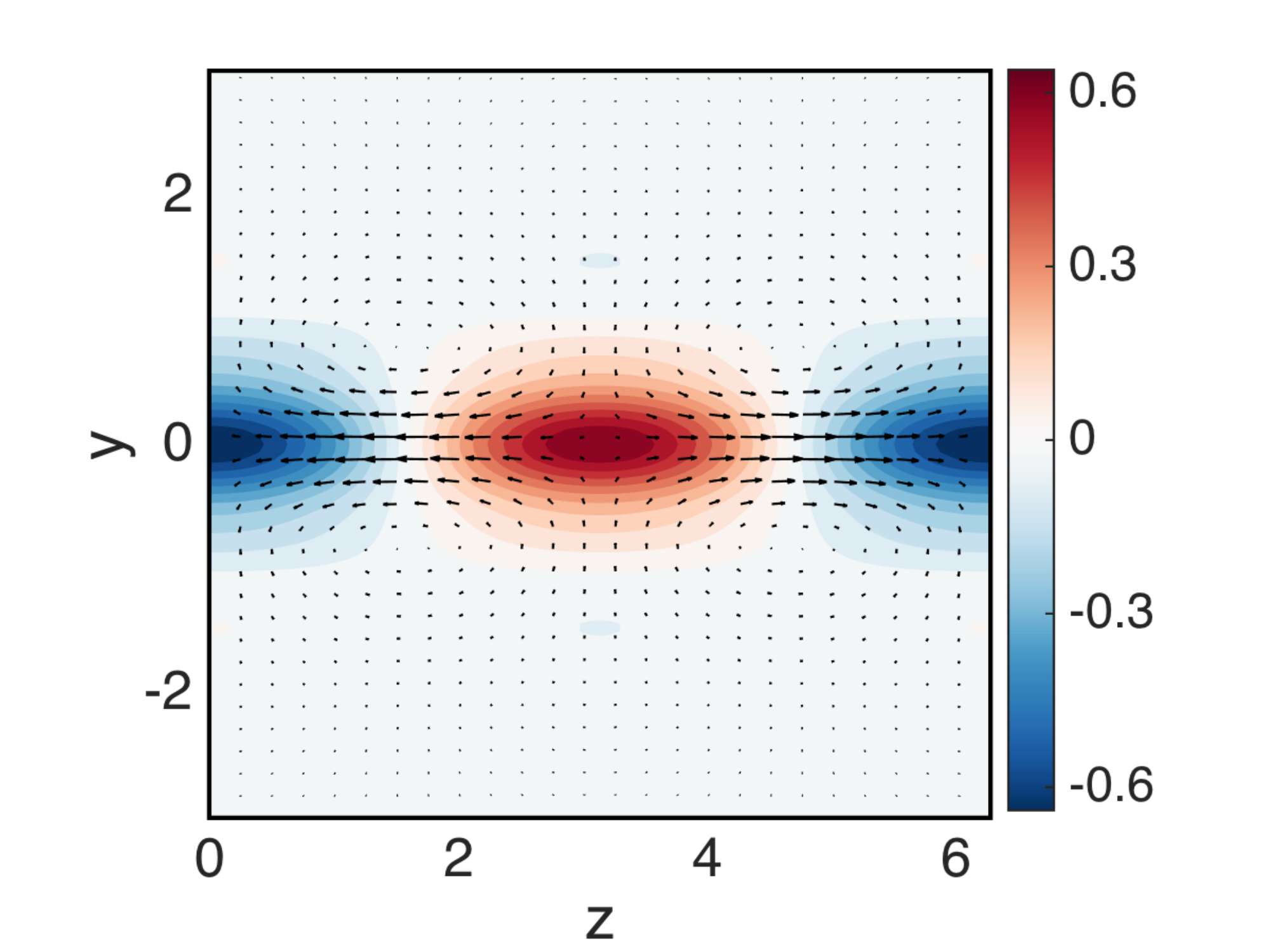}
   \end{overpic}   
}
%
%--- plot_mono_C1_Q1.m
%--- plot_mono_C1_Q1_Q2.m
\centerline{   
   \begin{overpic}[trim=5mm 14mm 0mm 17mm, clip, height=4.1cm,tics=10]{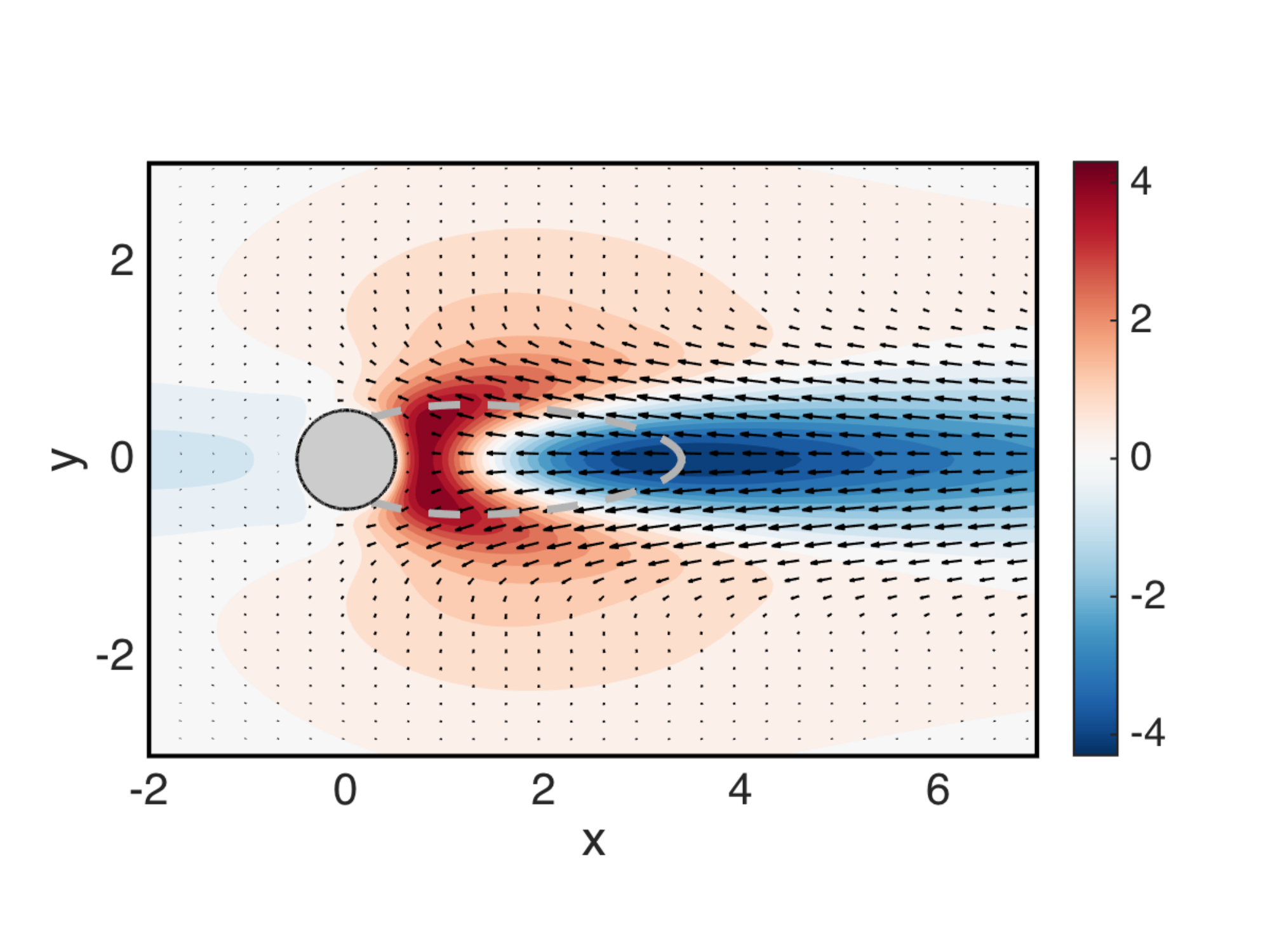}
      \put(-2,52){$(c)$} 
   \end{overpic}      
   \begin{overpic}[trim=23mm 0 5mm 0, clip, height=4.05cm,tics=10]{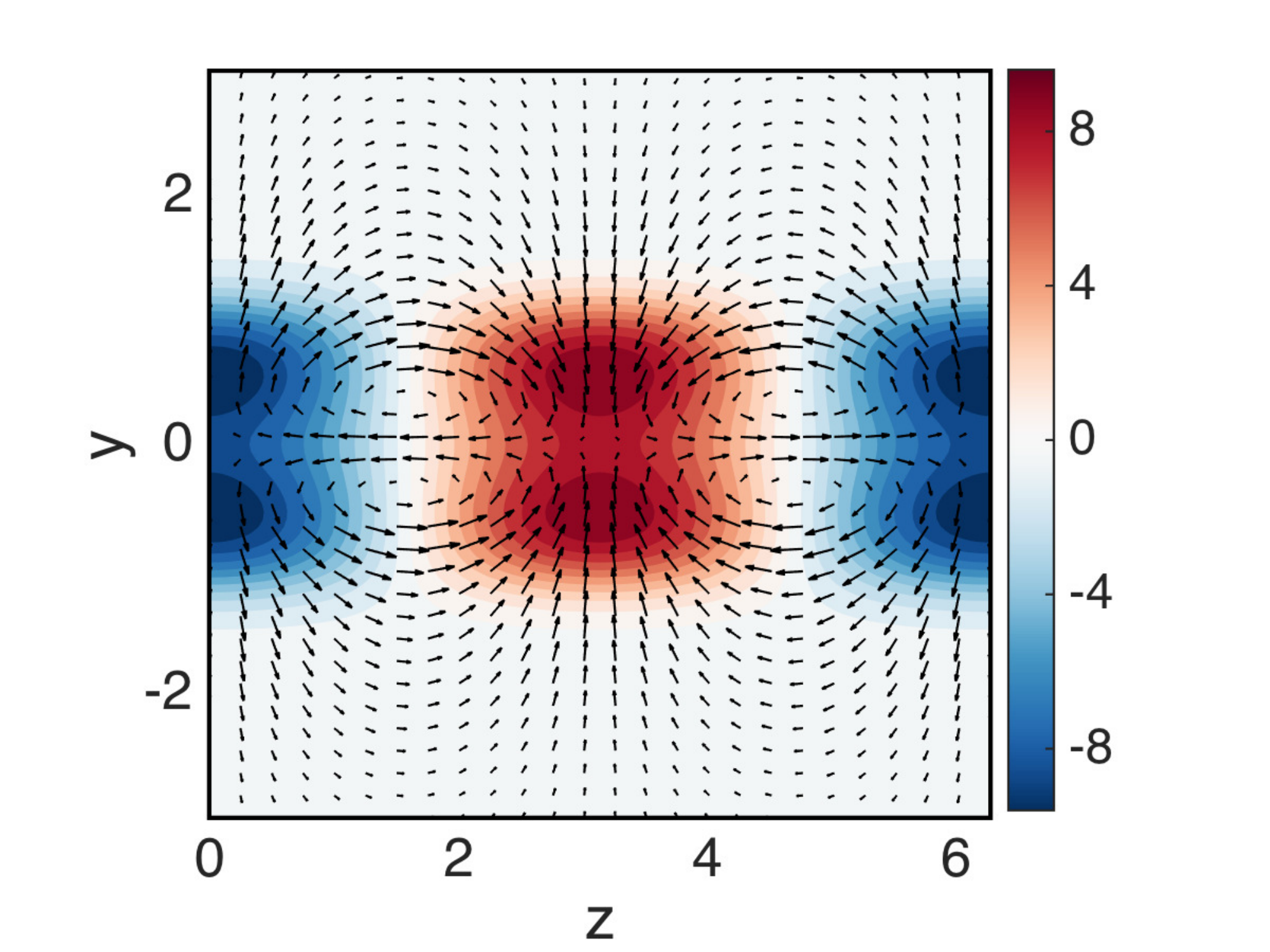}
   \end{overpic} 
   \begin{overpic}[trim=15mm 14mm 5mm 17mm, clip, height=4.1cm,tics=10]{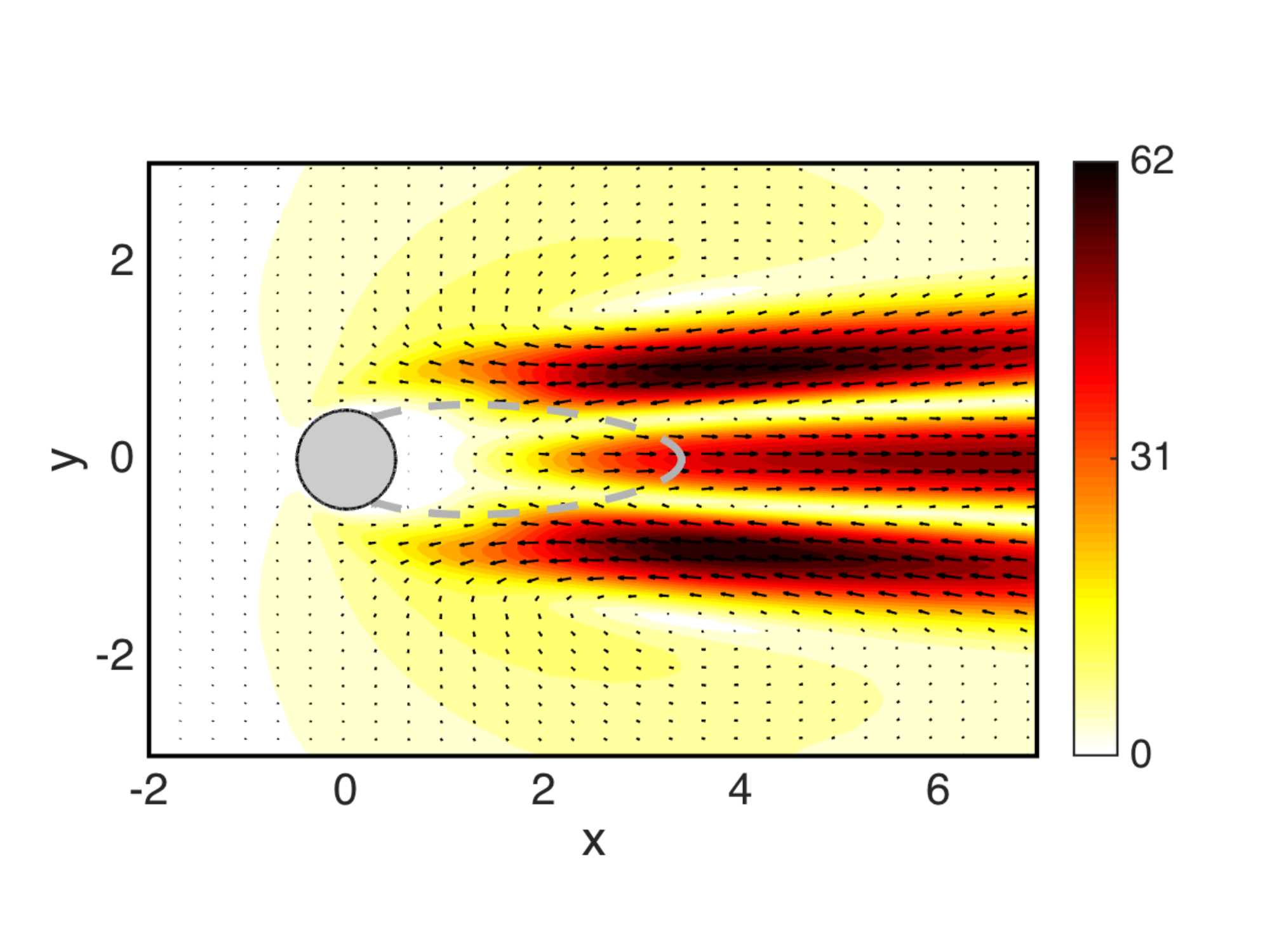}     
   \end{overpic}   
}
\caption{
$(a)$
2D sensitivity of the leading eigenvalue's growth rate to 2D (spanwise-invariant) volume control (arrows show the $x$ and $y$ components, while the $z$ component is zero by definition).
$(b)$ 
Optimal stabilizing spanwise-periodic volume control $\CC$ at $\Rey=50$, $\beta=1$.
Left, right: cuts at $z=0$ and $x=2$, respectively, showing the in-plane vector field and contours of the out-of-plane component.
$(c)$
Flow modification induced by 
the optimal stabilizing  volume control of  panel $b$.
Left, middle: first-order modification $\UU_1(x,y,z)$ (spanwise harmonic) at $z=0$ and  $x=2$, respectively, shown with in-plane velocity vector fields and contours of out-of-plane velocity.
Right: induced mean flow correction $\UU_2^{2D} = ( U_2^{2D}(x,y), V_2^{2D}(x,y), 0)^T$ (spanwise invariant), shown with vector field $(U_2^{2D},V_2^{2D})^T$ and contours of velocity magnitude.  
}  
\label{fig:opt_C1_stab_and_2D_sens}
\end{figure}

For the sake of completeness, we briefly comment here on volume control for stabilization.
Figure~\ref{fig:opt_C1_stab_and_2D_sens}$(b)$ shows the optimal volume control for $\beta=1$.
Interestingly, the $x$ and $y$ components are  reminiscent of the 2D growth rate sensitivity [Fig.~\ref{fig:opt_C1_stab_and_2D_sens}$(a)$; see also \cite{Marquet08cyl}], suggesting that similar flow regions are sensitive to control and are involved in similar stabilization mechanisms. 
This is remarkable, considering that the spanwise-harmonic optimal control is harmonic in $z$ and is therefore alternatively similar to the optimal \textit{stabilizing} (at $z=2n\pi/\beta$, like in Fig.~\ref{fig:opt_C1_stab_and_2D_sens}$(b)$), and similar to the optimal \textit{destabilizing} 2D control (at $z=(2n+1)\pi/\beta$).

The flow modification [Fig.~\ref{fig:opt_C1_stab_and_2D_sens}$(c)$] induced by the optimal stabilizing spanwise-periodic volume control has a spatial structure very similar to that of the flow modification induced by the optimal stabilizing wall control [Fig.~\ref{fig:Q1_from_opt_Uw_stab}$(b)$]. This suggests that optimal stabilizing varicose streaks are a robust feature of the cylinder flow.

%----------------------------------------
%----------------------------------------
%----------------------------------------
\section{Small-$\beta$ limit}
\label{sec:small_beta}

%--- plot_eigs3D_unctrl2D_nice.m
\begin{figure}
\centerline{   
\begin{overpic}[trim=18mm 82mm 23mm 85mm, clip=true, width=14cm, tics=10]{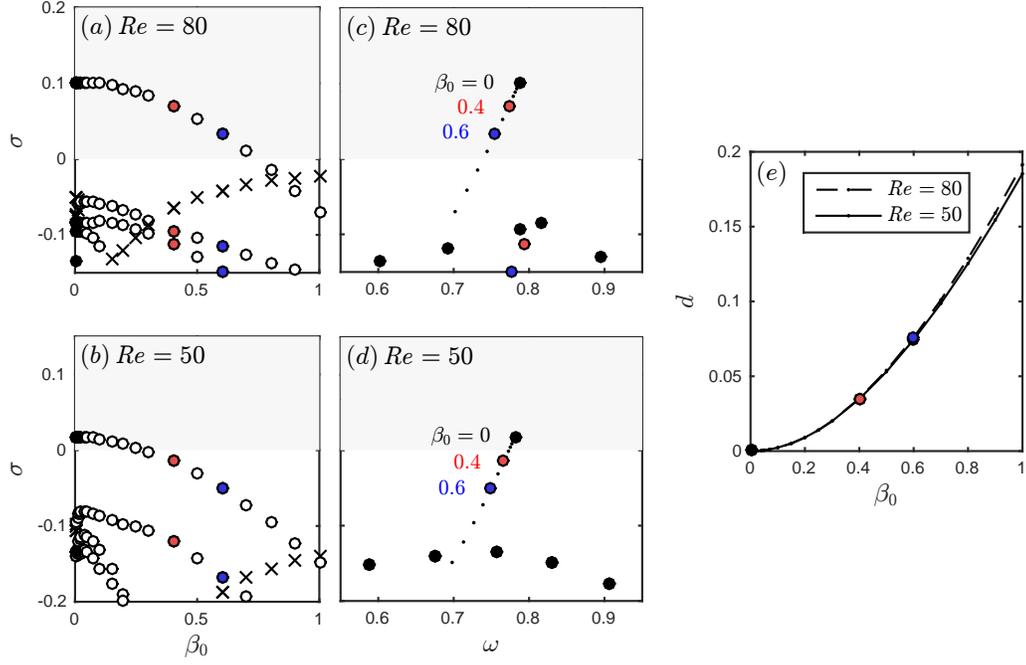}     
   \end{overpic}      
}
\caption{
$(a-d)$~3D eigenvalues of the uncontrolled 2D base flow:
$(a,b)$~growth rate vs. spanwise wavenumber 
(circles: unsteady eigenmodes $\omega\neq0$; 
crosses: steady eigenmodes $\omega=0$);
$(c,d)$~growth rate vs. frequency (close up of the leading eigenvalue).
$(a,c)$~$\Rey=80$;
$(b,d)$~$\Rey=50$.
$(e)$~Minimum distance in the complex plane $(\sigma,\omega)$ from 3D eigenvalues ($\beta_0>0$) to the 2D leading eigenvalue ($\beta_0=0$).
In all panels, spanwise wavenumbers $\beta_0=0$, 0.4 and 0.6 are shown in black, red and blue, respectively. 
} 
\label{fig:small_beta}
\end{figure}

Figure~\ref{fig:small_beta} shows the 3D eigenvalues of the uncontrolled 2D flow at two different Reynolds numbers, $\Rey=50$ and $80$, calculated 
for small-amplitude perturbations of the form
$ \qq(x,y) e^{i\beta_0 z} e^{\ev t} $ 
(numerical method similar to that described in section~\ref{sec:num2D}).
As the spanwise wavenumber $\beta_0$ increases, the leading unstable eigenmode can be followed continuously. 
Along this branch, the growth rate and frequency decrease.
The minimal distance $d$ from all 3D eigenvalues to the 2D leading eigenvalue increases with $\beta_0$ (approximately like $\beta_0^2$), and takes sensibly the same values for $\Rey=50$ and $80$, as shown in Fig.~\ref{fig:small_beta}$(e)$.
Because of the requirement $\epsilon<d$ for the expansion (\ref{eq:exp}) to remain valid,
this curve gives the maximal control amplitude $\epsilon$ allowed for a given control wavenumber $\beta$ (or equivalently the minimal $\beta$ allowed for a given $\epsilon$).
For instance, at $\Rey=50$, one can read 
$\epsilon \lesssim 0.035$ for $\beta=0.4$,  
$\epsilon \lesssim 0.075$ for $\beta=0.6$ and
$\epsilon \lesssim 0.185$ for $\beta=1$.

%\newpage %Just because of unusual number of tables stacked at end
%---------------------------------------
%---------------------------------------
%---------------------------------------
\bibliographystyle{apsrev}

%\bibliography{test/apssamp}
%\bibliography{../2ndorder_bibli\bibliography{/Users/eboujo/Documents/_PAPERS_and_THESIS/ALL.bib}

\end{document}